# Influence of metal nanoparticles on whispering gallery modes of a microcavity

*Tulika Agrawal*[1], *Venkata R. Dantham*[2], *Shubhayan Bhattacharya*[3], *Aneesh V. Veluthandath*[4], *Stephen Arnold*[5,6] *and Prem B. Bisht*[1*]

[1]Department of Physics, Indian Institute of Technology Madras, Tamil Nadu, India
[2]Department of Physics, Indian Institute of Technology Patna, Bihar, India
[3]Applied Materials Inc., Bangalore, India
[4]Optoelectronics Research Centre, University of Southampton, Southampton, UK
[5]New York University Tandon School of Engineering, New York, USA
[6]Visiting Faculty Fellow, Indian Institute of Technology Madras, Tamil Nadu, India
*Email Address: bisht@iitm.ac.in

**Abstract:** Photonic microcavities offer high energy density and low mode volume and attract special attention due to wide range of applications. Subsequent developments in the area of plasmonic nanocavities have also opened up applications despite their high absorptive properties. In the last two decades, photonic-plasmonic hybrid microcavities have been investigated extensively. In this article, we have conducted a review of the literature on photonic and plasmonic systems individually as well as in hybrid systems. Along with the underlying physics, applications such as cavity quantum electrodynamics, lasing, sensing, nonlinear optics and Raman spectroscopy have been reviewed. The outputs of the systems for each applications have been discussed and compared for photonic and hybrid systems. This review will provide a thorough understanding of the effects of optical cavities and plasmonics and their applications.

## 1 Introduction

Depending upon the inherent properties of a material and the strength of the interacting electromagnetic (EM) radiation, the properties of the system can be modified. The light-matter interaction [1] can be enhanced by using various techniques [2, 3, 4]. This is achieved by either trapping the material or light for a longer period. For example, an optical tweezer is used to trap and manipulate the motion of the particles/biological cells [5], while a cavity is used to trap the light [6]. By trapping light in a cavity, one can enhance the binding energy for particle trapping by orders of magnitude [7]. Several types of the cavities have been explored including Fabry-Perot (FP) cavity [8], whispering gallery mode (WGM) cavity [9], and photonic crystal [10] (Fig. 1).

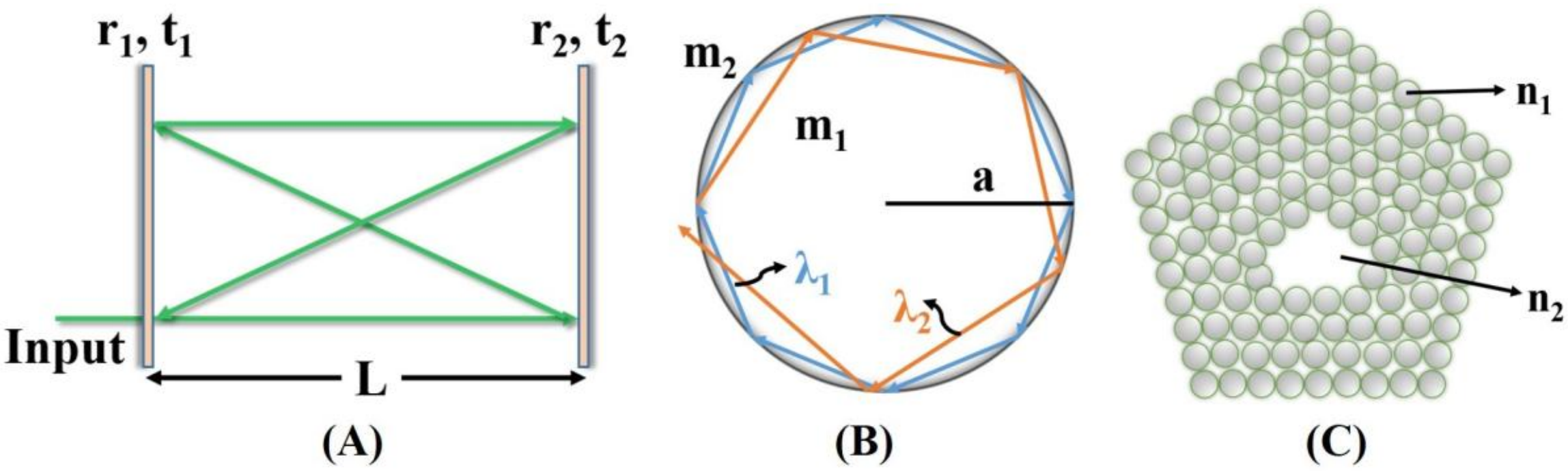

Figure 1: (A) Fabry-Perot cavity. $L$ is the length of the cavity, $r_1$, $t_1$ and $r_2$, $t_2$ are the reflectivity and transmittivity of the mirror 1 and 2, respectively. (B) WGM cavity. $m_1$ and $m_2$ are refractive indices of medium and cavity, respectively, $a$ is the radius of the cavity, $\lambda_1$ and $\lambda_2$ are the resonating and non-resonating wavelengths, respectively. (C) Photonic crystal. $n_1$ and $n_2$ are the refractive indices of the holes and core, respectively.

The term WGM was first introduced by British physicist John William Strutt (Lord Rayleigh) in his paper "The problem of the whispering gallery" with reference to the propagation of the acoustic waves along the peripheral walls of symmetric dome of the cathedral [11]. Analogous to this effect, light can also get trapped in a structure having at least one axis of symmetry of the dimension of the order of the wavelength. These microstructures are known as optical microcavities. Light bouncing inside the microcavity and returning to its original position in phase gives rise to resonant modes. These modes can also appear in the form of ripples in the photoluminescence, Raman, transmission, and elastic scattering spectra. Symmetric shapes such as spheres, disks and rings [12, 13, 14], capillaries and fibers [15, 16], bubbles [17], toroids [18] and bottles (MBR) [19] exhibit WGMs. A cavity without any emitter is known as a passive cavity and is mainly used in integrated photonics [20]. When it is coupled with the emitter, it is known as the active cavity. It is used in lasing [13, 21] and as single photon sources [22]. A number of research articles exist on the applications based on WGMs [23, 24, 25, 26].

Metal nanoparticles (NPs) exhibit resonant modes due to the coherent oscillations of free electrons present on their surface, known as localized surface plasmon resonances (LSPRs) [27]. Unlike bulk metal which generate propagating plasmon modes, metal NPs generate localized modes leading to strong EM field confinement. This confinement is known as hotspots present in the vicinity of the NPs. LSPR modes are also sensitive to the material, geometry, orientation, and surrounding [28, 29]. As a result, plasmonics has applications in various fields such as sensing [30], surface enhanced Raman spectroscopy (SERS) [31], photodynamic therapy (PDT) [32], and photocatalysis [33]. Some other applications of plasmonics have been reviewed earlier [34, 35, 36, 37, 38].

When the metal NPs are coupled with WGM cavity, the cavity modes interact with the LSPR mode. This resultant photonic-plasmonic hybrid cavity has ultralow mode volume, even though it has lower Q value. Recent studies have been mainly focussed on the hybrid cavity as they combine the benefits of WGMs and LSPRs [39, 40, 41, 42, 43]. Although various review articles already exist in literature for hybrid cavity [44, 45, 46, 47], only a handful of articles have combined the usage of photonic and hybrid cavities for applications along with the theory [48].

In view of the above, we have compiled the works done in some of the important applications using bare and hybrid microcavities. The paper has been organized as follows. Section 2 reviews the established theory of WGMs and plasmonics. Section 3 gives the applications of bare and hybrid microcavities. Section 3.1 gives the CQED aspect and section 3.2 gives the lasing aspect of the bare and hybrid cavities. The three areas of sensing, namely, biological, chemical, and physical sensing are given in section 3.3. Nonlinear optical and Raman spectroscopic studies in microcavities are given in sections 3.4 and 3.5, respectively. Section 4 gives the summary of the paper along with the future outlook of this field.

# 2 Theory

## 2.1 Photonic microcavity

Classically, WGMs can be explained by using geometrical optics. A ray passing from a denser medium to a rarer medium at an incident angle greater than the critical angle gets totally reflected. Similar to the FP cavity [49], the ability of the cavity to confine light can be determined by reflection ($r$) and transmission ($t$) coefficient given by [50]

$$r = \frac{\chi m_1 cos\theta_i - m_2 cos\theta_t}{\chi m_1 cos\theta_i + m_2 cos\theta_t} \tag{1}$$

$$t = \frac{2[\chi m_1 m_2 cos\theta_i Re(cos\theta_t)]^{\frac{1}{2}}}{\chi m_1 cos\theta_i + m_2 cos\theta_t} \tag{2}$$

where $\theta_i$ and $\theta_t$ are the angles of incidence and refraction, respectively, $m_1$ and $m_2$ are the refractive indices of the two media. Depending upon the polarization, the value of $\chi$ is 1 for transverse electric (*TE*) and $(m_2/m_1)^2$ for transverse magnetic (*TM*) modes.

For $m_1 > m_2$, $\theta_t$ is real for $\theta_i$ that follows the condition $sin\theta_i < m_1/m_2$. For other incident angles, $\theta_t$ is complex and the wave vector in the rarer medium becomes imaginary [50]. Therefore, instead of propagating, the ray will decay in the rarer medium and get trapped in the denser medium. To satisfy the boundary conditions at the surface of the cavity, the decaying field, known as the evanescent field cannot enter the cavity. Constructive overlapping of this trapped ray in the cavity forms a resonant mode. These modes are formed on the surface of the cavity of radius a and are denoted by mode number *n*, mode order *l*, and the polarization. The first order modes, appearing at the surface of the cavity, occur at the size parameter $x_{n,l} = \Lambda/m_1$ where $\Lambda = n+1/2$. The complex wave vector in the surrounding medium is given as [51]

$$k_2(r) = \pm i \frac{(\Lambda^2 - k_0^2 r^2)^{\frac{1}{2}}}{r} \tag{3}$$

The above equation is imaginary if the term in the bracket is positive i.e. $\Lambda > k_0 r$ or $x < \Lambda$. Writing directly the expressions for the resonance positions for the first order mode ($x_{n,l}$), free spectral range ($\Delta\nu_{FSR}$) and mode linewidth ($\delta\nu$) are derived by using geometrical optics [51] as

$$x_{n,l} = \frac{\Lambda}{m_1}\left[\Lambda + 1.84\Lambda^{-\frac{2}{3}}\left(1 - \frac{2}{3\pi}(\pi + \phi)\right)\right] \tag{4}$$

$$\Delta\nu_{FSR} = \frac{x}{\Lambda}\frac{\tan^{-1}\left(\left(\frac{m_1 x}{\Lambda}\right)^2 - 1\right)^{\frac{1}{2}}}{\left(\left(\frac{m_1 x}{\Lambda}\right)^2 - 1\right)^{\frac{1}{2}}} \tag{5}$$

$$\delta\nu = \frac{2(\Lambda^2 - x^2)^{\frac{1}{2}}}{(m_1^2 - 1)x\chi}\exp\left(2\psi_r(x)\right) \tag{6}$$

Here $x = 2\pi a m_2/\lambda$ is the size parameter, $\phi$ is the boundary phase shift, $\psi_r(x)$ is the phase propagation. For *TE* polarization, the value of $\chi$ is 1 and $(\Lambda/x)^2+(\Lambda/m_1x)^2-1$ for *TM* polarization. The above equation gives zero linewidth at $x = \Lambda$ which is not practically possible. So, the equation has been modified for $x = \Lambda$ as

$$\delta\nu = \frac{2.12\chi^2}{(m_1^2-1)x^{\frac{1}{3}}}\exp\big(2\psi_r(x)\big) \tag{7}$$

where $\chi$ takes the value as 1 for *TE* and $m_1$ for *TM*. Geometrical optics provides fairly approximate solutions of these modes far away from the ray confinement region. But it fails to explain the leakage of the field even after the total internal reflection of the rays or the near field solutions. This has been explained by using the beam theory where a bunch of rays enter the cavity at different incident angles and hence diverges while exiting the cavity [52]. This beam diverges in the far field depending upon the distance of the field from the cavity surface. The properties of WGMs such as the radial behaviour of these resonances are understood by considering the scattering scheme using the Maxwell's equations.

#### 2.1.1 Mie scattering theory

Scattering of a plane EM wave through homogeneous spherical particles was studied by Mie in 1908 for optical cavities [53]. The second order differential equation is solved for the radial, polar and azimuthal behavior of the whispering gallery wave. The solutions of these equations are found by satisfying inside, outside, and at the surface of the cavity. The radial equation has a form of

$$\frac{d^2X}{dr^2}+\frac{2}{r}\frac{dX}{dr}+\left(k^2-\frac{n(n+1)}{r^2}\right)X=0 \tag{8}$$

The solution of the above Bessel equation can be written as the sum of two linearly independent solutions. Since the solution inside the surface should have a finite value, it can be defined by the Bessel function of the first kind. On the other hand, the solution outside the sphere should vanish gradually because no wave can exist infinitely. Hence, this can be defined by using the Hankel function of the first kind. So the solution of the equation (Eq. 8) can be written as

$$X(r) = C_1J_n(km_1r)+C_2H_n^{(1)} \tag{9}$$

where $C_1$ and $C_2$ are normalization constants and can be determined using boundary conditions. $J_n$ and $H_n^{(1)}$ are the Bessel and Hankel functions of the first kind, respectively. The scattering cross-section of the above mentioned wave due to a sphere can be written as [54]

$$Q_{sca} = \frac{2\pi}{|k|^2} \sum_{n=1}^{\infty} (2n+1)(|a_n|^2 + |b_n|^2) \tag{10}$$

where $a_n$ and $b_n$ are *TM* and *TE* coefficients and can be expressed as

$$a_n = \frac{m_1 J_n(m_1 x) J_n'(x) - J_n(x) J_n'(m_1 x)}{m_1 J_n(m_1 x) H_n^{(1)}(x) - J_n'(m_1 x) H_n^{(1)}(x)} \tag{11}$$

$$b_n = \frac{J_n(m_1 x) J_n'(x) - m_1 J_n(x) J_n'(m_1 x)}{J_n(m_1 x) H_n^{(1)}(x) - m_1 J_n'(m_1 x) H_n^{(1)}(x)} \tag{12}$$

The free spectral range (*FSR*) of the cavity is defined as

$$\Delta\nu_{FSR} = \frac{1}{2\pi a m_2} \frac{\tan^{-1}\left(\left(\frac{m_1}{m_2}\right)^2 - 1\right)^{\frac{1}{2}}}{\left(\left(\frac{m_1}{m_2}\right)^2 - 1\right)^{\frac{1}{2}}} \tag{13}$$

and the quality factor ($Q$) is given as

$$\frac{1}{Q} = \frac{\delta\nu}{\nu} = \sum_i \frac{1}{Q_i} \tag{14}$$

where the summation over $i$ includes various possible losses. These losses include the internal $Q$ value, the absorptive ($Q_{abs}$) and scattering ($Q_{sca}$) losses of the light by the sphere, coupling ($Q_{cou}$) loss of the excitation light to the cavity, impurity or roughness ($Q_{mat}$) loss present in the sphere during the fabrication process, and the bending ($Q_{ben}$) loss due to the curvature of the sphere. Another important parameter of a mode is its effective volume ($V_{eff}$) given as

$$V_{eff} = \frac{\iiint \varepsilon(r)|E(r)|^2 dV}{\max[\varepsilon(r)|E(r)|^2]} \tag{15}$$

where $\varepsilon(r)$ and $V$ are the dielectric constant and volume of the cavity, respectively and $E(r)$ is the electric field strength. All the above equations are valid for small cavities. For large cavities, Mie scattering problem needs to determine the asymptotic values of the FSR, position and linewidth of the modes [55]. The mode for large cavities can be observed as an analogy to the quasi-bound states known as the shape resonances in quantum mechanics [56].

#### 2.1.2 Coupling mechanisms

Aforementioned discussion deals with the free space coupling - the least efficient coupling mechanism. This is due to the different phase velocities in the cavity and the surrounding medium resulting in the loss of the input and output signals, especially in large cavities. The coupling efficiency can reach up to 30% with free space which is low [57]. This type of coupling is generally used for small sized fluorescent cavities. Another coupling mechanism uses a prism in which an incident light is reflected into the cavity (Fig. 2(a)). Depending upon the angle of incidence of the incoming light and the material and shape of the prism, the light can be coupled to the cavity via a phase-matched evanescent field. The coupling efficiency, in this case, can go up to 80% [58]. The use of the low refractive index prism makes the system bulky which is improved either by using a high refractive index prism or by using a polished fiber tip. In this technique, the end of the fiber is cut at an inclination angle which behaves as the surface of the prism (Fig. 2(b)). This technique eliminates the need for the adjustment of the coupler to the source and the cavity.

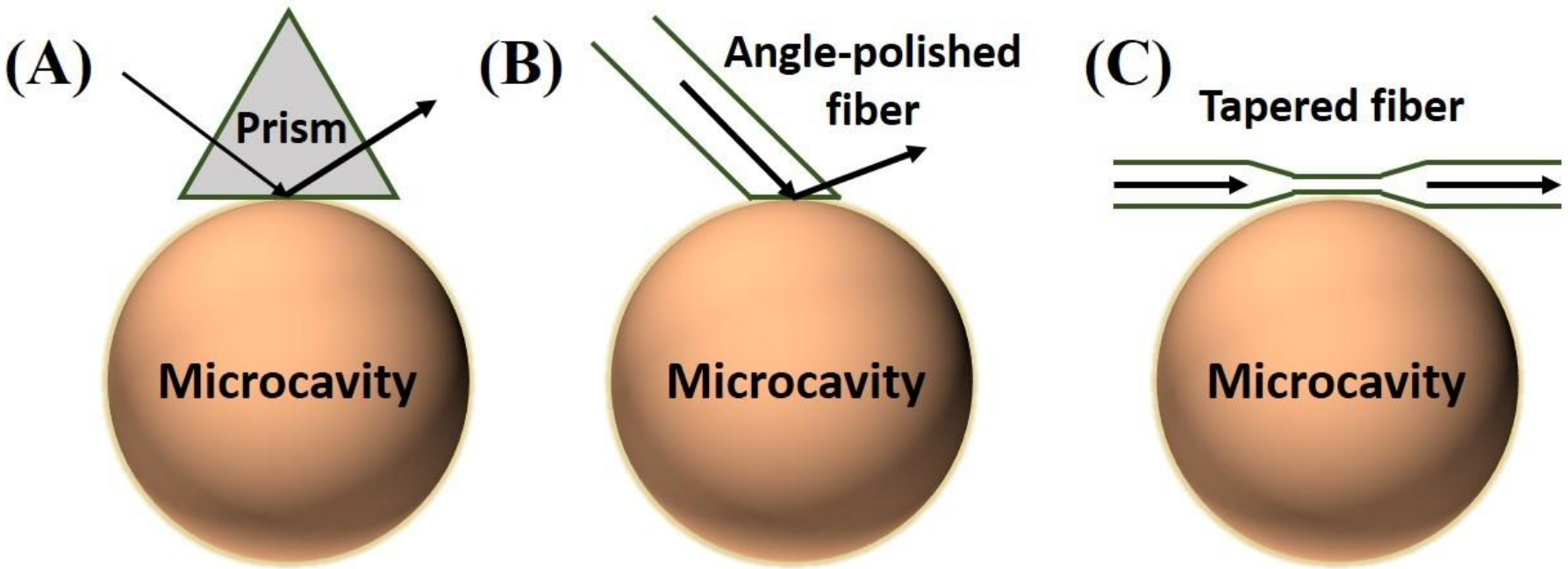

Figure 2: Different coupling mechanisms. (a) Prism coupling, (b) Angle fiber coupling, and (c) Tapered fiber coupling.

One of the widely used coupling techniques is tapered fiber coupled microcavities (Fig. 2(c)) [59]. They are ultra-thin optical fibers ($\sim 1 - 2\mu m$) and can be used to excite high Q single as well as multimodes. It has been shown that the ideal mode matching via tapered fiber can be described using a single coupling factor ($t$) and the cavity round-trip factor ($\beta$). Using these parameters, the power transmission ($T$) at the output end can be written as [60]

$$1 - T = \frac{C}{1 + f sin^2\left(\frac{\pi\nu}{\Delta\nu_{FSR}}\right)} \tag{16}$$

Here $f$ is the coefficient of finesse and $C$ is the cavity coupling parameter and are given as follows

$$f = \frac{4\beta t}{(1 - \beta t)^2}; C = 1 - \left(\frac{t - \beta}{1 - \beta t}\right)^2 \tag{17}$$

The coupling efficiency via this technique can go up to 99% [61].

#### 2.1.3 Geometry of microcavities

**Microsphere**: Microspheres are easy to fabricate as they rely on the surface tension of molten material and hence can be made of a wide variety of materials. The dense mode spectrum having equatorial, radial, and polar dependence provides a broad range of selection of modes. They are easy to be functionalized by doping or coating. The $Q$ value up to $10^{11}$ has been demonstrated experimentally for microspheres [62].

**Microtoroid, microring and disk**: Due to the acquired planer structures, toroids, rings, and disks are compact and compatible with the integrated circuits. They are generally fabricated on-chip using lithographic technique and have a Q value as high as $\sim 10^8$ [63]. Toroids have a wide mode separation than the sphere due to the lesser symmetry making them better for sensing. It also has a lower mode volume as compared to the sphere due to its planar structure. Rings and disks are widely used in biochemical sensing due to their compactness, high sensitivity, and compatibility. But they suffer from a loss of $Q$ value ($\sim 10^6$) as compared to the toroids [64, 65]. The relatively low $Q$ value of rings and disks can be attributed to the surface roughness while fabrication.

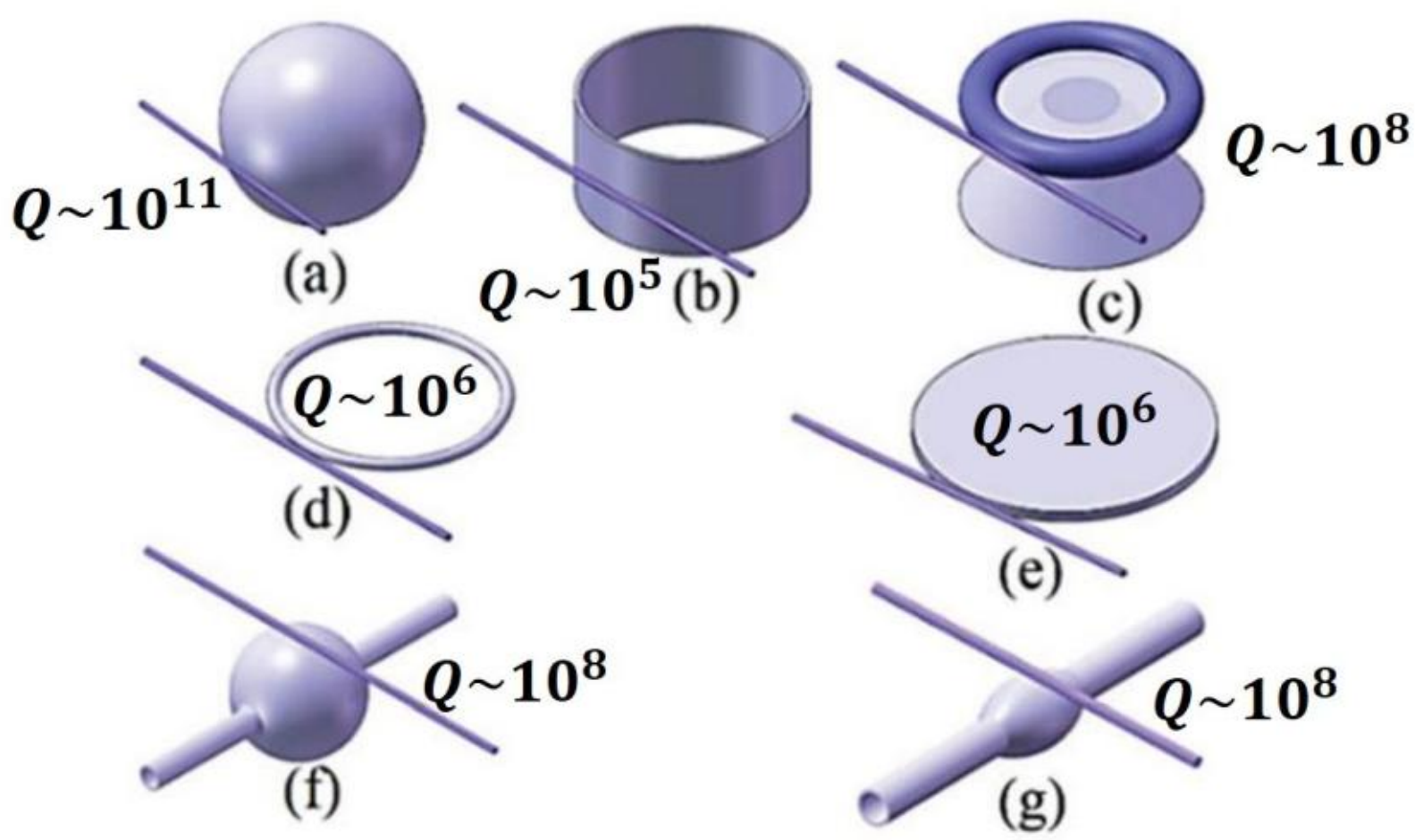

Figure 3: Geometries of microcavity. (a) microsphere, (b) microcapillary, (c) microtoroid, (d) microring, (e) microdisk, (f) microbubble, and (g) microbottle. Adapted from ref. [66]. Q values of the cavities are adapted from [62] (a), [15] (b), [63] (c), [65] (d), [64] (e), [67] (f), and [68] (g).

**Microbottle (MBR) and bubble**: Bottle cavities are prolate spheroidal shaped structures made generally by splicing two fibers or by self-assembly. These cavities take advantage of their shape and exhibit a very high $Q$ value ($\sim 10^8$) [68]. Such cavities possess a new species of modes, known as axial modes or bottle modes. The axial modes oscillate along the cavity axis between two turning points. These modes specifically depend only on the curvature of the cavity and can be tuned easily. The parameters of the modes can be tuned by simply changing the position of the cavity. Another interesting geometry is microbubble. This shape combines the properties of a spherical cavity with optofluidics. A capillary tip is melted using the localized heating and then surface tension plays its role to make a bubble such that it is hollow. A purely liquid bubble is also used for biochemical and pressure sensing. A $Q$ value of more than $10^7$ has been obtained for microbubbles [67].

**Microfiber and capillary**: Fiber and capillary are easy to fabricate and do not require a sophisticated setup for coupling. Due to their cylindrical structure, they are generally used as liquid and gas sensors. But due to the longitudinal degree of freedom in their structure, the light confinement is poor and the energy is eventually lost, even without any material loss. A $Q$ value of around $10^4 - 10^5$ has been found for these structures [15, 16]. Figure 3 shows the various geometries mentioned in this section.

#### 2.1.4 Atom-cavity system

When an emitting atom is embedded into the cavity, the photon emitted by the atom and the cavity modes can interact. Assuming that the cavity mode frequency overlaps with the transition frequency of the atom, the strength of atom-cavity interaction can be determined by three parameters. First is the photon decay rate ($\kappa$) of the cavity which is related to the $Q$ value inversely (Eq. 18). Second is the non-resonant decay rate ($\gamma$), which is inversely proportional to the dephasing time, $T_2$ (Eq. 18). Finally, third is the atom-photon coupling rate ($g$), whose strength relative to the previous two rates determines the overall coupling strength.

$$\kappa = \frac{\omega}{Q}; \gamma \propto \frac{1}{T_2} \tag{18}$$

If $g \gg \kappa, \gamma$, then the atom-cavity coupling falls into the strong coupling regime. In this condition, the photon emitted by the atom is re-absorbed by the atom itself before it is lost from the cavity. On the other hand, if $g \ll \kappa$ and/or $\gamma$, then the coupling is weak coupling. In this case, the emission of the photon by the atom in a cavity is similar as the free space spontaneous emission, but the emission rate is affected by the presence of the cavity.

##### 2.1.4.1. Weak coupling regime (Purcell effect)

In the weak coupling regime i.e. $g \ll \kappa$ and/or $\gamma$, due to the small effect of the cavity, the atom-cavity interaction can be analyzed by using the first order perturbation theory. The cavity modifies the radiative decay rate by modifying the photon density of states of an atom as compared to the free space. This density of states can be calculated by using Fermi's golden rule [69]. The modification of the radiative decay, known as Purcell effect [70], is given by Purcell factor $F_P$ as

$$F_P = \frac{3}{4\pi^2}\left(\frac{\lambda}{n}\right)^3 \frac{Q}{V_{eff}} \tag{19}$$

In the above equation, $F_P > 1$ suggests the decay rate enhancement while $F_P < 1$ shows the rate inhibition due to the cavity. The radiative rates of an atom inside the cavity can also be derived from classical theory for radial as well as the tangential oscillations [71, 72].

#### 2.1.4.2. Strong coupling regime (Rabi splitting)

Strong coupling regime is generally referred to as cavity quantum electrodynamics (CQED). This problem was first studied by Jaynes and Cummings where they described the interaction of a two-level atom with a single quantized mode of the field [73]. In this model, the atom is resonant with the cavity photon but is uncoupled, known as undressed states. The $q^{th}$ energy state of the system where the atom is unexcited with no photon in the cavity can be written as

$$E_q = \left(q + \frac{1}{2}\right)\hbar\omega \tag{20}$$

which are the degenerate excited levels. When the system is coupled, this degeneracy is removed and the levels are split into two. Since the ground state is a singlet state, it will remain unchanged in the coupled system. The $q^{th}$ state energy of the coupled system can be written as

$$E_q^{\pm} = \left(q + \frac{1}{2}\right)\hbar\omega \pm \sqrt{q}\hbar g \tag{21}$$

This is called the Rabi splitting and its magnitude ($\Delta E$) is given as

$$\Delta E = 2\hbar g \equiv \left(\frac{2\mu_{12}^2\hbar\omega}{\varepsilon_0 V_{eff}}\right)^{1/2} \tag{22}$$

## 2.2 Nanoplasmonic resonator

When the EM wave interacts with a metal NP, there is a collective shift of the conduction electrons with respect to the positive ion cores due to the Coulomb force. Due to this, the negative surface charge accumulates on one side and the positive charge on the other side (Fig. 4). In other words, the incident electric field induces the electric dipoles in the NP. The restoring force always exists between the separated charges and that opposes incident electric force. Since the incident field has oscillating nature, it sets the dipoles into oscillatory motion which therefore emits radiation. On the account of the size of the NP being comparable to the skin depth of EM waves in metals, the EM field penetrates through the NP and the induced dipoles oscillate almost in the same phase. When the frequency of the incident field is equal to the frequency of collective oscillation of dipoles then there is a formation of resonance and this is popularly known as localized surface plasmon resonance (LSPR). Due to this reason, the metal NPs are called as metal nanoresonators or nanoplasmonic resonators. For a metal NP of radius a and dielectric function $\varepsilon(\omega)$ in a medium with dielectric constant $\varepsilon_m$, its polarizability ($\alpha$) is given as

$$\alpha = 4\pi a^3 \frac{\varepsilon(\omega) - \varepsilon_m}{\varepsilon(\omega) + 2\varepsilon_m} \tag{23}$$

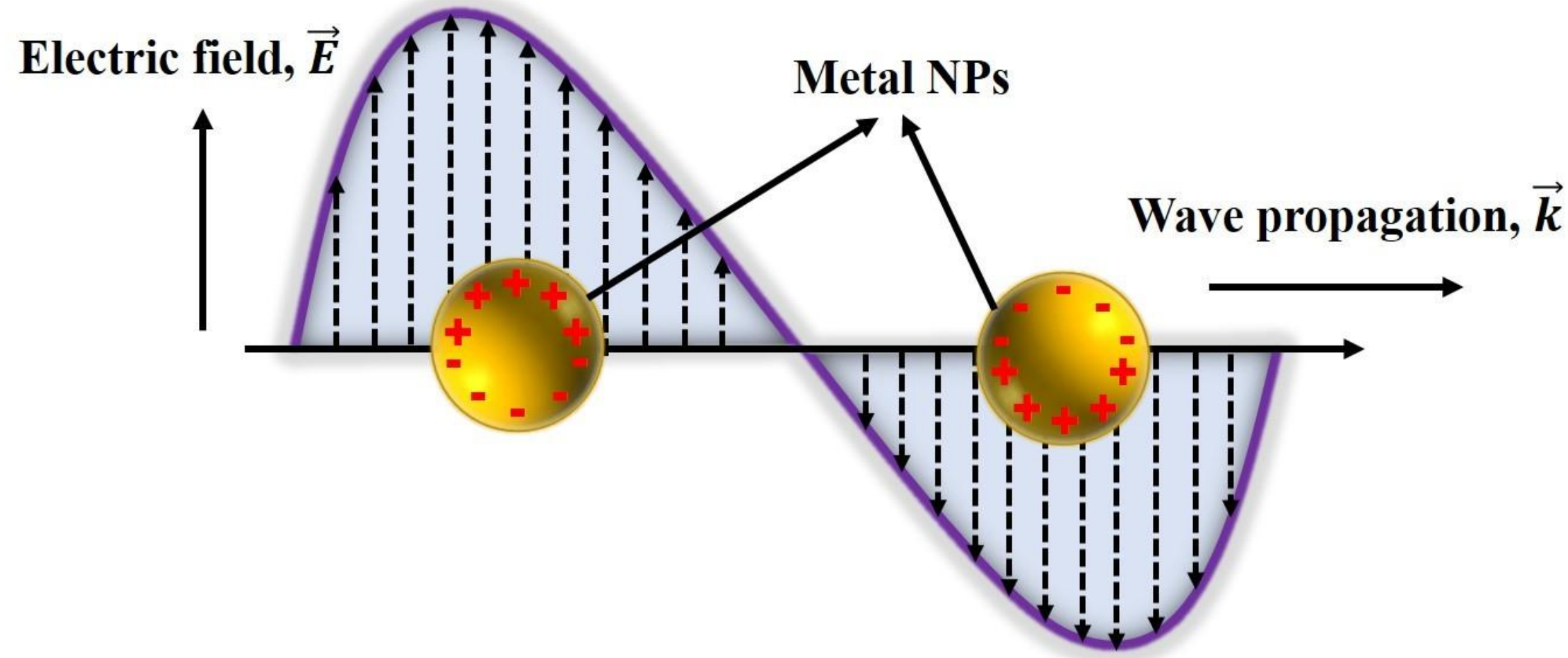

Figure 4: Schematic of LSPR generation due to interaction of EM field with a metal NP.

From the above equation, it can be observed that the polarizability is maximum or an enhancement will occur if the denominator is minimum i.e. $Re[\varepsilon(\omega)] = -2\varepsilon_m$ which is the *Frohlich*¨ resonance condition and the mode will be dipolar plasmon mode. In general, the LSPR frequency depends on the strength of the restoring force which further depends on the separation between the charges (or size of the nanoparticle), electric permittivity of the NP, polarization of the incident light, and electric permittivity of the surrounding medium. The scattering ($C_{sca}$) and absorption ($C_{abs}$) cross-sections of an incident wave by the NP can be defined as [74]

$$C_{sca} = \frac{k^4}{6\pi}|\alpha|^2; C_{abs} = kIm[\alpha] \tag{24}$$

From the above equations, it is observed that the absorption process dominates for the small NPs over the scattering process. In the plasmonic NPs, as a result of the resonance, there is a large enhancement in the electric field in a very small volume near the surface of the NP, also known as the hot spot. It is important to note that the electric field at any of the hot spots would be much larger as compared with the electric field of the incident light.

It is important to note that the electric field enhancement at the hot spot of the elongated NPs (or NPs with sharp tips) is larger due to the non-resonant or geometrical enhancement ($\eta_g$) factor in addition to the resonant enhancement factor ($\eta_r$) [75]. The value of $\eta_g$ purely depends upon the geometry of the NP. Its value is 1 for the spherical NPs. However, it is much larger than 1 for NPs of nanotriangle, nanorod, nanoellipsoid, and nanostars shapes. In the case of nanodimers also, the electric field enhancement is large at the nanogaps due to the constructive coupling of the plasmon modes of the nanoparticles forming the nanodimer.

In the case of larger nanoparticles, the spatial separation between the negative and positive surface charges increases significantly and this decreases the interaction between the charges or weakens the restoring force, and eventually, the resonance frequency decreases. The retardation effects in larger particles are responsible for the breakdown of the quasi-static approximation beyond a certain limit. It is well known that upon increasing the nanoparticle size, the variation in the phase of the incident light across its surface becomes significant. Due to this, the higher-order plasmon modes are excited along with the fundamental dipole mode. However, one can use the Lorenz-Mie theory to understand the plasmonic properties of the nanospheres of arbitrary size. The optical properties of various geometries of NPs could be studied numerically using different approaches like discrete dipole approximation (DDA), finite element method (FEM), and finite difference time domain method (FDTD).

For the case of very small NP ($a < 10nm$), an additional effect known as chemical interface damping [76] is observed. This is an inelastic scattering of the electrons of the surface of the NPs due to which the energy of the electrons is transferred to the attached surface of the molecules. This damping has been found to vary inversely with the size of the NP. For NPs with size $a < 1nm$, quantum effects start appearing. This can be explained by the fact that such small NPs have a limited amount of electrons and the energy of each electron is comparable to the room temperature energy. Therefore, instead of the collective electron oscillation, this problem has to be treated as multi-particle excitations [77].

Experimentally, the LSPR frequency of simple as well as complex shaped nanoparticles can be estimated by recording the absorption/extinction spectra of an ensemble of nanoparticles using the UV-Visible spectrophotometer. Moreover, the high-resolution dark field microscopes would allow us to record the scattering spectra of single monometallic/bimetallic nanoparticles. This means, using these microscopes one can find out the heterogeneity in optical properties of nanoparticles.

### 2.3 Hybrid microcavity

When a plasmonic field is made to interact with the dielectric cavity field, the configuration is known as hybrid microcavity. The high Q-mode of cavity couples with the LSPR mode of the metal NP and gives rise to cascaded field with low mode volume. The evanescent field of the excited cavity mode acts as an excitation field for the LSPR modes of the metal NP kept in its vicinity. Depending upon the coupling strength and the position of the NP, the resultant hybrid mode differs from its parent modes [39]. The hybrid mode greatly depends on the size, shape and material of the dielectric cavity as well as those of the metal NP. Several interactions including cavity-NP as well as NP-NP coupling occur when multiple NPs are attached to the cavity. The major disadvantage of multiple NPs coupled cavity is the increased absorption due to NPs resulting in broadening of the hybrid mode.

## 3 Applications of photonic and hybrid cavities

### 3.1 Cavity quantum electrodynamics (CQED)

#### 3.1.1 Emitter-cavity coupling

As mentioned in section 2.1.4, CQED falls into two regimes: weak and strong coupling. Originally, the modification of the radiative rate of a fluorophore has been demonstrated by Drexhage for a dye molecule kept in front of mirror [78]. Kleppner demonstrated the inhibition of spontaneous emission in Rydberg atoms [79, 80, 81]. Lin et al have demonstrated the enhancement and inhibition of lifetime (Fig. 5A-C) of chelated europium ions in microdroplets [82]. The enhancement of emission cross-section of R6G in microdroplets due to CQED effects has been estimated [83].

Purcell effect for R6G in microdroplets has also been studied in literature [84, 89, 90] (Fig. 6A). A Purcell factor of an order of magnitude has been reported in QD coupled microdisk [91, 92] and semiconductor microdisk laser [93]. A splitting in the cavity modes and reduction in the lifetimes have been observed for QD coupled microspheres [94]. Sandeep and Bisht have demonstrated the effects of adsorbed concentration of dye and the size of the microcavity on the radiative rates [95, 96]. Purcell effect in luminescent nanocrystals coupled microcavity systems has been observed previously [97, 98, 99]. Dantham and Bisht have conducted a detailed study to observe the effects of coating and doping of organic dyes in the microspheres [6]. The inhibition of decay rates for coated microspheres have been observed. They have also demonstrated the effect of the refractive index of the microsphere on the radiative rates [100]. A strong dependence of fast lifetime of the dye doped in microdroplets on the size of the droplet (Fig. 6B) has been demonstrated [85]. An increase of lifetime has been observed with the size of the cavity. They have obtained distinguished lifetimes for doped and coated microspheres.

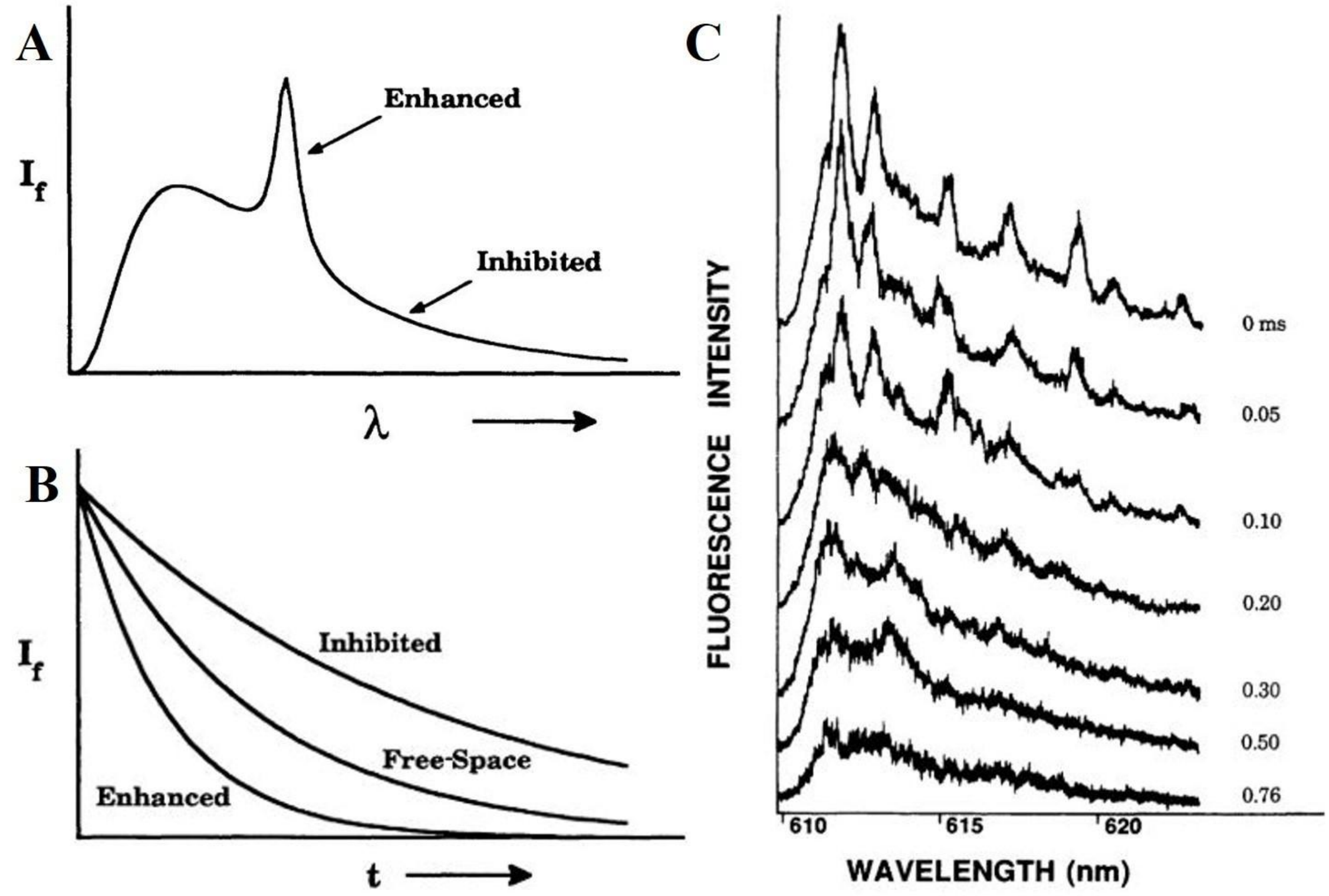

Figure 5: CQED effects on a microdroplet in emission (A) and time-resolved (B) spectra. Panel C shows the enhancement and inhibition of WGMs and broad background, respectively. Reproduced with permission [82].

Recently, CQED effects on the rotational and vibrational relaxation times of levitated microdroplets have been demonstrated using polarized Raman spectroscopy [101]. The rotational relaxation time of droplet have been shown to depend on the size of the droplet, unlike vibrational relaxation time. A theoretical analysis to determine the roles of stimulated emission, Purcell effect and leaky modes has been carried out for ZnO microspheres [102]. Purcell effect in hexagonal CdS microflakes has been observed as a function of size as well as the input power [103]. Nitrogen vacancy (NV) embedded diamond-based cavities and emitters have also been reported to demonstrate Purcell effect [104, 105, 86] (Fig. 6C). Purcell effect in systems such as 2D transition metal dichalcogenides (TMDCs) [106, 107, 108, 109] and pervoskites [110, 111] coupled to microcavities have also been exploited due to their tunable properties. The homogeneous linewidth ($\gamma$) and dephasing time ($T_2$) of an emitter are determined using the following equations [112, 113]

$$\frac{\Gamma_{cav}}{\Gamma_0} = \frac{\Delta_{FSR}}{\gamma}; T_2 = \frac{2\hbar}{\gamma} \quad (25)$$

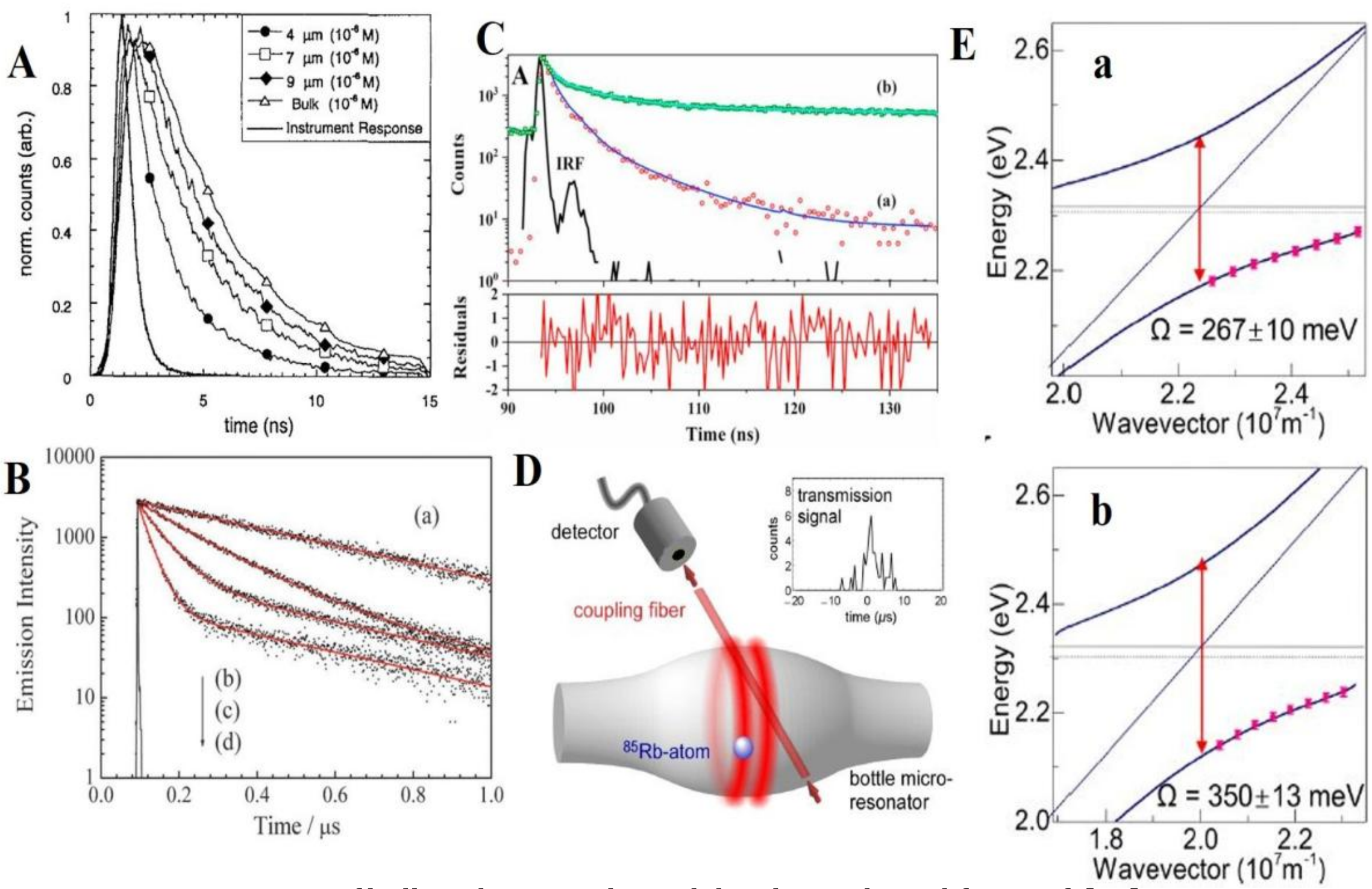

Figure 6: A: Decay curves of bulk and R6G in glycerol droplets. Adapted from ref. [84]. B: Decay curves of bulk (a) and dye in water droplets of diameters 5μm (b), 4μm (c) and 3μm (d). Adapted from ref. [85]. C: Decay curves of (a) drop casted nanodiamond (ND) and (b) ND in microsphere (Top panel). IRF denotes the instrument response function. The bottom panel shows the residuals for the curve a. Adapted from ref. [86]. D: Schematic of MBR coupled with a tapered fiber. Reproduced with permission [87]. E: Energy-wavevector dispersion curve of nanowire sitting on glass (a) and $SiO_2/Ag$ substrates (b). The horizontal solid and dashed lines indicate the longitudinal and the transverse resonance energies, respectively. Adapted from ref. [88].

Mode splitting under the strong coupling of modes has been demonstrated by Kippenberg et al [114]. They have also studied the toroidal cavities for strong coupling with a coupling rate of 86 MHz [115]. Rabi splitting due to strong coupling in QDs coupled micropillar [116, 117], microdisk [118, 119, 120, 121] and microring [122]. Aoki et al have showed the strong coupling between the single caesium atoms and toroidal cavities [123]. A coupling rate of 40 MHz has been experimentally achieved. They also demonstrated the use of strong interaction between atom and cavity to locate the caesium atoms at a distance from the cavity [124]. Mode splitting has been reported by Park et al for NV centers embedded nanodiamonds (ND) coupled to microsphere [125]. Rabi splitting has been demonstrated in semiconductor microrods [126, 127] and microwires [128, 129, 130, 131]. Vasista and Barnes have studied the evolution of strong coupling for J-aggregrate dye molecules of different layers coupled with microcavity [132]. They observed a change in Rabi splitting of 56$meV$ for single layer to 94$meV$ for four layers. They have also demonstrated the coating of multiple dyes and obtained different Rabi splitting energies due to multimolecular coupling [133]. The effect of the longitudinal polarization component of WGM on the light-matter interaction has been demonstrated under coupling 85Rb atoms to MBR [87] (Fig. 6D). They have further used 2-color laser beams to trap and demonstrate vacuum Rabi splitting in 85Rb atoms coupled MBR [134]. Farr et al have studied the coupling between cavity mode photons and $Cr^{3+}$ spins in a ruby [135]. A strong coupling strength of $610 MHz$ has been obtained. Hu et al have numerically investigated the coupling between the magnetic dipole transitions of the quantum emitters and WGM cavity [136]. A Rabi splitting has been observed due to the high magnetic field enhancement of WGMs. Table 1 gives the values of radiative rates modification and Rabi splitting energy for using different microcavity systems.

Table 1: Radiative rates modification and Rabi splitting energy exhibited by bare and hybrid cavities.

| Gain medium | Geometry | Radiative rates modification |
|---|---|---|
| Rare earth ion | Microdroplet | 2.5 times enhancement, 1.5 times inhibition [82] |
| | Microdisk | 4.5 times enhancement [137] |
| Rydberg atom | Parallel planes | 20 times inhibition [80] |
| Dye | Microdroplet | 10, 120 times enhancement [84], [83] |
| | Microsphere | 10 times inhibition [96] |
| QD | Microdisk | 6 times enhancement [91], [92] |
| | Nanowire | 16 times inhibition, 1.5 times enhancement [138] |
| Nanocrystal | Microdisk | 25 times enhancement [97] |
| | Microtoroid | 171 times enhancement [99] |
| | Microsphere | 8 times inhibition [86] |
| Semiconductor | Hexagonal flake | 4.7 times enhancement [103] |
| | Microdisk | 10 times enhancement [93] |
| | Metal film coated microwire | 5 times enhancement [139] |
| | Metal-photonic core-shell nanowire | > $10^3$ times enhancement [140] |
| Perovskite | Hemisphere | 14 times enhancement [111] |
| | Microsphere | 190 times enhancement [110] |
| Dipole | Metal antenna-microdisk | 76 times enhancement [141] |
| 2D | 2D-coated nanodisk | ~ $10^7$ times enhancement [142] |
| | | **Rabi splitting energy** |
| Nanocrystal | Microsphere | 0.05 meV [124] |
| QD | Micropillar | 0.14 meV [116] |

| | Microdisk | 0.40 meV [118, 119, 120] |
|---|---|---|
| | Microring | 0.70 meV [122] |
| 2D | 2D coated nanodisk | 37.1 meV [143] |
| Semiconductor | Microrod | 300 meV [126], 400 meV [127] |
| | Microwire | 180 meV [128], 195 meV [129], 294 meV [130], 300 meV [131] |
| | Metal nanocubes-microwire | 515 meV [144] |
| Dye | Microsphere | $\sim 100 meV$ [132, 133] |
| Perovskite | Metal substrate-nanowire | $\sim 564 meV$ [88] |

#### 3.1.2 Hybrid cavity for enhanced emitter-cavity coupling

Purcell effect in a surface plasmon ring resonator has been investigated theoretically [145]. Purcell factor of order $10^3$ has been estimated in the energy range of $1.0-1.8eV$ . Frimmer and Koenderink have studied the Purcell effect for a fluorophore placed in between an AgNP dimer coupled with a WGM cavity [146]. A Purcell enhancement factor as high as 1000 has been estimated. Purcell enhancement in a cavity-antenna hybrid system has been investigated [141, 147, 148]. Chao et al have demonstrated a high Purcell enhancement factor of $10^3$ and sub-picosecond lifetimes in a core-shell semiconductor-metal hybrid cavity [140]. A graphene monolayer coated semiconductor nanodisk has been studied with varying chemical potential and relaxation time of graphene [142]. At $0.9eV$ chemical potential and $1.4ps$ relaxation time, a Purcell factor of $10^7$ has been obtained. Gu et al have calculated Purcell factor for a hybrid nanowire and a hybrid pseudo-ring cavities and observed an improvement by an order of magnitude [149]. A detailed rate equation analysis has been conducted to investigate the Purcell effect in stimulated and spontaneous emission rates of semiconductor-plasmonic hybrid cavity [150]. Recently, a hybrid microcavity based on dielectric microsphere on the metal substrate has been studied for confinement and Purcell factors [151]. Srinivasan and Ramamurthy have proposed a ceramic based nanocavity on the metal substrate for varying Purcell enhancement based on composition [152]. Zhao et al have studied the Purcell effect for the ND-AuNP system coupled microcavity at different cavity positions [153]. A comparatively new geometry of hybrid microcavity which is a spiral plasmonic structure embedded into a hollow cylinder has been proposed. Numerical studies have shown an enhancement in Purcell factor more than $10^6$ [154].

The enhancement of the stimulated emission of R6G in a metallic optofluidic resonator has been observed [155]. The effect has been explained on the basis of the high Q value and the spontaneous emission coupling factor. Strong coupling with a coupling rate of $9GHz$ has been obtained from a hybrid cavity consisting of metal NPs on a microtoroid [156]. Shang et al have demonstrated a Rabi splitting of energy up to $\sim 564 meV$ in a hybrid perovskite nanowire [88] (Fig. 6E). Enhancement of Rabi splitting energy for ZnO microwire has been observed from $\sim 400 meV$ to $515 meV$ by depositing metal nanocubes on the microwire [144].

## 3.2 Lasing

### 3.2.1 Photonic cavity for microlasing

One of the major applications of optical microcavity is microlasing. The bandwidth of the lasing mode can be controlled by optimizing the cavity conditions. Also, the continuous oscillation of the light in the microcavity builds up the energy density leading to low value of threshold power. The rate equation analysis of the microlaser has been derived by Yokoyama and Brorson in 1989 [157]. Microlasers can be now fabricated using self-assembly technique due to their symmetric shapes [158, 159, 160, 161, 162, 163] (Fig. 7A). Single mode lasing in MBRs with a spectral tunability of more than $8 nm$ [164] has been demonstrated (Fig. 7B). A low threshold tunable dye-doped polymer microbubble laser has been demonstrated attaining a threshold of $15 \mu J/pulse$ [165]. On chip microlasers fabricated using direct laser writing and lithography have also been reported [21, 166, 167, 168]. Guo et al have demonstrated the continuous-wave operation of a microcavity cascaded laser at $8\mu$ emission wavelength [169].
Coupled microcavities have also been used for single mode and low threshold lasing [172, 173, 174]. The coupled microcavity is sometimes referred to as photonic molecule and leads to the formation of supermodes due to the coupling between WGMs of different radial orders [170] (Fig. 7C). Tunable lasing emission power from $0.25 \mu W$ to $0.41 \mu W$ has been obtained from a dual erbium-doped microspheres [175]. Apart from coupled microlasers, core-shell microlasers have also been demonstrated and are widely used due to their chemical stability and biocompatibility [176, 177]. The effect of excitation and light collection techniques on lasing threshold have also been explored. Lu et al have reported different lasing threshold obtained for an active MBR without and with tapered fiber [178]. Siegle et al have shown four different combinations of light excitation and collection [179].

Starch based [180], protein based [181, 182] and egg white based [183, 184] biolasers have shown the lasing threshold from few $nJ$ to few tens of $\mu J$ which can be endured by the cells. Soft and hard microlasers have also been studied to determine the dynamics of the live cells [171, 185, 186] (Fig. 7D). Microlasers have been proved to be used as cell trackers and diagnostic tools using biocompatible and bioresponsive cavities [187, 188, 189, 190, 191, 192, 193]. Fluorescence resonance energy transfer (FRET) using optofluidic lasers have been used to study the protein-protein interactions and protein-drug interactions [194]. A microdisk resonator made up of titanium doped sapphire has shown a lasing threshold of 14.2 mW [195].

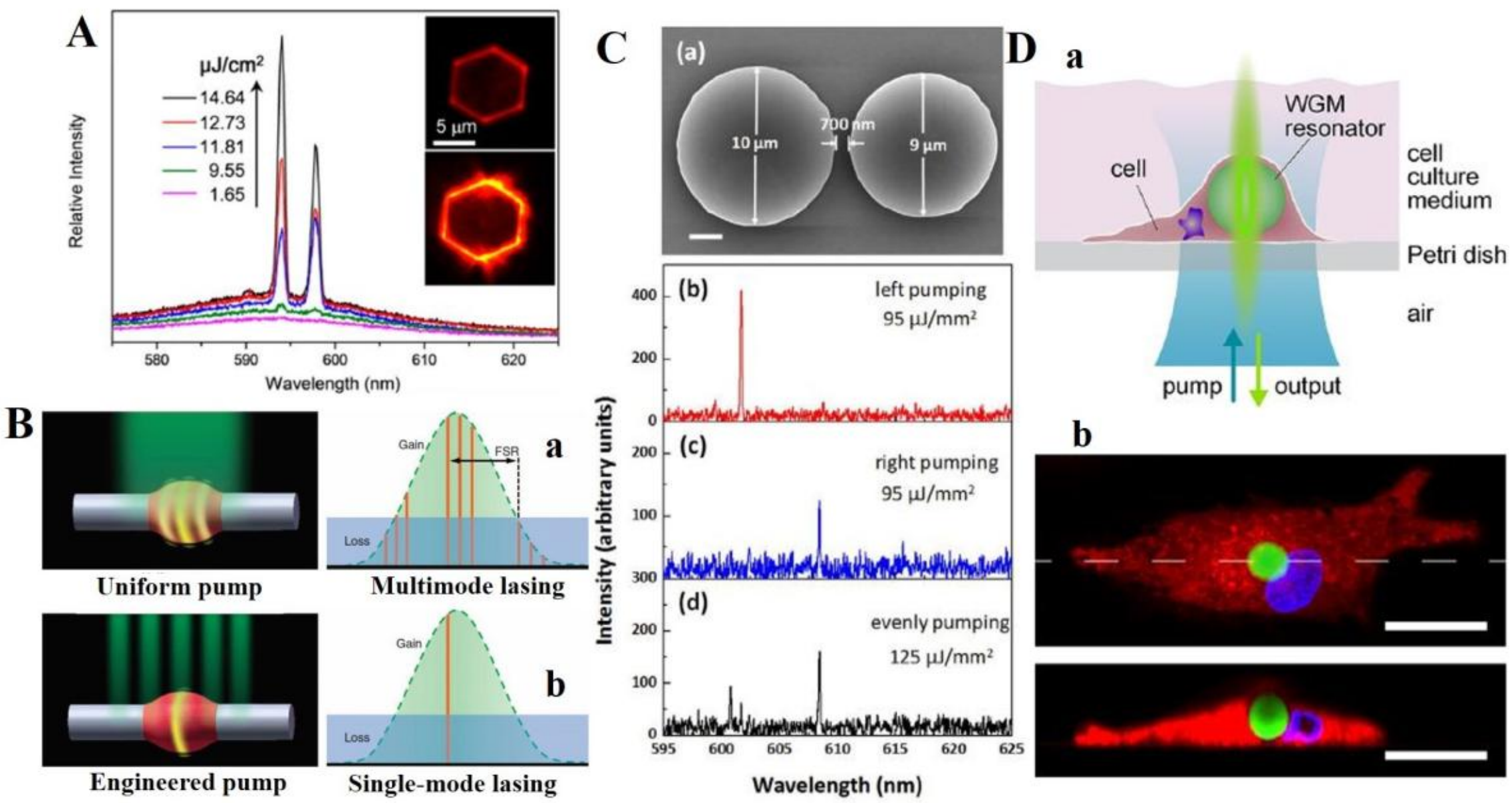

Figure 7: A: Emission spectra of a single hexagonal disk under different pump pulse densities at room temperature. Reproduced with permission [163]. B: (a) Multimode lasing using uniform pump and (b) single mode lasing using engineered pump in MBR. Reproduced with permission [164]. C: (a) SEM images of the coupled microdisks. The scale bar is 2μm. The laser spectra under the three pumping schemes (b, c, d). Reproduced with permission [170]. D: Illustration of the intracellular laser (a) and confocal laser scanning microscopy data of a macrophage with cytoplasm (red), cell nucleus (blue), and internalized microsphere cavity (green) (b). Maximum intensity projection (top) and cross section along the dashed line (bottom). Reproduced with permission [171].

As in conventional lasers, microlasing also depends on the gain medium. Organic media such as fluorescent dyes [198, 199, 196], polymers [178] and conjugated polymers [202] are typically excited with pulsed lasers to reduce photo bleaching and the formation of non-radiative decay channels. On the other hand, inorganic materials such as semiconductors [150, 213, 214, 207, 215, 208], rare earth elements [216, 217, 218], perovskites [209, 210, 201] and 2D materials [219], although with reduced effects, are being encouraged for stable emission and tunable properties. These materials even allow the continuous wave laser excitation due to the photostability and reduced losses. An application of multicolor WGM lasing has been demonstrated in textiles [220]. Flexible multicolor microfibers have been woven into color textiles. Table 2 gives the consolidated values of lasing threshold achieved using different microcavity systems and gain media.

#### 3.2.2 Hybrid cavity for low threshold microlasing

The use of metal NPs along with the microcavity further reduces the threshold power because of the high field density near the NPs. The simplest configuration of WGMLSPR lasing is the coupling of metal NPs to the WGM cavity. Shi et al have demonstrated the coupling of silica microtorodial cavity to Au nanorods by coating NPs on the toroid. The threshold power as low as $20\mu W$ could be reached with $1nm$ linewidth [221] (Fig. 8A). Another configuration reported is the AuNPs doped active microsphere and AuNPs doped active microsphere covered by a polymer layer [40]. The lasing threshold was found to be 75% lower with AuNPs. The influence of metal on the upconversion lasing characteristics of a hexagonal microrod has been investigated by Wang et al [212] (Fig. 8B) with a reduced lasing threshold by 50% after

Table 2: Lasing threshold of various bare and hybrid cavities.

| **Gain medium** | **Geometry** | **Lasing threshold** |
|---|---|---|
| Dye | Hemisphere | $2.6\mu J$ [161] |
| | Coupled fiber | $2.6\mu J/mm^2$ [174] |
| | Microtoroid | $15kW/cm^2$ [167] |
| | Microbubble | $15\mu J/pulse$ [165] |
| | Microdisk | $0.39\mu J$ [162], $3\mu J/mm^2$ [196] |
| | Coupled microdisk | $125\mu J/mm^2$ [170] |
| | Core-shell microwire | $1.54\mu J/cm^2$ [176] |
| | Core-shell microsphere | $0.78\mu J/cm^2$ [177], $3.26\mu J/cm^2$ [197] |
| | Microbottle | $49nJ/mm^2$ [178] |
| | Microellipsoid | $1.26nJ/\mu m^2$ [180] |
| | Microsphere | $3.2nJ$ [185], $2\mu J/mm^2$ [189], $7.8\mu J/mm^2$ [181], $23.2\mu J/mm^2$ [184], $26\mu J/mm^2$ [183], $0.6mJ/cm^2$ [171] |
| | Microdroplet | $0.9\mu J$ [198] |
| | Hollow microcylinder | $12nJ$ [199] |
| | Metal nanowire embedded microdisk | $\sim 2.1\mu J/cm^2$ [200] |
| | Metal NP doped microsphere covered with polymer | $\sim 3.2\mu J/cm^2$ [40] |
| Semiconductor | Microdisk | $790nJ/cm^2$ [159], $10.8\mu J/cm^2$ [163], $15mJ/cm^2$ [168] |
| | Nanoplatelets | $\sim 2.0\mu J/cm^2$ [201] |
| | Microring | $30\mu J/cm^2$ [202] |
| | Microrod | $20kW/cm^2$ [203] |
| | Hexagonal microrod on metal substrate | $0.45kW/cm^2$ [204] |
| | Metal NPs on coupled hexagonal microteeth | $86kW/cm^2$ [205] |
| | Metal nanodisk on nanoplate | $128kW/cm^2$ [206] |
| QD | OFRR | $60\mu J/mm^2$ [207] |
| | Microbubble | $\sim 10.8\mu J$ [208] |
| Perovskite | Microrod | $2.37\mu J/cm^2$ [209] |
| | Microdisk | $3.6\mu J/cm^2$ [210] |
| Crystal | Microdisk | $14.2mW$ [195], $\sim 600nJ/cm^2$ [211] |
| Rare-earth ion | Metal coating on hexagonal microrod | $0.89mJ/cm^2$ [212] |

using the metal coating.

Polystyrene microspheres are doped with various laser dyes with different optical gain and Au nanowires are attached to the microspheres [200] (Fig. 8C). By varying the proportion of the dyes, different lasing modes were excited simultaneously. Other than dielectric microcavities, semiconductor microcavities forming WGMs are also utilized because of the variable bandgap of semiconductors [206, 204, 205, 222, 223, 224]. Moiseev et al have reported the lasing mode intensity enhancement by about 20 times from a quantum dot based microdisk attached with platinum-carbon plasmonic wire nanoantenna [225]. The threshold of nonlinear multiphoton absorption-based laser has been observed to reduce by an order due to hybrid cavity as compared to bare cavity [139].

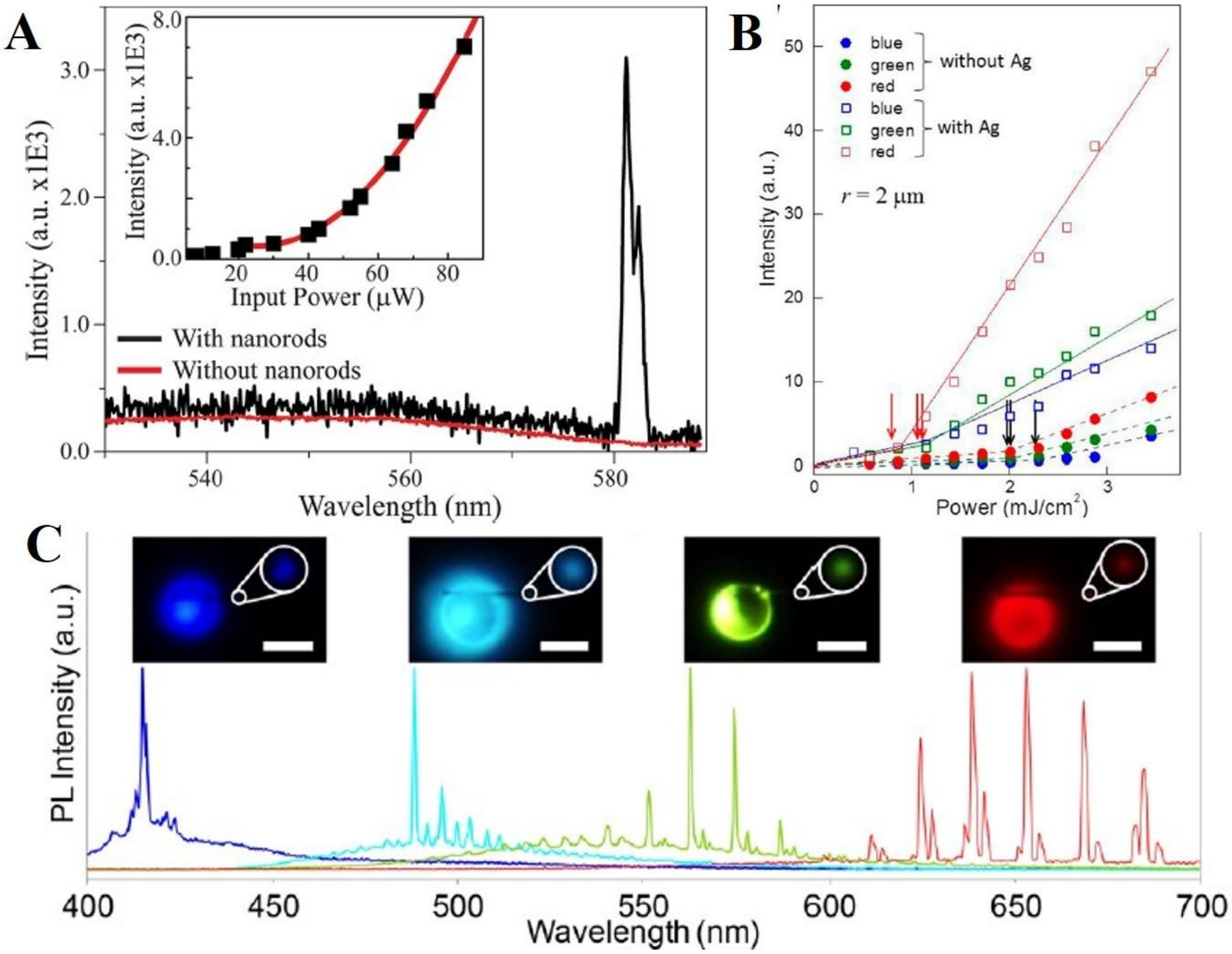

Figure 8: A: Lasing spectra and threshold plot (inset) of the Au nanorod-coated microtoroid. Reproduced with permission [221]. B: Intensity curves of the hexagonal microrods without and with deposition on the Ag-coated substrate for the three-color emission peaks. Reproduced with permission [212]. C: PL spectra of the Ag nanowire tips. Insets: PL images of the corresponding composite microdisks doped with different dyes. Scale bars are 5$\mu m$. Reproduced with permission [200].

### 3.3 Sensing

Due to the enhanced light-matter interaction in a WGM cavity, a small amount of perturbation can cause a significant change in the resonant mode characteristics. The biomolecules or NPs that are attached to the microcavity induce perturbations in terms of changes in the refractive index of the system. Depending upon the dielectric properties of the analyte, the sensing mechanism can be broadly classified as reactive sensing mechanism and dissipative sensing mechanism. The former relies upon the scattering properties of the analyte and shall be used for dielectric NPs and biomolecules. The presence of an analyte leads to either shift [226, 227, 228], split [229] or broadening of the modes. The latter however, depends on the absorptive properties of the analyte and applies to lossy materials such as metal NPs. The absorption of some part of light by the particles leads to the broadening of the modes. This broadening is different from the type observed in the reactive sensing in the sense that it is characterized by the absorptive losses instead of the scattering losses. There are some parameters that decide the quality and hence the usage of the sensor. These are sensitivity, limit of detection (LOD), resolution, specificity, and figure of merit (FOM).

#### 3.3.1 Biological sensors

The major areas in biosensing include detection and dynamical study of bacteria, virus, protein, DNA and fluid analysis. The merit of WGM based biosensor over other existing biosensors is the determination of the mass, size and orientation of the biomolecules. Adsorption of bovine serum albumin (BSA) protein and determination of its molecular weight, surface density and anisotropic polarizability have been studied thoroughly using microspherical cavity [226, 230, 231, 232, 233] (Fig. 9A). Optofluidic ring resonator (OFRR) has also been used to detection BSA protein [234].

##### 3.3.1.1. Photonic cavity for biosensing

Frustaci and Vollmer have established the use of WGM cavities to monitor the protein dynamics in various time scales and to visualize the movements of protein and single atoms [242]. Chao et al have demonstrated the performance of a ring resonator for sensing small and large biomolecules [14] of glucose, protein and vitamin. Fluorescent microspheres on the tip of an optical fiber are utilized to detect biotin D (or vitamin B7) and neutravidin protein [243]. Specific adsorption and detection of biomolecules in complex environment have been demonstrated by using functionalized cavities [244, 245, 246, 235] (Fig. 9B). Detection and identification of various blood components have been demonstrated using WGMs [247, 248, 249]. A nematic liquid crystal microdroplet is used to detect the urea from the urine samples [250]. The change in the pH value causes the reorientation of the droplet which leads to the change in the mode position. WGMs have also been used to detect the mutation and hybridization of DNA [236, 251, 252, 253] (Fig. 9C).

Detection of various kinds of bacteria has been demonstrated experimentally and analysed theoretically [254, 255, 256, 257, 258]. Hua-Jun Chan proposed a biomolecule mass sensor based on WGM cavity using pump-probe technique [259]. The label free detection of ovarian cancer biomarkers has been reported by Huckabay et al using WGM imaging [237] (Fig. 9D). They have demonstrated the multiplexed detection of various biomarkers mixed in a single microcavity. Chan et al have investigated the detection of vitamin D3 in different solvents using a high refractive index QD coated capillary [260]. Detection of virus-sized NPs and Influenza A virus have been reported using WGMs [227, 261] (Fig. 9E).

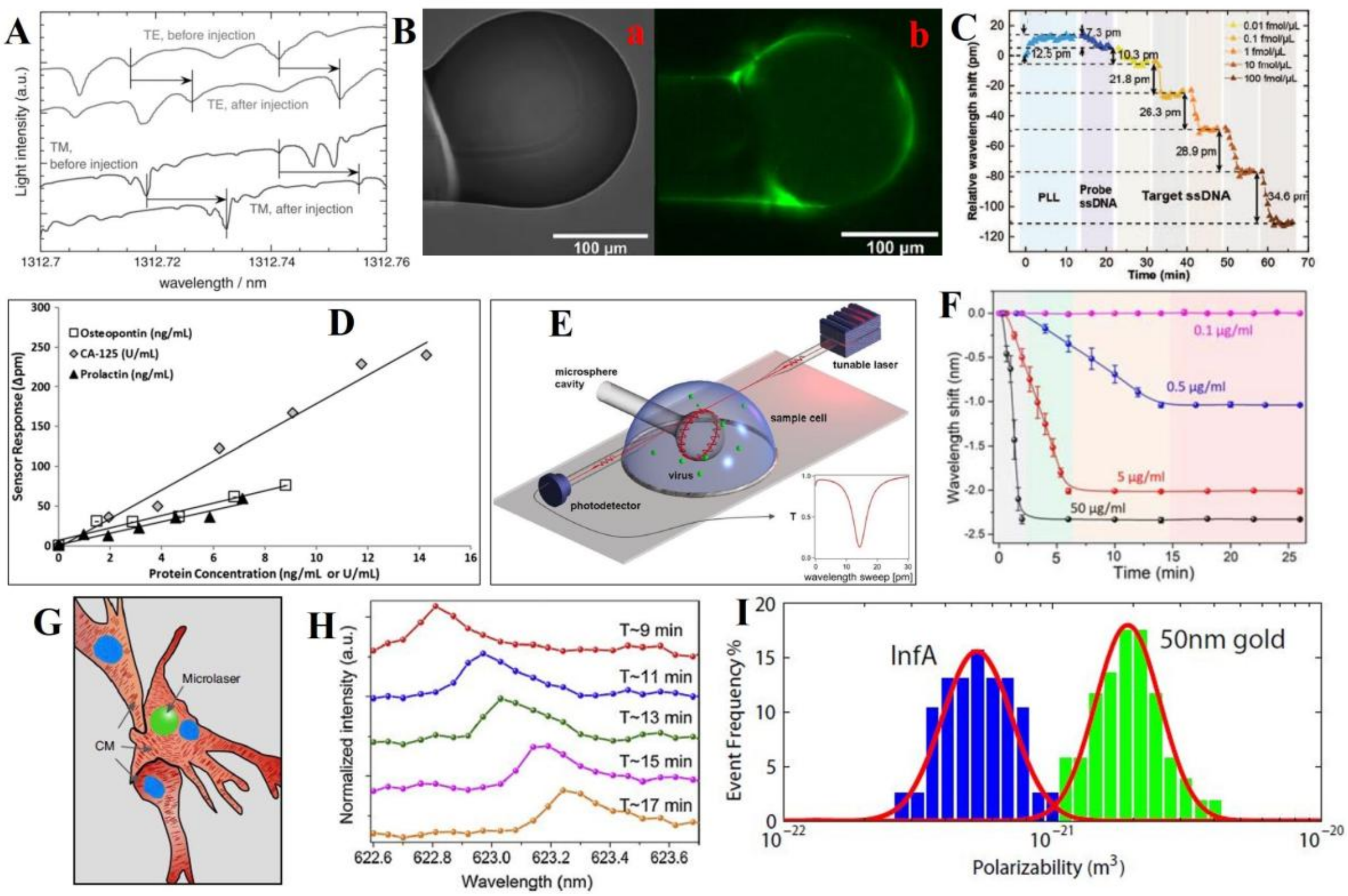

Figure 9: A: Light intensity spectrum before and after injection of BSA. The arrows indicate the shifts for some dips. Adapted from ref. [233]. B: Brightfield (a) and fluorescent (b) images of a functionalized microsphere. Reproduced with permission [235]. C: Real-time relative wavelength shift indicating surface functionalization and DNA hybridization processes. Adapted from ref. [236]. D: Detection of multiple ovarian cancer biomarkers in a single assay in PBS. Adapted from ref. [237]. E: Excitation of WGM of a microsphere by evanescent coupling to a guided wave in a tapered optical fiber. Reproduced with permission [227]. F: Temporal dependence of WGM wavelength shift for different concentrations of urease. Reproduced with permission [238]. G: A schematic illustration of a group of nuclei (blue) with a microlaser (green sphere). Adapted from ref. [239]. H: Emission spectra of microdroplets in PBS containing penicillin G after adding penicillinase. Adapted from ref. [240]. I: Measured polarizability distributions of Influenza A and AuNPs (a) using a size spectrometry system. Red curves are Gaussian fits to the experimentally obtained distributions. Reproduced with permission [241].

Microcavities have also been used to observe the dynamics of a cell or a tissue inside the body. Himmelhaus and Francois have measured the biomechanical stress induced by a live cell on an injected microsphere [262]. The contraction of the cardiac cells have been monitored using WGM lasers by observing the mode shifts due to change in the local refractive index [239] (Fig. 9G) as a result of the changes in sarcomeric protein density.

Table 3: Detection limit of various biological analytes using bare and hybrid cavities.

| Analyte | Geometry | Detection limit |
|---|---|---|
| Protein | Microring | $\sim 250 pg/mm^2$ [14] |
| | Microsphere | $10^{-4} Units/mL$ [263] |
| | Microtoroid | $2.45 nM$ [264] |
| | Microdroplet | $\sim 0.5 \mu g/mL$ [238], $0.1 pg/mL$ [267] |
| | Microfiber | $0.01 \mu g/mL$ [268] |
| | Metal nanoshell-dielectric microsphere | $\sim 0.008 ag$ [269] |
| DNA | Microdroplet | $1.32 \mu g/ml$ [270] |
| | Microsphere | $6 pg/mm^2$ [252] |
| | Microtoroid | $2.32 nM$ [251] |
| | Microfiber | $0.01 fmol/\mu L$ [236] |
| Bacteria | Microsphere | $1.2 \times 10^2 E.coli/mm^2$ [254], $10^4 cells/mL$ [257] |
| | Microdisk | $5 pg/mL$ [256] |
| Biomarker | Microsphere | $\sim 1.8 U/mL$ [237] |
| Hemoglobin | Microring | $361.3 nm/RIU$ [248] |
| Dopamine | Metal film on semiconductor microrod | $1.0 pM$ [271] |

Hanumegowda et al have demonstrated the proteolytic activity of BSA in the presence of protease trypsin [263]. Toren et al have conducted the selective detection of Exotoxin A enzyme [264]. The protein (glucose oxidase) adsorption and denaturation on the various chemically modified glass surfaces are demonstrated by Wilson et al to monitor the enzyme's catalytic activity [265]. A hydrogel with an extremely low contrast in water environment used for adsorbing biomolecules, is characterized by using a WGM cavity [266]. The enzymatic reactions of urease [238] (Fig. 9F), acetylcholinesterase [267], lipase [268] and penicillinase [240] (Fig. 9H) have been monitored using WGM sensors.

NPs undergoing the Brownian motion have been detected by monitoring the resonance fluctuations of a WGM cavity [272]. Su et al have demonstrated the detection of NPs with radii from $100 nm$ to $2.5 nm$ in aqueous using laser-frequency locking in microtoroids. They have showed the detection of exosomes from human mesenchymal stem cells, yeast ribosomes, human interleukin-2 and mouse immunoglobulin G [273]. Similar analytes have been detected by Zhu et al using split

WGM cavities [241] (Fig. 9I). Detection of icosahedral shaped RNA virus $MS_2$ and DNA virus have been reported by Arnold et al [274]. Table 3 gives the consolidated values of detection limit of various analytes.

#### 3.3.1.2. Enhanced biosensing with hybrid cavity

The sensors based on hybrid microcavities, with strong field confinement of plasmonics and high Q values of WGMs are promising. Swaim et al have performed numerical simulations on the resonance enhancement of a single BSA molecule using Au nanorod coupled microtoroid [277]. Shopova et al have estimated an increase in enhancement by 101 for $MS_2$ and 199 for BSA as the size of the analyte decreases [278]. The detection of the various concentrations of BSA using BSA mixed AuNPs assay near silica microsphere has been reported by Santiago-Cordoba et al [275] (Fig. 10A). Dantham et al have reported the detection of single thyroid cancer marker using WGM-nanoshell cavity [269] (Fig. 10B). A mechanism has been proposed to optically trap the BSA proteins at the plasmonic sites to achieve fM sensitivity [279]. Arbabi et al [280] have identified three different coupling regions: weak symmetric, weak anti-symmetric and strong coupling for Tg molecule in a periodic plasmonic epitope attached to the equator of the microsphere. Nadgaran and Garaei have obtained the wavelength shift of $26.45fm$ and field enhancement of 1469 for a triangular Au nanoprism coupled to microtoroid when a BSA molecule binds to the surface [281]. To increase the field enhancement, Garaei et al have explored the usage of a bimetallic nanoshell coupled with a WGM cavity [42] (Fig. 10C) leading to an abrupt increment of the field ( 863%) to detect a single Tg protein. Hybrid cavities have also been utilized for the detection of viruses and enzymatic reactions with a periodic array of AuNPs at the equator of the cavity [276] (Fig. 10D). Real-time detection of the smallest RNA virus $MS_2$ of mass as low as $6ag$ has been reported using a dielectric-plasmonic core-shell cavity [282].

Dopamine, a neuro-moderator in the brain has been detected by using the Raman signals of Ag NPs-decorated ZnO microrod [271]. A glucose oxidase (GOx) based glucose sensor has been fabricated to monitor the enzymatic oxidation of the glucose [283] (Fig. 11A). Detection of nucleic acid hybridization down to 8-mer oligonucleotides and enzymatic interactions have been demonstrated by using glass microspheres with Au nanorods [284, 285] (Fig. 11B). Kim et al have monitored the evolution of the reaction in varying environmental conditions such as change in pH and the electrolyte concentration [286].

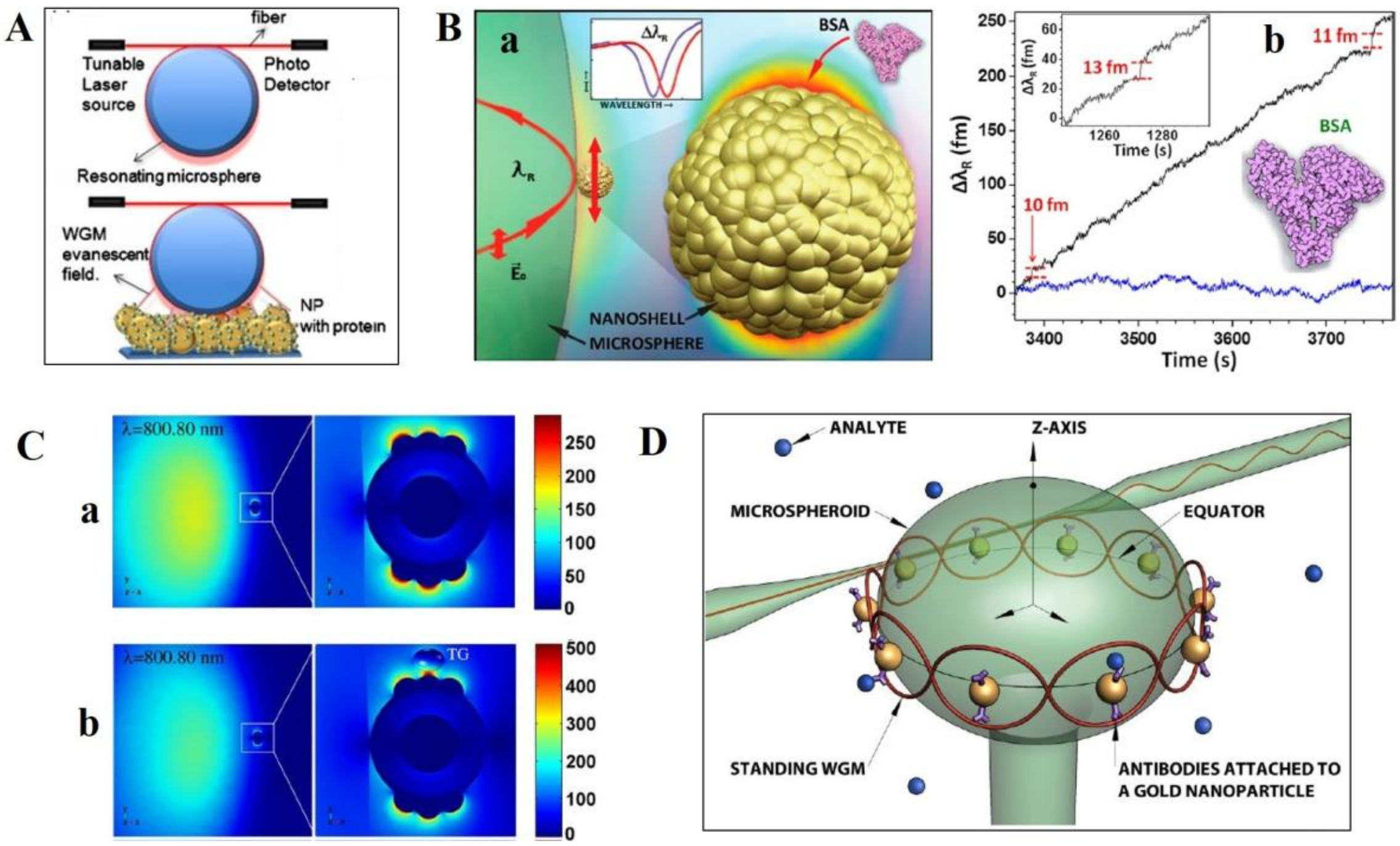

Figure 10: A: Schematic diagram of the evanescently couple WGMs to the AuNP layer. Adapted from ref. [275]. B: Illustration of a BSA protein adsorbing at the surface of a bump on a nanoshell, attached to a dielectric microcavity (a) and wavelength shift curve (upper) associated with single BSA adsorbing to the nanoshell attached at the equator of cavity (b). Insets show the maximum step associated with this protein. The lower trace shows the background without protein or the nanoshell. Reproduced with permission [269]. C: Field enhancement of bumpy bimetallic nanoshell (a) and a cylindrically shaped Tg molecule located (b) near the hybrid plasmonic-microsphere sensor for $TE_{1,387}$ mode. Right panels show the magnified image of the dipole plasmon resonance. Adapted from ref. [42]. D: Illustration of an oblate spheroidal WGM cavity with functionalized periodic nanoplasmonic epitopes. Reproduced with permission [276].

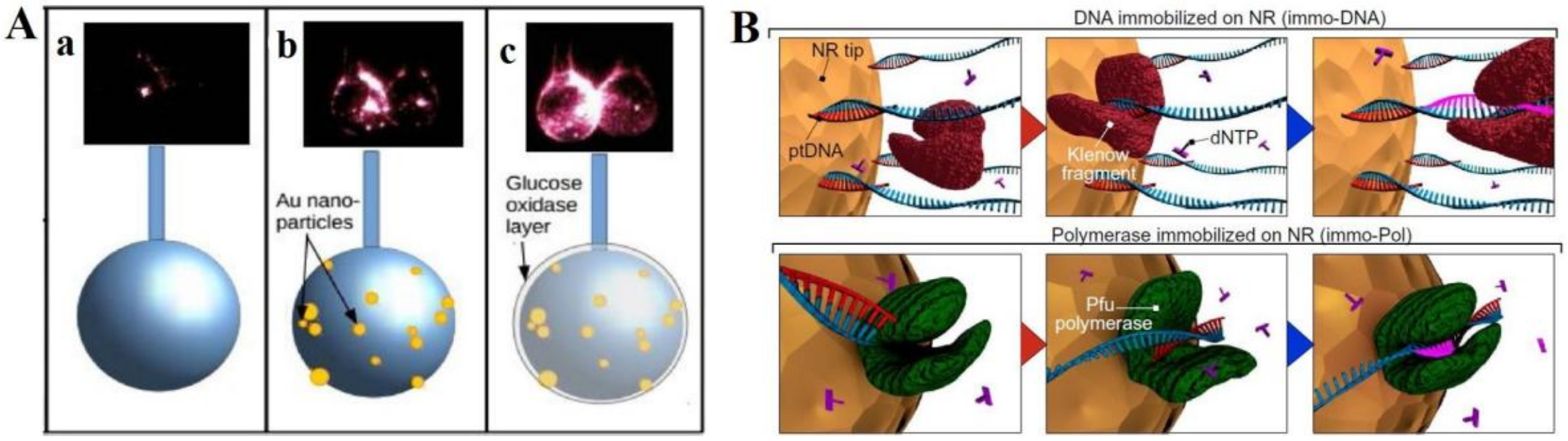

Figure 11: A: WGM-cavity before surface modification (a), after the modification by AuNPs (b), and WGMR/AuNPs cavity after the coating with the GOx (c). Adapted from ref. [283]. B: Conceptual representation of the two different approaches used for monitoring DNA/polymerase interactions. Reproduced with permission [285].

### 3.3.2 Chemical sensors

#### 3.3.2.1. Photonic cavity for chemical sensing

In the area of chemical industry, WGMs are used to detect the heavy metal ion present in the water and food, chemical vapors present in air and the ongoing chemical reactions. The resonance wavelength shifts in the presence of chemical vapors due to the change in the size and/or the refractive index of the cavity (Eq. 26).

$$\frac{\Delta\lambda}{\lambda} = \frac{\Delta m}{m} + \frac{\Delta a}{a} \tag{26}$$

Specific detection of heavy metal ions such as mercuric ion Hg(II) [287, 288] and lead ion Pb(II) [289, 290] have been demonstrated by using functionalized WGM cavities. Duan et al have demonstrated the use of a liquid crystal microdroplet for the detection of copper ion Cu(II) and other heavy metal ions [291]. WGM cavities have also been used in food and pharmaceutical industries. The detection of vitamin D3 has been investigated using QD coated optofluidic cylindrical microcavity in ethanol and food-grade soy oil [260]. Detection of salt [292] and sucrose [293, 294] (Fig. 12A) have been demonstrated in aqueous solutions. A new geometry, rolled-up microtubes has been proposed for solvent sensing by Song et al [295]. These microtubes are made up of strain-engineered flexible nanomembranes. They have used SiOx/SiNx bilayer nanomembranes to fabricate microtubes and DI water and ethanol.

Stoian et al have reported the pH sensor based on silica hollow MBR modified with swellable pH-sensitive polymer particles [300]. The surfactant sensing has been demonstrated by Humar and Musevic using nematic liquid crystal microdroplet laser [301]. The orientation of liquid crystal molecules has been shown to be affected by even mM concentration of sodium dodecyl sulfate in water resulting in the spectral variations. Similar kind of the liquid crystal based WGM cavity has been utilized to demonstrate the real time detection of hydrogen peroxide ($H_2O_2$) and catalase [296] (Fig. 12B). An optofluidic chip based on the silicon microring cavity has been proposed and demonstrated for the real time detection of a herbicide acid [302]. WGM cavities are also advantageous in agriculture as they are generally used to detect the pesticides and other harmful chemicals. An organophosphorus pesticide, parathionmethyl, has been detected by using an OFRR [303]. The results obtained have been further compared to the standard pesticide detection system to verify the sensing performance. Lee et al have reported the detection of a plant growth bioregulator, 6-benzylaminopurine, using an OFRR [304].

The shift of the WGM peak of the fused silica due to the variation of refractive index and the thickness of the polymer layer due to chemical analyte has been used [305]. Passaro et al have proposed a silicon-on-insulator microring coupled with a waveguide having an ammonia sensitive polymer cladding [306]. The similar structure has been tested to observe the sensitivity and the time response of the ammonia with respect to $CO_2$ [307]. Silica gel coated microsphere [297] (Fig. 12C) and graphene coated MBR [308] have also been used to detect the ammonia gas with varying levels of sensitivities. Hydrogen ($H_2$) gas detection has been done using a polymer microdisk coated with a palladium (Pd) layer [309]. Other chemical vapors such as methane [310], ethanol [311, 312], and acetone [313, 314] have also been used for the test analytes to observe the performance of the sensing devices. The OFRR sensor have been used for the detection of the explosive, DNT [315] as well as the volatile organic vapors of benzene, toluene, ethylbenzene, m-xylene, and n-octane [316].

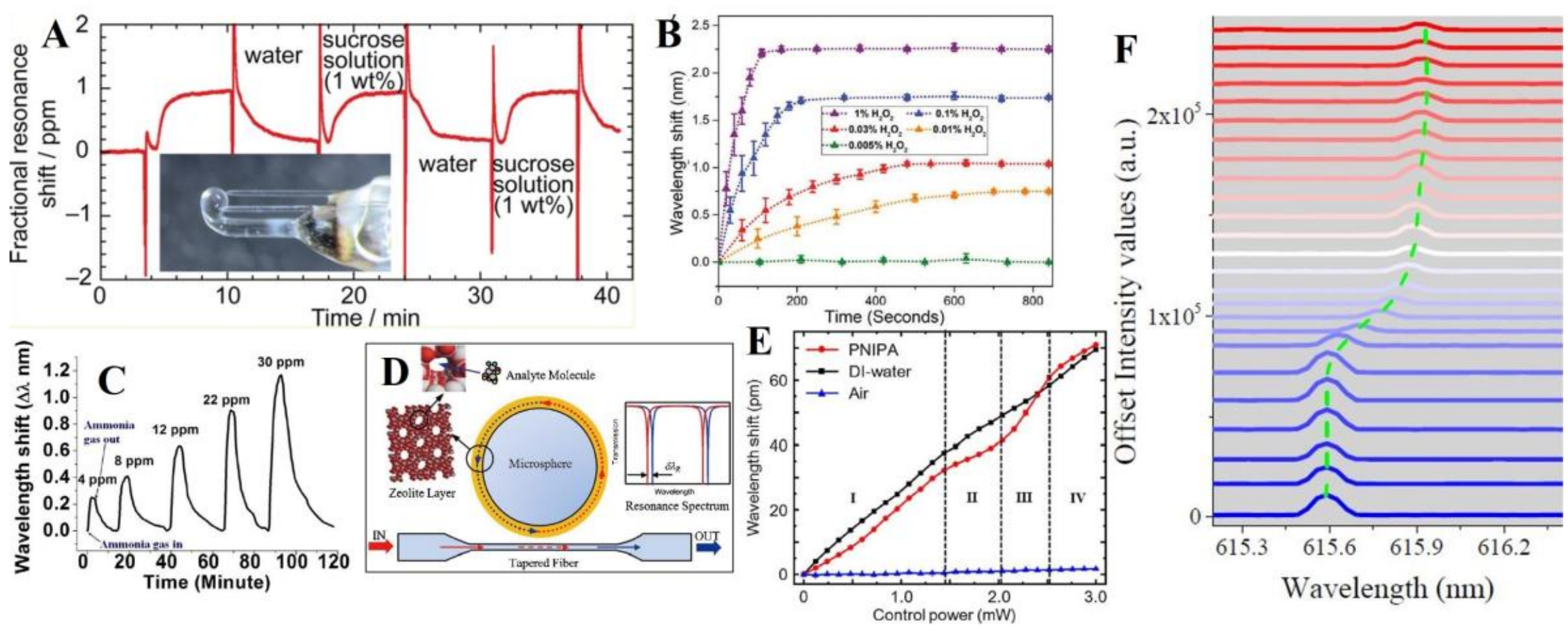

Figure 12: A: Resonance shift as a function of time during repeated transfer of the sensor head (inset) between water and sucrose solution. Reproduced with permission [294]. B: Temporal dependence of WGM wavelength shift with varying $H_2O_2$ concentrations. Adapted from ref. [296]. C: WGM spectral shift during sensor's exposure to different concentrations of NH3. Reproduced with permission [297]. D: Illustration of a zeolite-coated microsphere sensor. Adapted from ref. [298]. E: WGM wavelength shifts as a function of control power of the irradiation light, when the MBRs are filled with air, DI water, and hydrogel during four transition states: pure hydrophilic (I), subtransition (II), transition state (III) and pure hydrophobic state (IV). Reproduced with permission [67]. F: Dynamic change of lasing spectra in the process of dissolution of salt in water. The blue line (initial position) to red line (end position) is guided by the green dotted line. Reproduced with permission [299].

Applications of microcavities in multi-gas sensing and gas chromatography have also been explored [319]. Shopova et al have exploited the capillary based optical ring resonator to differentiate and detect various chemical vapors [323]. Determination of the gas concentration by measuring the changes in its thermal conductivity has been demonstrated using helium-argon mixture as test samples [324]. To increase the adsorption selectivity and interaction between the analyte and the cavity modes, Lin et al have proposed the usage of a zeolite layer, a porous aluminosilicate material, on the surface of a microsphere [298] (Fig. 12D). Depending upon the pore size of the zeolite, the specific analyte can settle into the pores and changes the refractive index of the porous layer. Similar to this work, the detection of various chemical vapors using porous glass microspheres have been demonstrated [325, 320, 326].

Due to the fast response time, it has been made possible to use microcavity as a real time chemical adsorption and reaction sensor. Ahmed et al have analysed microring with periodically arranged iron nanodisks to measure the oxidation of iron metal to iron-oxide for corrosion sensing numerically [322]. Furthermore, the chemical reaction, gelation dynamics of hydrogels has been reported by Huang et al and the results are compared with the rheology-based technique [327]. Similarly, the characterization of the dynamic phase transition of a thermosensitive hydrogel has been conducted using a microbubble resonator [67] (Fig. 12E). The mode shift and broadening have been monitored to determine a hydrophilic to hydrophobic transition. Lu et al have utilized a water immersion objective pumped microdisk laser to investigate the dissociation and diffusion rates of the salt in water [299] (Fig. 12F). Table 4 gives the consolidated values of detection limit of various chemical analytes.

Table 4: Detection limit of various chemical analytes using bare and hybrid cavities.

| Analyte | Geometry | Detection limit |
|---|---|---|
| Metal ion | Microring | 50$ppb$ [287], 58$ppb$ [289] |
| | Microbubble | 15$fM$ [290] |
| | Microdroplet | 40$pM$ [291] |
| | Microsphere | 50$nM$ [288] |
| | Metal NP decorated microsphere | 0.05$nM$ [317] |
| Explosive | OFRR | 200$pg$ [315] |
| Pesticide | OFRR | 38$pM$ [303] |
| Plant growth bioregulator | OFRR | 0.05$ppm$ [304] |
| Gas vapor | Microsphere | 34.46$pm/ppm$ [297] |
| | Microring | 5$ppm$ [307] |
| | Graphene on microbottle | 200$kHz/ppm$ [308] |
| | Microdisk | 32$pm/\%H_2$ [309] |
| | Metal nanoantenna coupled microfiber | $\sim$ 2.22 [318] |
| Hydrogen peroxide | Microfiber | $\sim$ 0.26$\mu M$ [296] |
| Salt solution | Microfiber | 0.372$nm/\%$ [292] |
| Organic compound vapor | OFRR | $5.6 \times 10^{-6} RIU$ [311], 0.5$ppm$ [316] |
| | Porous film coated microring | 25$ppm$ [312] |
| | Hemisphere on Bragg reflector | 130$nm/RIU$ [313] |
| | Microring | 1.5$ppm$ [319] |
| | Capillary coupled porous microsphere | 20$ppm$ [320] |
| | Metal NP coated OFRR | 38*ng* [321] |
| Corrosion | Microring | 517*nm/RIU* [322] |
| Acid | Optofluidic chip | 4.5*pg/mL* [302] |

#### 3.3.2.2. Hybrid cavity for static and dynamic chemical sensing

Panich et al have showed the detection of Pb(II) of concentrations down to 0.5 nM using gluthatione modified AuNPs-glass microspheres in the presence of several alkaline and heavy metals [317]. They have also demonstrated the real time binding of Pb(II) ions to the modified AuNPs through the resonant wavelength shifts. The detection of $H_2$ gas using palladium (Pd) nanorods coupled with silica microfiber has been reported by Gu et al [318] (Fig. 13A). Scholten et al have introduced OFRR coated with thiolate-monolayer-protected AuNPs to detect and distinguish various volatile organic chemical vapors [321].

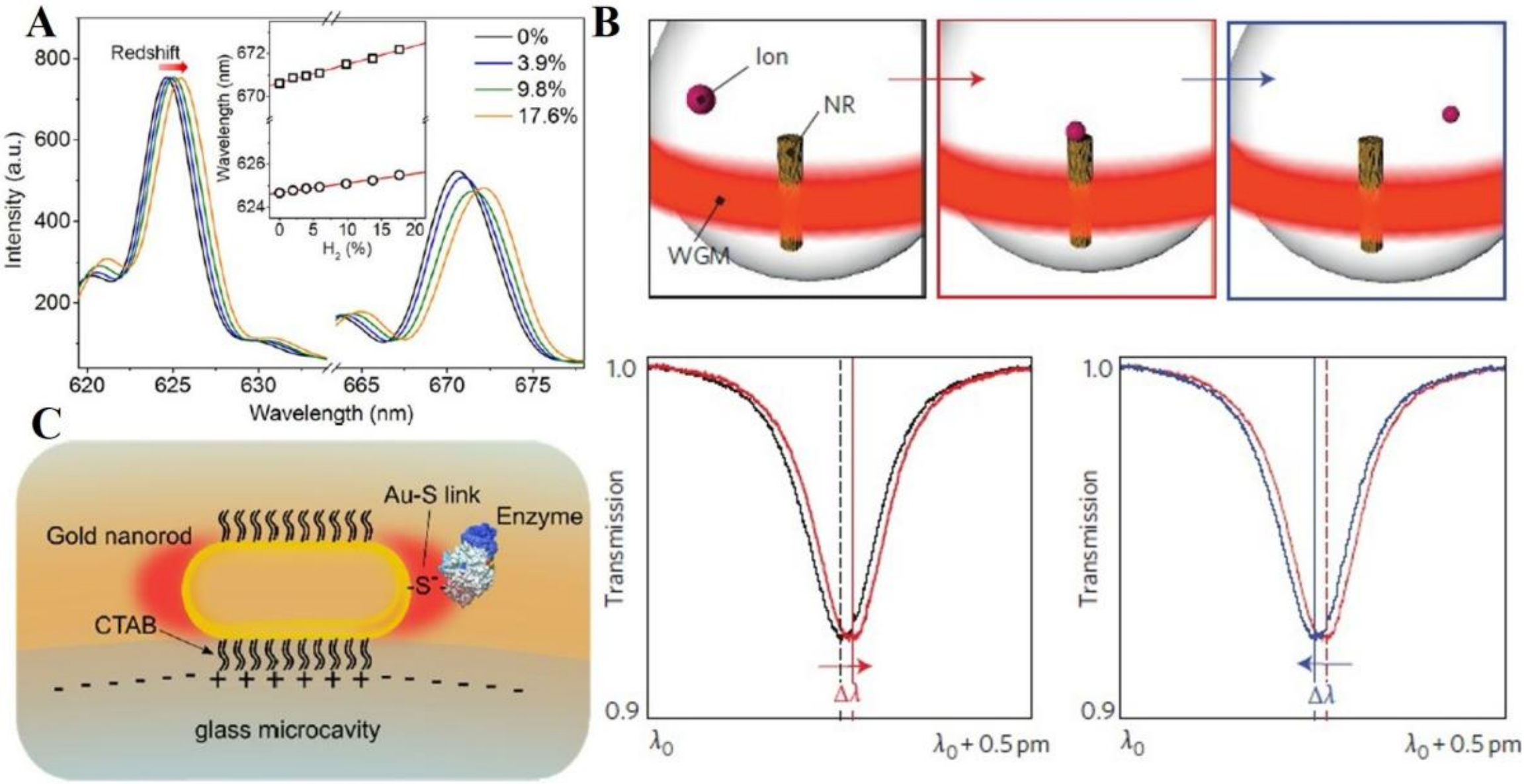

Figure 13: A: Concentration of $H_2$ dependent scattering of the Pd nanoantennas. Inset shows the wavelengths shift of the resonant peaks at 624.6*nm* (squares) and at 670.6*nm* (circles). Adapted from ref. [318]. B: Interaction of single zinc and mercury ions with the nanorods excited at their plasmon resonance. Adapted from ref. [328]. C: Schematic of the gold (*Au*) nanorod-based optoplasmonic sensor assembly with an enzyme attached via $Au - S$ link. Reproduced with permission [329].

Fast response and short recovery time make these sensors suitable for real-time visualization of chemical interactions and reactions. Baaske and Vollmer [328] have studied the interaction of zinc and mercury ions with Au nanorods (Fig. 13B) have been used to differentiate between permanent binding and transient interaction for the ion species. The effect of ionic strength of the surrounding medium on the ion-nanorod interaction have also been studied. Vincent et al have quantified the reactions between sub-kDa thiolated species at sub-fM concentrations using microspherical cavity attached with Au nanorods [330]. In a recent study, the immobilization, kinetics and thermodynamics of single enzymes have been investigated using glass microspheres coupled to AuNPs [329] (Fig. 13C).

#### 3.3.3 Physical sensors

##### 3.3.3.1. Photonic cavity for sensing external parameters

WGMs have been used for environmental sensing such as refractive index (RI) [331, 332, 333], temperature [334, 335, 336], pressure [337, 338, 339, 340], humidity [341, 342, 343, 344] and external fields [345, 346, 347]. Multiple parameters sensing with a single device has been reported [348, 349, 350, 351]. Hanumegwoda et al have demonstrated a RI sensitivity of 20 nm/RIU using silica microspheres [352] and liquid core optical ring resonator [353, 354, 355]. A microbubble coupled with a tapered fiber and coated with a polymer has been used for bulk RI and surface sensing measurements [356]. RI sensor based on microcavity with broken symmetry resulting in mode splitting have been demonstrated by Kang et al [357]. The magnitude of the splitting increases with RI contrast between the surface of the cavity and the surrounding medium. Coupled microcavities have also been used in various works to achieve high sensitivities [174, 358, 359] (Fig. 14A). The shift in the resonant wavelength due to the temperature depends upon the thermo-optic coefficient ($dm/dT$) and thermal expansion ($da/dT$) of the cavity (Eq. 27).

$$\frac{\Delta\lambda}{\lambda} = \frac{1}{m}\frac{dm}{dT}\Delta T + \frac{1}{a}\frac{da}{dT}\Delta T \tag{27}$$

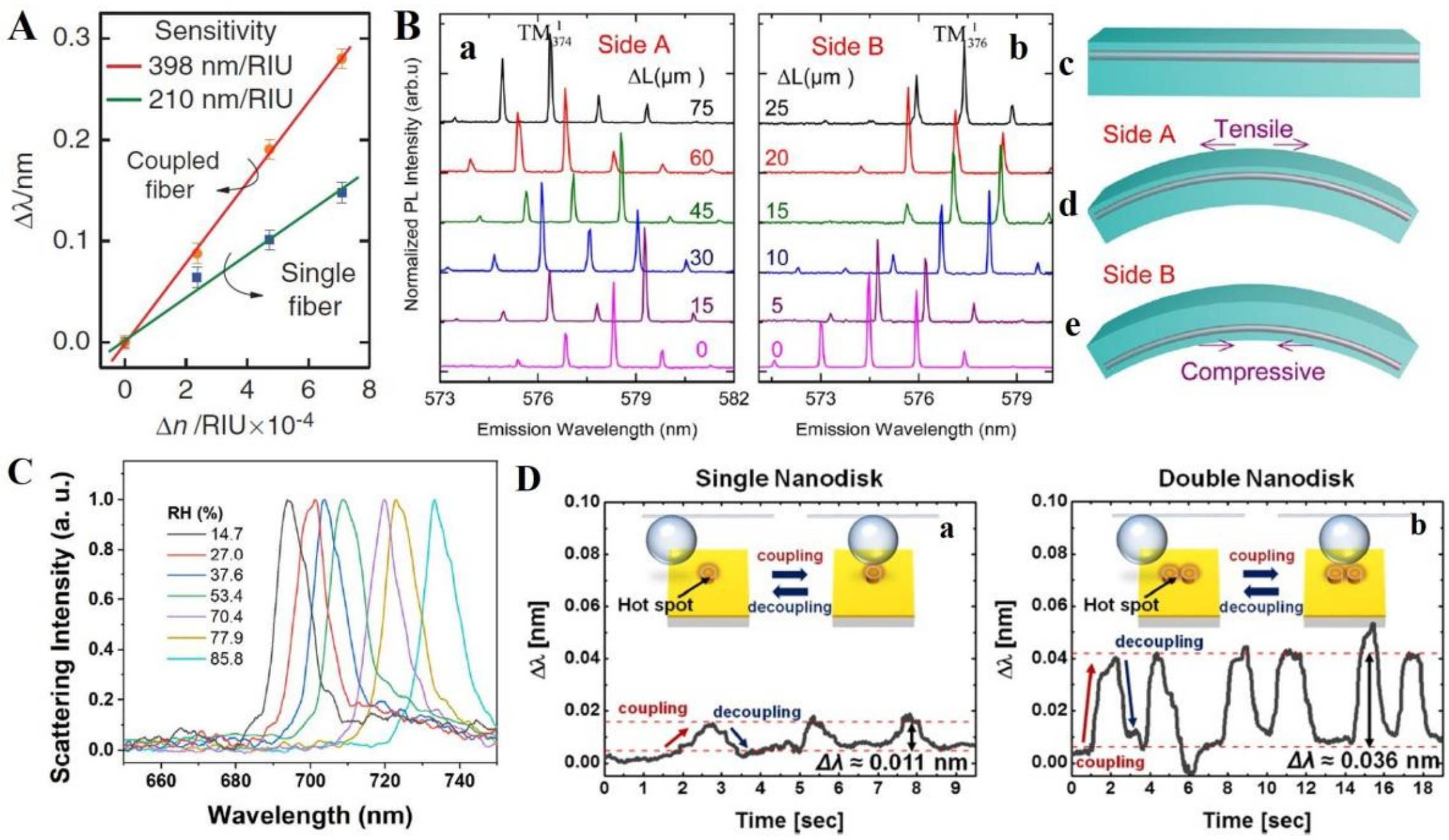

Figure 14: A: Resonance shift of the coupled and single fiber with change in RI. Adapted from ref. [174]. B: Lasing emissions (a,b) from fiber under different bending (c-e). Reproduced with permission [360]. C: Normalized scattering spectra of a humidity sensor with increasing RH. Reproduced with permission [361]. D: Experimental measurements of WGM resonance wavelength shifts due to coupling and decoupling of a microsphere to gold film on BK7 glass substrate with (a) single and (b) double nanodisks. Reproduced with permission [362].

The shift in the resonant frequency occurs with the change in the temperature due to variation in size and RI of the microcavity [363]. The temperature sensitive material polydimethylsiloxane (PDMS) has been used to fabricate [364] and coat [365, 366, 367] the microcavity to enhance the sensitivity. Xu et al have fabricated a portable phone-sized thermal sensing system by integrating the required components [368]. Owing to high thermo-optic coefficient and birefringence, temperature sensors based on liquid crystals have been demonstrated [369, 370]. Another device based on thermo-optic birefringence has been made of Z-cut lithium niobate microdisk [371]. Due to polarization dependent thermo-optic coefficient, TE and TM modes have relative resonance shifts with change in temperature. Bhattacharya et al have reported the temperature sensitivity of quantum dots doped polymer self-assembled MBRs to be 9.2 pm/°C with the resolution of 2°C [372].

Ioppolo et al fabricated a force sensor based on the silica and PMMA microspheres [373]. The effect of pressure and shear stress on cylindrical microlasers embedded in a polymeric slab [374] have shown the linear dependence of external pressure and quadratic dependence of shear stress on the mode shifts. Strain induced bidirectional tuning of laser peaks have been demonstrated using a dye doped microfibers with PDMS elastomer [360] (Fig. 14B). A refractometric air pressure sensor based on glass microsphere reported by Bianchetti et al [375] has been tested in the range of ±1.8 kPa to ±0.5 kPa.

Bhola et al have introduced a sol-gel clad polymer microring based humidity sensor with a dynamic range up to 72% RH [376]. Humidity sensitive polymers [377] and hydrogels [378]. Labrador-Pa´ez et al have reported a humidity sensor based on R6G doped glycerol droplet as a humidity sensor[379]. Armani et al have reported the tuning of microtoroid modes with external electric field with tuning rate of $85 GHz/V^2$ and can be capable of tuning more than 300 GHz [380]. The detection of electric field as small as $\sim 500 V/m$ has been demonstrated by using water filled PDMS hollow spheres [345]. An optomechanical magnetometer has been proposed by Forstner et al with sensitivity in 100nT range at 2 Hz frequency [401]. Several other magnetometers based on microcavities have also been reported previously [347, 400, 399]. A novel magnetic sensor based on the integration of WGMs of a microfiber with photonic crystal has been demonstrated [397]. Table 5 gives the consolidated values of sensitivity of the physical parameters.

#### 3.3.3.2. Hybrid cavity for physical sensing

Chamanzar et al numerically demonstrated a sensitivity of 150 nm/RIU of a ring resonator coated with an Ag layer [402]. A sensitivity higher than 200 nm/RIU and NP trapping power with gradient force of 48 pN/W for a 5 nm radius particle has been reported [382]. Similarly, Zhang et al have investigated the silicon based hybrid microcavity consisting of a silicon microring with an Ag nanoring on the top of it for RI sensing [383, 384]. Graphene based hybrid microsensor has also been proposed and investigated numerically [385]. With electrically controlled different layers of graphene, the sensitivity can be tuned. Zhang et al have also proposed a new type of hybrid microcavity for RI sensing [386]. The cavity consists of a silica strip and air-filled regions rather than a homogeneous low RI material and a metallic strip on the top of it. Urbonas et al have presented a RI sensor based on silicon microring patterned with periodically arranged Au nanodisks [387]. The hybrid cavity has shown 2-fold higher sensitivity as compared to the cavity without Au nanodisks. A double slot hybrid plasmonic ring resonator has been developed by Sun et al [388].

Table 5: Detection limit of various physical parameters using bare and hybrid cavities.

| Parameter | Geometry | Sensitivity |
|---|---|---|
| Refractive index | Microring | $20 nm/RIU$ [353], $75 nm/RIU$ [333] |
| | Microfiber | $300 nm/RIU$ [348] |
| | Microsphere | $30nm/RIU$ [352] |
| | Microfluidic chip integrated microsphere | $\sim 40nm/RIU$ [381] |
| | Microbubble | $18.8nm/RIU$ [356] |
| | Coupled OFRR | $2510nm/RIU$ [358] |
| | Coupled capillary | $3874nm/RIU$ [359] |
| | Metal microdisk coupled semiconductor microdisk | $200nm/RIU$ [382] |
| | Metal nanoring on insulator microring | $497nm/RIU$ [383] |
| | Metal nanoring on semiconductor microring | $529.6nm/RIU$ [384] |
| | Graphene sandwiched in dielectric microdisk | $1000nm/RIU$ [385] |
| | Metal microring on dielectric microdisk | $100nm/RIU$ [386] |
| | Metal nanodisk on dielectric microring | $176nm/RIU$ [387] |
| | Double slot hybrid microring | $687.5nm/RIU$ [388] |
| Temperature | Microbubble | $39pm/^{\circ}C$ [334] |
| | Microtoroid | $8.13pm/K$ [368], $0.151nm/K$ [365], $-1.17nm/K$ [335] |
| | Microfiber | $0.13nm/^{\circ}C$ [389], $0.212nm/^{\circ}C$ [390], $0.624nm/^{\circ}C$ [336], $1.79nm/^{\circ}C$ [367], |
| | Microdroplet | $0.377nm/^{\circ}C$ [391], $0.96nm/^{\circ}C$ [370] |
| | Microring | $0.11nm/^{\circ}C$ [392] |
| | Microsphere | $0.245nm/^{\circ}C$ [364] |
| | Dome shape | $\sim 0.06nm/^{\circ}C$ [366] |
| | Microdisk | $0.834GHz/K$ [371] |
| Ultrasound | Microring | $215mPa/Hz^{1/2}$ [346] |
| Pressure | Microsphere | $6.674pm/Pa$ [337] |
| | Microbubble | $58GHz/MPa$ [393], $38GHz/bar$ [394] |
| Force | Microsphere | $7.664 nm/N$ [373] |
| Humidity | Microring | $16 pm/\%RH$ [376] |
| | Microbottle | $0.107 dB/\%RH$ [344] |
| | Two microspheres | $0.23 dB/\%RH$ [395] |
| | Microdisk | $108 pm/\%RH$ [341] |
| | Microcylinder | $389.1 pm/\%RH$ [342] |
| | Microsphere on semiconductor nanorod | $0.0142 nm/\%RH$ [343] |
| | Metal nanorod coupled microfiber | $0.51 nm/\%RH$ [361] |
| Electric field | Microsphere | $\sim 500 V/m$ [345] |
| Magnetic field | Microsphere | $7.73 pm/mT$ [396] |
| | Microfiber coupled photonic crystal | $53 pm/mT$ [397] |
| | OFRR | $75.7 pm/mT$ [347] |
| | Microtoroid | $26 pT/Hz^{1/2}$ [398], $585 pT/Hz^{1/2}$ [399], $880 pT/Hz^{1/2}$ [400], $150 nT/Hz^{1/2}$ [401] |

The concentrated plasmonic field enhancement in the narrow slots is assumed to be the reason for this enhanced sensitivity. A miniature sensor has been reported using diamond nanoring coated with Au layer [403]. Experimentally, various configurations have been tried successfully to enhance the sensitivity. RI sensor based on a crescent shaped cavity with a full layer coating of metal has been numerically studied [404].

A hybrid cavity humidity sensor has been realised experimentally with a single Au nanorod coupled with the microfiber [361] (Fig. 14C). Another configuration has been shown by Kang et al in which they have used single and double plasmonic nanodisk to tune the sensitivity [362] (Fig. 14D). Experiments have showed different shifts in resonant wavelength due to coupling and decoupling of single and double microdisks. A hybrid cavity has been realised for fluid analysis consisting of a silicon disk and a coating of metal with a channel between them [405]. The transmission spectrum of a waveguide coupled to a $0.9 \mu m$ radius cavity has been observed to shift by 30 nm when RI of fluid increases by 0.141.

## 3.4 Nonlinear optics

### 3.4.1 Photonic cavity for nonlinear optical processes

WGM cavities made up of nonlinear optical materials exhibit enhanced nonlinearity even at low pump power. Braginsky et al have observed the optical bistability in a high $Q$ and low $V_{eff}$ fused quartz microspheres with a threshold power of $10\mu W$ [406]. The optical bistability of a waveguide-nonlinear hemisphere coupled system has been analyzed [407]. Raman threshold of $86\mu W$ with Raman differential conversion more than 35% for $40\mu m$ diameter sphere has been achieved. Following this, a Raman threshold of $74\mu W$ and Raman differential conversion with 45% was reported for silica toroid by Kippenberg et al [408]. Similarly, Del'Haye et al have reported various nonlinear processes observed in fused quartz cavities of diameters ranging from $170\mu m$ to $8mm$ [409] (Fig. 15A). These processes include FWM induced frequency comb generation, Brillouin forward and backward scattering. Spillane et al have reported the generation of ultralow threshold Raman laser from a spherical silica microsphere [410] (Fig. 15B). Several nonlinear optical processes such as stimulated Raman scattering (SRS), Raman assisted four-wave mixing (FWM) and stimulated Brillouin scattering (SBS) have been observed in a $70\mu m$ diameter silica microsphere. Kumar and Biswas have provided an analytical model to determine the impact of Kerr nonlinearity and SRS on WGMs of a microsphere [411, 412].

The parametric nonlinear interactions in a WGM cavity also requires the phase matching condition to be fulfilled. The cavity has rotational symmetry with orthogonal momenta. However, the dispersion due to cavity material and the geometrical mode structure do not permit efficient nonlinearity. This bottleneck has been removed with the WGM cavity with a periodically poled lithium niobate ($LiNbO_3$) [415]. Haertle studied different poling patterns of $LiNbO_3$ cavities in terms of their effective nonlinearity, spectral bandwidth and design tolerances [416]. An order of magnitude of efficiency increased when the cavity was poled along all the three axes.

Higher harmonics generation has also been studied in various systems [417, 418, 419, 420]. Dominguez-Juarez et al have reported the detection of crystal violet molecules adsorbed on silica microsphere [421]. The detection was observed by recording the intensity of the second harmonic signal with time by eliminating the inversion symmetry at the surface of the cavity. Chen et al have observed a conversion efficiency of third harmonic generation (THG) $\sim 1680\%/W^2$ at $\sim 2.90mW$ pump power after functionalizing a silica microsphere with organic molecules [413] (Fig. 15C). The continuous wave UV emission through fourth harmonic generation (FHG) from the $LiNbO_3$ cavity has been demonstrated [414] (Fig. 15D). Four equally spaced spectral lines each at ultraviolet, visible, near-infrared and infrared were observed due to cascaded $\chi^{(2)}$ harmonic process.

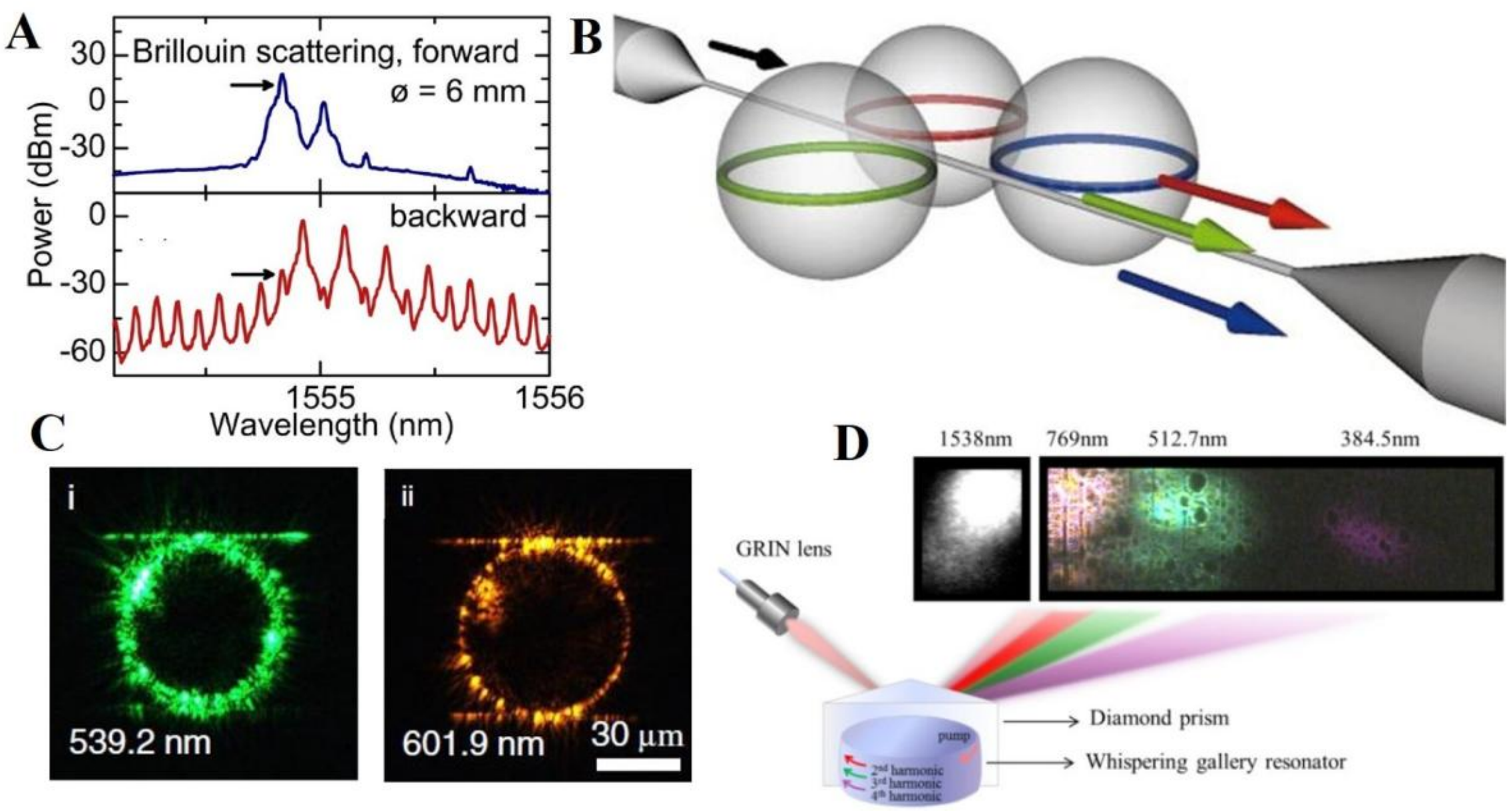

Figure 15: A: Cascaded Brillouin scattering in a 6*mm* diameter cavity. Arrows in the figure show the position of the wavelength of the pump laser. Adapted from ref. [409]. B: Cascading multiple Raman lasers along a single fibre. Black arrow represents input wave and coloured arrows show the output. Adapted from ref. [410]. C: Observation of multicolour light emissions due to third-order sum frequency generation from a functionalized microsphere. Reproduced with permission [413]. D: Cascaded-harmonic generation. Reproduced with permission [414].

Frequency comb [422, 427, 428, 429, 430, 431, 432, 433] (Fig. 16A) connects the radiowave frequencies to optical frequencies through Fourier transform which make it advantageous in different spectroscopies and sensing. Suh et al have demonstrated the dual-comb spectroscopy to obtain the absorption spectra of a sample in a microcavity [423] (Fig. 16B). The system consists of the combination of two comb trains used as the sample and the reference with slightly different repetition rates from two different microcavities. The obtained periodic signal was converted into radio frequency by Fourier transform and normalized with the reference to obtain the absorption spectrum of the sample. Other types of frequency comb techniques are direct comb spectroscopy [434] and Vernier spectroscopy [435]. Gas phase spectroscopy based on frequency comb has been demonstrated for acetylene to measure the absorption spectra of its rovibrational transition bands [436].

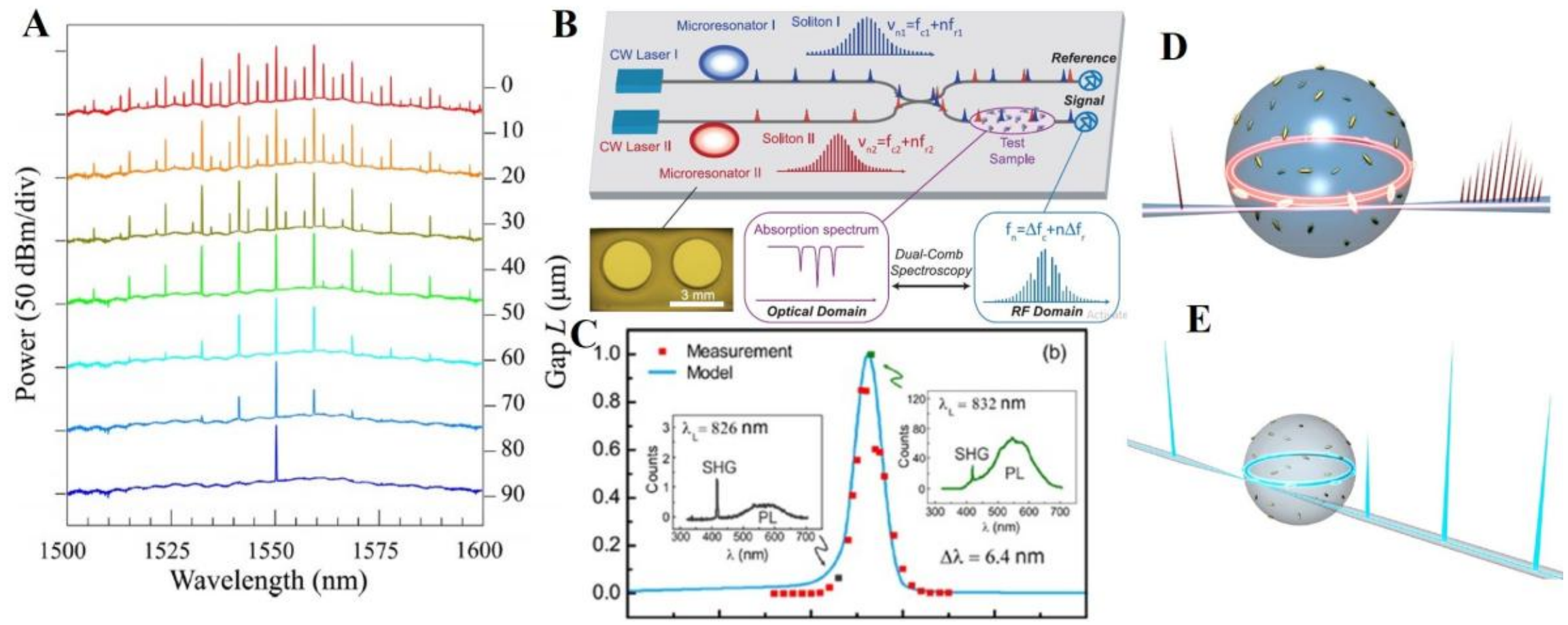

Figure 16: A: Fine tuning of optical frequency comb generation as the fiber probe approaches the microsphere. Adapted from ref. [422]. B: Microcavity based dual-comb spectroscopy. Adapted from ref. [423]. C: SHG efficiency as a function of excitation laser wavelength of gold nanorod coated microfiber. Reproduced with permission [424]. D: Schematic of Au nanorods decorated on microsphere for frequency comb generation. Reproduced with permission [425]. E: Schematic of Au nanorods decorated on microsphere for Stimulated Anti-Stokes Raman scattering. Reproduced with permission [426].

#### 3.4.2 Hybrid cavity for improved nonlinear optics

The high electric field generated by NPs trapped in the WGM cavity gives rise to a high energy density. This energy density eliminates the need for high input power for nonlinear phenomena. Pu et al have reported the enhancement of second harmonic generation (SHG) by a factor of 500 using a 100*nm* $BaTiO_3/Au$ core-shell cavity [437], although the theoretical estimates have shown enhancements to be as high as 3500. Similarly, Ren et al have demonstrated 1000 times higher efficiency of SHG from a CdS nanowire integrated with Ag nanocavities [438]. Weber et al have utilized ZnO microdisk combined with Au nanoantennas [439] to observe an increase in SHG and 3-photon PL. Ai et al have reported the multiphoton PL and SHG with linewidth as narrow as 6.4*nm* using Au nanorod coated tapered silica microfiber [424] (Fig. 16C) on pumping with a tunable fs laser.

The reduction of frequency comb generation threshold using Au nanorods coated silica microcavity has been demonstrated by Castro-Beltran et al [425] (Fig. 16D). The threshold has been found to be 15 times lower for coated cavities as compared to the uncoated cavities. Stimulated anti-Stokes' Raman Scattering with 14*mW* input power [426] (Fig. 16E) and two-photon excited fluorescence (TPF) has been observed under a weak continuous wave excitation condition from a microsphere coupled Au tip [440] have also been reported.

## 3.5 Raman spectroscopy

### 3.5.1 Photonic cavity for enhanced Raman scattering

The importance of Raman spectroscopy lies in the characterization of the trace materials based on vibrational and rotational energy levels. Due to a weak cross-section, the Raman signal is smaller by a factor of $10^{-6}$ than the incident intensity. Therefore, to enhance it, several approaches such as surface enhanced Raman scattering (SERS) [441], tip enhanced Raman scattering [442], resonance Raman scattering [443] and photonic nanojet (PNJ) from dielectric microparticles [444] have been explored till now.

When the WGMs of a microcavity are used to enhance the Raman signal from the analyte, the technique is referred to as cavity enhanced Raman scattering (CERS). The enhanced interaction of incident photon with the analyte leads to the increase in the Raman signal. Hopkins et al have used CERS to determine the size and composition of ethanol/water droplets [445]. While Mie resonances provide the size of the droplet, to probe its evaporation dynamics, the Raman peaks are used to characterize the analyte [446, 447, 448]. Rafferty and Preston have obtained the value of the imaginary refractive index as low as $5.91 \times 10^{-9}$ [449] with CERS method. Ausman and Schatz have performed a numerical investigation [450] to calculate the Raman enhancement factor of $10^8$ when incident light and Stoke shifted radiation are resonant with WGMs. Anderson has utilized silica microspheres as WGM-SERS sensors [451] by detecting the monolayer of films and subpicograms of organic materials coated on microspheres. Huang et al have observed the Raman enhancement factor of $1.4 \times 10^4$ from R6G molecules coated on the silica microspheres [452] (Fig. 17A). Ogura has observed an anomalous enhancement of Raman signal of silicon nitride when microspheres are excited at periphery instead of the center [453]. The effect of the coating on the Raman spectra of silica microspheres [454] and Raman enhancement from silicon NPs due to electric and magnetic resonances [455] have also been reported. Veluthandath and Bisht have observed the enhancement of luminescence and Raman modes of $MoS_2$ absorbed on PMMA microspheres [456]. Raman active materials $SiO_2/TiO_2$ with core-shell microsphere geometry have been used to enhance the detection of $TiO_2$ and determine the number of the coating layer [457] and to detect environmental $CO_2$ [458]. CERS of a deformed aqueous NaCl aerosol has been reported recently by Vennes et al, theoretically and experimentally [459]. Inoue and Kohno have fabricated a droplet Raman spectrometer and demonstrated enhanced Raman signal by controlling the shape of the droplet [460] (Fig. 17B).

### 3.5.2 Hybrid cavity involving SERS

Raman enhancement of silicon has been reported by placing silica microspheres on the silicon/metal/substrate structure [464]. An improved Raman signal sensitivity of kinetin, a plant bioregulator, from 841 to 4449 has been demonstrated using microspheres coated on a metal nanopillars SERS substrate [461] (Fig. 17C).

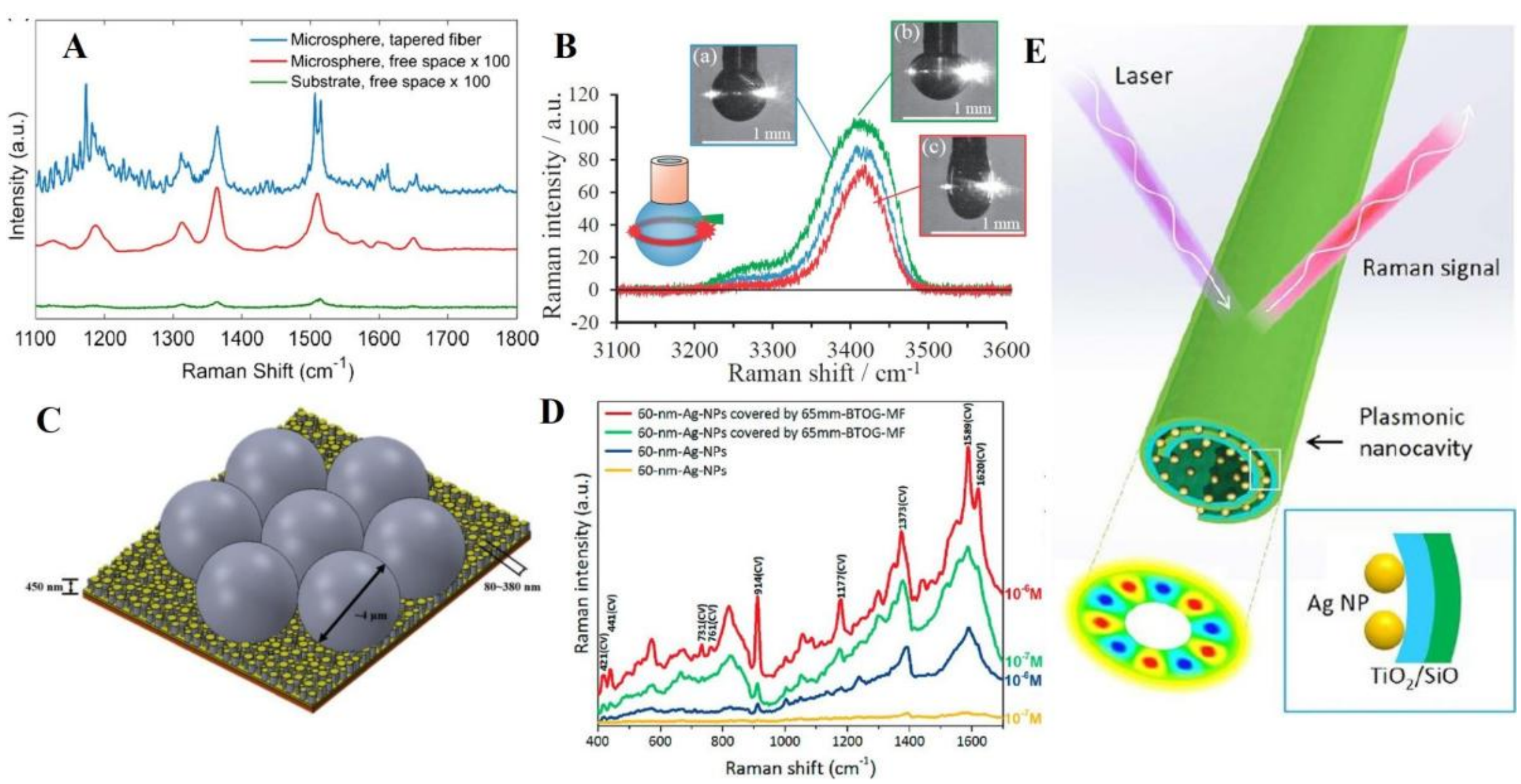

Figure 17: A: Raman spectra of dye for free-space excitation on the substrate (green) and a microsphere (red) and tapered fiber coupled microsphere excitation (blue). Reproduced with permission [452]. B: Raman spectra obtained from spherical (a) and deformed (b), (c) droplets. Reproduced with permission [460]. C: Illustration of a layer of polystyrene microspheres on metal nanopillars substrate. Adapted from ref. [461]. D: Raman spectra of dye by AgNPs covered by $BaTiO_3$ glass microsphere embedded film. Reproduced with permission [462]. E: Schematic of a whispering-gallery plasmon rolled-up nanocavity. Reproduced with permission [463].

White and Fan have demonstrated the enhanced Raman detection of R6G on the AgNPs decorated on the microsphere [465]. An enhancement of approximately 300 has been obtained. SERS based OFRR platform was used [466] to reach a detection limit of $400pM$. SERS based detection of dopamine, a neuro-modulator, has been reported with a detection limit as low as $10^{-12}M$ [271]. For this work, a SERS active hybrid microcavity made of ZnO hexagonal microrod with AgNPs decoration. Xing et al have fabricated a flexible microsphere embedded film in which an array of microspheres is embedded in the PDMS layer [462] (Fig. 17D) for Raman identification of low dimensional materials. This flexible film has been coupled with AgNPs for further enhancement. Crystal violet and Sudan I molecules have been identified at concentrations down to $10^{-7}M$. Similar to this work, a flexible system consisting of microsphere array on random AuNPs capped with PDMS has shown an enhancement factor of the order $10^7$ [467]. The microspheres coupled to AuNPs have been studied for the dependence of Raman enhancement on the relative positions of microspheres, NPs and analyte [468]. SERS using core-shell micro and nanoparticles have also been reported [469, 470, 471]. Yang et al have demonstrated SERS enhancement factor of $\sim 2 \times 10^8$ using a large area orderly arranged $SiO_2@Si$ core-shell NPs decorated with AuNPs [472]. An enhancement in Raman signal of the order of $10^5$ for R6G has been demonstrated using rolled-up tubular nanocavities decorated with AgNPs [463] (Fig. 17E). Multifunctional core-shell particles made up of lanthanide doped microspheres and AuNPs have shown enhanced Raman signal when excited at the surface of the microspheres [473]. A cascaded structure of large microsphere-small microsphere-metal NPs has been demonstrated for SERS detection with detection limit as low as $10pM$ [474]. Similar kind of sandwich structure of microsphere-metal NPs-microsphere has observed detection limit as low as $10fM$ [475].

# 4 Conclusions and future scope

The photonic mode density of a microcavity is significantly higher at its periphery to that of the free space due to WGMs at specific frequencies. Theoretical and experimental CQED studies have been performed by placing dye molecules, quantum dots, and other quantum emitters in the presence of high $Q$ microcavities with various shapes. Brief introduction along with the theoretical aspects of CQED is given here. Some of the important CQED studies performed under strong and weak coupling regimes have been reviewed for bare as well as photonic-plasmonic hybrid microcavities in this article. In addition, some experimental studies related to the modification of the Raman scattering of molecules in the presence of bare and hybrid microcavities have also been reviewed. The high-$Q$ microcavities are excellent tools for demonstrating microlasing as well non-linear effects. Recently, the threshold power for observing these effects has been reduced significantly by enhancing the field intensities with the help of hybrid microcavities. Some important reports related to low-threshold microlasing and non-linear effects have been reviewed here.

From the biosensing point of view, the wavelength corresponding to a WGM is extremely sensitive to the size, shape, and relative refractive index of the microcavity. However, the microcavities have limited efficiency for detection of single protein molecules as well as viruses. The sensitivity has been boosted several orders of magnitude by making use of hybrid microcavities. This article therefore, reviews recent biosensing experiments. Finally, a review of chemical and physical sensing performed with the help of bare and hybrid microcavities has also been incorporated. The hybrid microcavities have been proved efficient for several applications providing higher quality results, and more statistically reliable conclusions.

Advancements in fabrication technologies and numerical methodologies have further enhanced the diversity of applications. Novel materials such as QDs, TMDCs, topological insulators, 2D materials, heterostructures and alloys have been widely incorporated in the cavities to characterize them as well as to enhance the sensing capability. Work have been done towards the integrated photonic circuits using WGM cavities. Continuous improvement in the fabrication processes will help replace the electronic and optoelectronic circuits by photonic or optoplasmonic circuits. One of the few challenges in optoplasmonic circuits is the increased losses which can be overcome by including low loss plasmonic materials other than metal NPs. This requires further development of the exotic materials which can be compatible with the photonic cavities and simultaneously provide access to tune their functionalities.

**Acknowledgements.** We acknowledge financial support from the Ministry of Education, New Delhi (Sanction No. 11/9/2019-U.3 (A)) and Science and Engineering Research Board (SERB), New Delhi (Sanction no. CRG/2021/000136). SA would like to thank IIT Madras for Visiting Faculty Fellow appointment.

## References

[1] D. S. Dovzhenko, S. V. Ryabchuk, Y. P. Rakovich, I. R. Nabiev, *Nanoscale* **2018**, *10* 3589.

[2] C. Kang, S. M. Weiss, Y. A. Vlasov, S. Assefa, *Optics Letters* **2012**, *37* 2850.

[3] K. H. Jensen, M. N. Alam, B. Scherer, A. Lambrecht, N. A. Mortensen, *Optics Communications* **2008**, *281* 5335.

[4] W. Qin, A. Miranowicz, P.-B. Li, X.-Y. Lu¨, J. Q. You, F. Nori, *Physical Review Letters* **2018**, *120* 93601.

[5] D. G. Grier, *Nature* **2003**, *424* 810.

[6] V. R. Dantham, P. B. Bisht, *Journal of the Optical Society of America B* **2009**, *26* 290.

[7] S. Arnold, D. Keng, S. I. Shopova, S. Holler, W. Zurawsky, F. Vollmer, *Optics Express* **2009**, *17* 6230.

[8] A. K. Bhowmik, *Applied Optics* **2000**, *39* 3071.

[9] K. J. Vahala, *Nature* **2003**, *424* 839.

[10] Y. Akahane, T. Asano, B.-S. Song, S. Noda, *Nature* **2003**, *425* 944.

[11] L. Rayleigh, *Phil. Mag.* **1910**, *20* 1001.

[12] P. B. Bisht, K. Fukuda, S. Hirayama, *Journal of Chemical Physics* **1996**, *105* 9349.

[13] S. L. McCall, A. F. J. Levi, R. E. Slusher, S. J. Pearton, R. A. Logan, *Applied Physics Letters* **1992**, *60* 289.

[14] C. Y. Chao, W. Fung, L. J. Guo, *IEEE Journal on Selected Topics in Quantum Electronics* **2006**, *12* 134.

[15] T. Hamidfar, K. V. Tokmakov, B. J. Mangan, R. S. Windeler, A. V. Dmitriev, D. L. P. Vitullo, P. Bianucci, M. Sumetsky, *Optica* **2018**, *5* 382.

[16] F. Luan, E. Magi, T. Gong, I. Kabakova, B. J. Eggleton, *Optics Letters* **2011**, *36* 4761.

[17] M. Sumetsky, Y. Dulashko, R. S. Windeler, *Optics Letters* **2010**, *35* 898.

[18] T. J. Kippenberg, D. K. Armani, S. M. Spillane, K. J. Vahala, *OSA Trends in Optics and Photonics Series* **2003**, *88* 181.

[19] A. V. Veluthandath, S. Bhattacharya, G. S. Murugan, P. B. Bisht, *IEEE Photonics Technology Letters* **2019**, *31* 226.

[20] Z. Yao, K. Wu, B. X. Tan, J. Wang, Y. Li, Y. Zhang, A. W. Poon, *IEEE Journal of Selected Topics in Quantum Electronics* **2018**, *24*.

[21] X. F. Jiang, Y. F. Xiao, C. L. Zou, L. He, C. H. Dong, B. B. Li, Y. Li, F. W. Sun, L. Yang, Q. Gong, *Advanced Materials* **2012**, *24* 260.

[22] B. Lounis, M. Orrit, *Reports on Progress in Physics* **2005**, *68* 1129.

[23] N. Toropov, G. Cabello, M. P. Serrano, R. R. Gutha, M. Rafti, F. Vollmer, *Light: Science and Applications* **2021**, *10*.

[24] X. Jiang, A. J. Qavi, S. H. Huang, L. Yang, *Matter* **2020**, *3* 371.

[25] T. Reynolds, N. Riesen, A. Meldrum, X. Fan, J. M. M. Hall, T. M. Monro, A. Franc¸ois, *Laser and Photonics Reviews* **2017**, *11* 1.

[26] G. Frigenti, D. Farnesi, G. N. Conti, S. Soria, *Micromachines* **2020**, *11* 1.

[27] Y. Chen, H. Ming, *Photonic Sensors* **2012**, *2* 37.

[28] A. Gonzalez, C. Noguez, J. Beranek, A. S. Barnard, *Journal of Physical Chemistry C* **2014**, *118* 9128.

[29] R. Jiang, F. Qin, Y. Liu, X. Y. Ling, J. Guo, M. Tang, S. Cheng, J. Wang, *Advanced Materials* **2016**, *28* 6322.

[30] S. Biswas, J. Chakraborty, A. Agarwal, P. Kumbhakar, *RSC Advances* **2019**, *9* 34144.

[31] P. Guo, D. Sikdar, X. Huang, K. J. Si, W. Xiong, S. Gong, L. W. Yap, M. Premaratne, W. Cheng, *Nanoscale* **2012**.

[32] C. Song, F. Li, X. Guo, W. Chen, C. Dong, J. Zhang, J. Zhang, L. Wang, *Journal of Materials Chemistry B* **2019**, *7* 2001.

[33] M. J. Kale, T. Avanesian, P. Christopher, *ACS Catalysis* **2014**, *4* 116.

[34] E. Hutter, J. H. Fendler, *Advanced Materials* **2004**, *16* 1685.

[35] M. I. Stockman, K. Kneipp, S. I. Bozhevolnyi, S. Saha, A. Dutta, J. Ndukaife, N. Kinsey, H. Reddy, U. Guler, V. M. Shalaev, A. Boltasseva, B. Gholipour, H. N. S. Krishnamoorthy, K. F. Macdonald, C. Soci, N. I. Zheludev, V. Savinov, R. Singh, P. Groß, C. Lienau, M. Vadai, M. L. Solomon, D. R. Barton, M. Lawrence, J. A. Dionne, S. V. Boriskina, R. Esteban, J. Aizpurua, X. Zhang, S. Yang, D. Wang, W. Wang, T. W. Odom, N. Accanto, P. M. D. Roque, I. M. Hancu, L. Piatkowski, N. F. V. Hulst, M. F. Kling, *Journal of Optics* **2018**, *20* 1.

[36] G. V. Naik, V. M. Shalaev, A. Boltasseva, *Advanced Materials* **2013**, *25* 3264.

[37] P. Yang, J. Zheng, Y. Xu, Q. Zhang, L. Jiang, *Advanced Materials* **2016**, *28* 10508.

[38] J. Boken, P. Khurana, S. Thatai, D. Kumar, S. Prasad, *Applied Spectroscopy Reviews* **2017**, *52* 774.

[39] C. Klusmann, J. Oppermann, P. Forster, C. Rockstuhl, H. Kalt, *ACS Photonics* **2018**, *5* 2365.

[40] M. Manzo, R. Schwend, *Journal of Engineering and Science in Medical Diagnostics and Therapy* **2019**, *2* 1.

[41] F. Pan, K. C. Smith, H. L. Nguyen, K. A. Knapper, D. J. Masiello, R. H. Goldsmith, *Nano Letters* **2020**, *20* 50.

[42] M. A. Garaei, M. Saliminasab, H. Nadgaran, R. Moradian, *Plasmonics* **2017**, *12* 1953.

[43] M. Aeschlimann, T. Brixner, M. Cinchetti, B. Frisch, B. Hecht, M. Hensen, B. Huber, C. Kramer, E. Krauss, T. H. Loeber, W. Pfeiffer, M. Piecuch, P. Thielen, *Light: Science and Applications* **2017**, *6* 17111.

[44] A. Bozzola, S. Perotto, F. D. Angelis, *Analyst* **2017**, *142* 883.

[45] J. Xavier, S. Vincent, F. Meder, F. Vollmer, *Nanophotonics* **2018**, *7* 1.

[46] Y. Chen, Y. Yin, L. Ma, O. G. Schmidt, *Advanced Optical Materials* **2021**, *2100143* 2100143.

[47] Z. Qian, L. Shan, X. Zhang, Q. Liu, Y. Ma, Q. Gong, Y. Gu, *PhotoniX* **2021**, *2* 21.

[48] Y. Hong, B. M. Reinhard, *Journal of Optics* **2019**, *21* 1.

[49] M. Vaughan, *The Fabry-Perot Interferometer: History, Theory, Practice and Applications*, Taylor & Francis, **1989**.

[50] M. L. Gorodetskii, A. E. Fomin, *Quantum Electronics* **2007**, *37* 167.

[51] G. Roll, G. Schweiger, *Journal of the Optical Society of America A* **2000**, *17* 1301.

[52] J. P. Braud, P. L. Hagelstein, *IEEE Journal of Quantum Electronics* **1991**, *27* 1069.

[53] G. Mie, *Annalen der Physik* **1908**, *330* 377.

[54] F. Vollmer, D. Yu, *Optical Whispering Gallery Modes for Biosensing*, **2020**.

[55] C. C. Lam, P. T. Leung, K. Young, *Journal of the Optical Society of America B* **1992**, *9* 1585.

[56] B. R. Johnson, *Journal of the Optical Society of America A* **1993**, *10* 343.

[57] S.-X. Zhang, L. Wang, Z.-Y. Li, Y. Li, Q. Gong, Y.-F. Xiao, *Optics Letters* **2016**, *41* 4437.

[58] R. Ulrich, *Journal of the Optical Society of America* **1971**, *61* 1467.

[59] A. Serpengu¨zel, G. Griffel, S. Arnold, *Optics Letters* **1995**, *20* 654.

[60] M. Cai, O. Painter, K. J. Vahala, *Physical Review Letters* **2000**, *85* 74.

[61] M. Cai, K. Vahala, *Optics Letters* **2000**, *25* 260.

[62] M. L. Gorodetsky, A. A. Savchenkov, V. S. Ilchenko, *Proceedings of SPIE - The International Society for Optical Engineering* **1996**, *2799* 389.

[63] T. J. Kippenberg, S. M. Spillane, K. J. Vahala, *Applied Physics Letters* **2004**, *85* 6113.

[64] E. S. Hosseini, S. Yegnanarayanan, M. Soltani, A. Adibi, *Optics Express* **2009**, *17* 14543.

[65] A. E. Dorche, B. Wei, C. Raman, A. Adibi, *Optics Letters* **2020**, *45* 5958.

[66] Y. nan Zhang, T. Zhou, B. Han, A. Zhang, Y. Zhao, *Nanoscale* **2018**, *10* 13832.

[67] D. Yang, A. Wang, J.-H. Chen, X.-C. Yu, C. Lan, Y. Ji, Y.-F. Xiao, *Photonics Research* **2020**, *8* 497.

[68] M. Po¨llinger, D. O'Shea, F. Warken, A. Rauschenbeutel, *Physical Review Letters* **2009**, *103* 1.

[69] M. Fox, *Quantum Optics: An Introduction*, OUP Oxford, **2006**.

[70] E. M. Purcell, Spontaneous emission probabilities at radio frequencies, **1946.**

[71] H. Chew, *Physical Review A* **1988**, *38* 3410.

[72] H. Chew, *The Journal of Chemical Physics* **1987**, *87* 1355.

[73] E. T. Jaynes, F. W. Cummings, *Proceedings of the IEEE* **1963**, *51* 89.

[74] C. Noguez, *Optical Materials* **2005**, *27* 1204.

[75] H. Chen, X. Kou, Z. Yang, W. Ni, J. Wang, *Langmuir* **2008**, *24* 5233.

[76] S. A. Lee, S. Link, *Accounts of Chemical Research* **2021**, *54* 1950.

[77] U. Kreibig, M. Vollmer, *Theoretical Considerations BT - Optical Properties of Metal Clusters*, 13–201, Springer Berlin Heidelberg, ISBN 978-3-662-09109-8, **1995.**

[78] K. H. Drexhage, *Journal of Luminescence* **1970**, *1-2* 693.

[79] D. Kleppner, *Physical Review Letters* **1981**, *47* 233.

[80] R. G. Hulet, E. S. Hilfer, D. Kleppner, *Physical Review Letters* **1985**, *55* 2137.

[81] S. Haroche, D. Kleppner, *Physics Today* **1989**, *42* 24.

[82] H.-B. Lin, J. D. Eversole, C. D. Merritt, A. J. Campillo, *Physical Review A* **1992**, *45* 6756.

[83] A. J. Campillo, J. D. Eversole, H.-B. Lin, *Physical Review Letters* **1991**, *67* 437.

[84] M. D. Barnes, W. B. Whitten, S. Arnold, J. M. Ramsey, *The Journal of Chemical Physics* **1992**, *97* 7842.

[85] S. Ishizaka, Y. Suzuki, N. Kitamura, *Physical Chemistry Chemical Physics* **2010**, *12* 9852.

[86] T. Agrawal, S. Bhattacharya, V. K. Sagar, P. B. Bisht, *Journal of Luminescence* **2021**, *234* 117963.

[87] C. Junge, D. O'Shea, J. Volz, A. Rauschenbeutel, *Physical Review Letters* **2013**, *110* 1.

[88] Q. Shang, S. Zhang, Z. Liu, J. Chen, P. Yang, C. Li, W. Li, Y. Zhang, Q. Xiong, X. Liu, Q. Zhang, *Nano Letters* **2018**, *18* 3335.

[89] M. D. Barnes, W. B. Whitten, J. Ramsey, *Chemical Physics Letters* **1994**, *227* 628.

[90] M. D. Barnes, W. B. Whitten, J. M. Ramsey, *Journal of the Optical Society of America B* **1994**, *11* 1297.

[91] A. Kiraz, P. Michler, C. Becher, B. Gayral, A. Imamoˇglu, L. Zhang, E. Hu, W. V. Schoenfeld, P. M. Petroff, *Applied Physics Letters* **2001**, *78* 3932.

[92] C. T. Yuan, Y. C. Wang, Y. C. Yang, M. C. Wu, J. Tang, M. H. Shih, *Applied Physics Letters* **2011**, *99* 53116.

[93] T. Baba, D. Sano, *IEEE Journal on Selected Topics in Quantum Electronics* **2003**, *9* 1340.

[94] A. V. Veluthandath, P. B. Bisht, *Journal of Applied Physics* **2015**, *118*.

[95] P. Sandeep, P. B. Bisht, *Chemical Physics Letters* **2005**, *415* 15.

[96] P. Sandeep, P. B. Bisht, *Journal of Chemical Physics* **2005**, *123* 1.

[97] R. D. Kekatpure, A. R. Guichard, M. L. Brongersma, *IEEE International Conference on Group IV Photonics GFP* **2007**, 264–266.

[98] A. Pitanti, M. Ghulinyan, D. Navarro-Urrios, G. Pucker, L. Pavesi, *Physical Review Letters* **2010**, *104* 1.

[99] T. J. Kippenberg, A. L. Tchebotareva, J. Kalkman, A. Polman, K. J. Vahala, *Physical Review Letters* **2009**, *103* 1.

[100] V. R. Dantham, P. B. Bishta, *European Physical Journal Applied Physics* **2010**, *51* 1.

[101] R. Nakajima, A. Miura, S. Abe, N. Kitamura, *Analytical Chemistry* **2021**, *93* 5218.

[102] C.-H. Chien, S.-H. Wu, T. H.-B. Ngo, Y.-C. Chang, *Physical Review Applied* **2019**, *11* 51001.

[103] Y. Mi, Y. Wu, J. Shi, S.-N. Luo, *Journal of Materials Chemistry C* **2020**, *8* 11201.

[104] A. Faraon, P. E. Barclay, C. Santori, K.-M. C. Fu, R. G. Beausoleil, *Nature Photonics* **2011**, *5* 301.

[105] A. S. Zalogina, R. S. Savelev, E. V. Ushakova, G. P. Zograf, F. E. Komissarenko, V. A. Milichko, S. V. Makarov, D. A. Zuev, I. V. Shadrivov, *Nanoscale* **2018**, *10* 8721.

[106] Y. Mi, Z. Zhang, L. Zhao, S. Zhang, J. Chen, Q. Ji, J. Shi, X. Zhou, R. Wang, J. Shi, W. Du, Z. Wu, X. Qiu, Q. Zhang, Y. Zhang, X. Liu, *Small* **2017**, *13* 1701694.

[107] H. Lee, V. T. Nguyen, J.-Y. Park, J. Lee, *Nanoscale* **2021**, *13* 4262.

[108] S. Bhattacharya, A. V. Veluthandath, P. B. Bisht, *Materials Research Express* **2018**, *5*.

[109] L. Eswaramoorthy, S. Mokkapati, A. Kumar, *Journal of Physics D: Applied Physics* **2022**, *55* 225103.

[110] Y. Yan, L. Yang, W. Liu, Q. Wang, S. Li, C. Xu, *Advanced Optical Materials* **2021**, *9* 2001932.

[111] C.-S. Wu, S.-C. Wu, B.-T. Yang, Z. Y. Wu, Y. H. Chou, P. Chen, H.-C. Hsu, *ACS Applied Materials & Interfaces* **2021**, *13* 13556.

[112] P. Sandeep, P. B. Bisht, *Indian Journal of Physics* **2001**, *75B* 539.

[113] T. Klar, M. Perner, S. Grosse, G. von Plessen, W. Spirkl, J. Feldmann, *Physical Review Letters* **1998**, *80* 4249.

[114] T. J. Kippenberg, S. M. Spillane, K. J. Vahala, *Optics Letters* **2002**, *27* 1669.

[115] S. M. Spillane, T. J. Kippenberg, K. J. Vahala, K. W. Goh, E. Wilcut, H. J. Kimble, *Physical Review A - Atomic, Molecular, and Optical Physics* **2005**, *71* 1.

[116] J. P. Reithmaier, G. Sek, A. L¨offler, C. Hofmann, S. Kuhn, S. Reitzenstein, L. V. Keldysh, V. D. Kulakovskii, T. L. Reinecke, A. Forchel, *Nature* **2004**, *432* 197.

[117] B. D. Jones, M. Oxborrow, V. N. Astratov, M. Hopkinson, A. Tahraoui, M. S. Skolnick, A. M. Fox, *Optics Express* **2010**, *18* 22578.

[118] E. Peter, P. Senellart, D. Martrou, A. Lemaˆıtre, J. Hours, J. M. G´erard, J. Bloch, *Physical Review Letters* **2005**, *95* 67401.

[119] E. Peter, J. Bloch, D. Martrou, A. Lemaˆıtre, J. Hours, G. Patriarche, A. Cavanna, J. M. G´erard, S. Laurent, I. Robert-Philip, P. Senellart, *physica status solidi (b)* **2006**, *243* 3879.

[120] E. Peter, P. Senellart, D. Martrou, A. Lemaˆıtre, J. Hours, J.-M. G´erard, J. Bloch, *physica status solidi (c)* **2005**, *2* 3825.

[121] K. Srinivasan, C. P. Michael, R. Perahia, O. Painter, *Physical Review A Atomic, Molecular, and Optical Physics* **2008**, *78* 1.

[122] J. Renner, L. Worschech, A. Forchel, S. Mahapatra, K. Brunner, *Applied Physics Letters* **2008**, *93* 151109.

[123] T. Aoki, B. Dayan, E. Wilcut, W. P. Bowen, A. S. Parkins, T. J. Kippenberg, K. J. Vahala, H. J. Kimble, *Nature* **2006**, *443* 671.

[124] D. J. Alton, N. P. Stern, T. Aoki, H. Lee, E. Ostby, K. J. Vahala, H. J. Kimble, *Nature Physics* **2011**, *7* 159.

[125] Y. S. Park, A. K. Cook, H. Wang, *Nano Letters* **2006**, *6* 2075.

[126] P. Ghosh, D. Yu, G. Li, M. Huang, Y. Liu, *Optik* **2021**, *240* 166829.

[127] P. Ghosh, D. Yu, T. Hu, J. Liang, Z. Chen, L. Yingkai, M. Huang, *Journal of Materials Science* **2019**, *54* 8472.

[128] S.-H. Gong, S.-M. Ko, M.-H. Jang, Y.-H. Cho, *Nano Letters* **2015**, *15* 4517.

[129] M. Jiang, P. Wan, K. Tang, M. Liu, C. Kan, *Nanoscale* **2021**, *13* 5448.

[130] M. Jiang, K. Tang, P. Wan, T. Xu, H. Xu, C. Kan, *Nanoscale* **2021**, *13* 1663.

[131] S. Luo, Y. Wang, L. Liao, Z. Zhang, X. Shen, Z. Chen, *Journal of Applied Physics* **2020**, *127* 25702.

[132] A. B. Vasista, W. L. Barnes, *Nano Letters* **2020**, *20* 1766.

[133] A. B. Vasista, W. L. Barnes, *The Journal of Physical Chemistry Letters* **2022**, *13* 1019.

[134] E. Will, L. Masters, A. Rauschenbeutel, M. Scheucher, J. Volz, *Physical Review Letters* **2021**, *126* 233602.

[135] W. G. Farr, M. Goryachev, D. L. Creedon, M. E. Tobar, *Physical Review B Condensed Matter and Materials Physics* **2014**, *90* 1.

[136] M.-L. Hu, Z.-J. Yang, X.-J. Du, L. Ma, J. He, *Optics Express* **2021**, *29* 26028.

[137] D. Timmerman, Y. Matsude, Y. Sasaki, S. Ichikawa, J. Tatebayashi, Y. Fujiwara, *Physical Review Applied* **2020**, *14* 64059.

[138] J. Bleuse, J. Claudon, M. Creasey, N. S. Malik, J. M. G´erard, I. Maksymov, J. P. Hugonin, P. Lalanne, *Physical Review Letters* **2011**, *106*.

[139] Z. Tang, H. Zheng, Y. Wang, R. Wang, Z. Qiu, Y. Shen, J. Zhou, S. Su, L. Li, H. Zhu, *Nanoscale* **2022**, *14* 7589.

[140] C. H. Cho, C. O. Aspetti, M. E. Turk, J. M. Kikkawa, S. W. Nam, R. Agarwal, *Nature Materials* **2011**, *10* 669.

[141] H. M. Doeleman, E. Verhagen, A. F. Koenderink, *ACS Photonics* **2016**, *3* 1943.

[142] J. Zhao, W. Qiu, Y. Huang, J.-X. Wang, Q. Kan, J.-Q. Pan, *Optics Letters* **2014**, *39* 5527.

[143] Z. Chai, W. Wang, Z. Tian, Y. Xu, *Journal of Optics* **2022**, *24* 64001.

[144] H. Xu, C. Miao, M. Jiang, Y. Liu, C. Kan, D. Shi, *Journal of Luminescence* **2021**, *235* 118016.

[145] E. J. R. Vesseur, F. J. G. D. Abajo, A. Polman, *Physical Review B - Condensed Matter and Materials Physics* **2010**, *82* 1.

[146] M. Frimmer, A. F. Koenderink, *Physical Review B - Condensed Matter and Materials Physics* **2012**, *86* 1.

[147] K. G. Cogn´ee, H. M. Doeleman, P. Lalanne, A. F. Koenderink, *Light: Science and Applications* **2019**, *8*.

[148] H. M. Doeleman, C. D. Dieleman, C. Mennes, B. Ehrler, A. F. Koenderink, *ACS Nano* **2020**, *14* 12027.

[149] Z. Gu, S. Liu, S. Sun, K. Wang, Q. Lyu, S. Xiao, Q. Song, *Scientific Reports* **2015**, *5* 1.

[150] B. Romeira, A. Fiore, *IEEE Journal of Quantum Electronics* **2018**, *54* 1.

[151] J. Yu, L. Yang, F. Gu, *Applied Physics Express* **2021**, *14*.

[152] V. Srinivasan, S. S. Ramamurthy, *Journal of Physical Chemistry C* **2016**, *120* 2908.

[153] J. Zhao, W. Zhang, T. Wen, L. Ye, H. Lin, J. Tang, Q. Gong, G. Lyu, *Chinese Physics B* **2021**, *30* 114215.

[154] H.-W. Wu, J.-Q. Quan, Y.-Q. Yin, Z.-Q. Sheng, *Journal of the Optical Society of America B* **2020**, *37* 98.

[155] B. Jiang, H. Dai, X. Chen, *Photonics Research* **2018**, *6* 597.

[156] Y. F. Xiao, Y. C. Liu, B. B. Li, Y. L. Chen, Y. Li, Q. Gong, *Physical Review A - Atomic, Molecular, and Optical Physics* **2012**, *85* 1.

[157] H. Yokoyama, S. D. Brorson, *Journal of Applied Physics* **1989**, *66* 4801.

[158] V. D. Ta, R. Chen, H. D. Sun, *Scientific Reports* **2013**, *3* 1.

[159] X. Wang, Q. Liao, Q. Kong, Y. Zhang, Z. Xu, X. Lu, H. Fu, *Angewandte Chemie* **2014**, *126* 5973.

[160] C. Wei, S. Y. Liu, C. L. Zou, Y. Liu, J. Yao, Y. S. Zhao, *Journal of the American Chemical Society* **2015**, *137* 62.

[161] V. D. Ta, R. Chen, H. D. Sun, *Advanced Materials* **2012**, *24*.

[162] G. Q. Wei, Y. Yu, M. P. Zhuo, X. D. Wang, L. S. Liao, *Journal of Materials Chemistry C* **2020**, *8* 11916.

[163] Z. Yu, Y. Wu, Q. Liao, H. Zhang, S. Bai, H. Li, Z. Xu, C. Sun, X. Wang, J. Yao, H. Fu, *Journal of the American Chemical Society* **2015**, *137* 15105.

[164] F. Gu, F. Xie, X. Lin, S. Linghu, W. Fang, H. Zeng, L. Tong, S. Zhuang, *Light: Science and Applications* **2017**, *6* 1.

[165] Z. Wang, Y. Ma, H. Zhou, G. Feng, G. Deng, S. Zhou, *Laser Physics Letters* **2020**, *17*.

[166] J. Lin, Y. Xu, J. Song, B. Zeng, F. He, H. Xu, K. Sugioka, W. Fang, Y. Cheng, *Optics Letters* **2013**, *38* 1458.

[167] P. Brenner, O. Bar-On, T. Siegle, T. Leonhard, R. Gvishi, C. Eschenbaum, H. Kalt, J. Scheuer, U. Lemmer, *Applied Optics* **2017**, *56* 3703.

[168] F. Tabataba-Vakili, L. Doyennette, C. Brimont, T. Guillet, S. Rennesson, E. Frayssinet, B. Damilano, J. Y. Duboz, F. Semond, I. Roland, M. E. Kurdi, X. Checoury, S. Sauvage, B. Gayral, P. Boucaud, *ACS Photonics* **2018**, *5* 3643.

[169] Q. Guo, J. Zhang, C. Ning, N. Zhuo, S. Zhai, J. Liu, L. Wang, S. Liu, Z. Jia, F. Liu, *ACS Photonics* **2022**, *9* 1172.

[170] H. Wang, S. Liu, L. Chen, D. Shen, X. Wu, *Scientific Reports* **2016**, *6* 1.

[171] M. Schubert, A. Steude, P. Liehm, N. M. Kronenberg, M. Karl, E. C. Campbell, S. J. Powis, M. C. Gather, *Nano Letters* **2015**, *15* 5647.

[172] L. Shang, L. Liu, L. Xu, *Optics Letters* **2008**, *33* 1150.

[173] M. A. Royz, A. N. Baranov, A. N. Imenkov, D. S. Burenina, A. A. Pivovarova, A. M. Monakhov, E. A. Grebenshchikova, Y. P. Yakovlev, *Semiconductors* **2017**, *51* 1224.

[174] V. D. Ta, R. Chen, H. Sun, *Advanced Optical Materials* **2014**, *2* 220.

[175] J. Yan, D. N. Wang, X. Liu, J. Chen, *Optics and Laser Technology* **2021**, *142* 107254.

[176] C. Feng, Z. Xu, X. Wang, H. Yang, L. Zheng, H. Fu, *ACS Applied Materials and Interfaces* **2017**, *9* 7385.

[177] Z. Lv, Z. Man, Z. Xu, C. Feng, Y. Yang, Q. Liao, X. Wang, L. Zheng, H. Fu, *ACS Applied Materials and Interfaces* **2018**, *10* 32981.

[178] Q. Lu, X. Wu, L. Liu, L. Xu, *Optics Express* **2015**, *23* 22740.

[179] T. Siegle, J. Kellerer, M. Bonenberger, S. Kra¨mmer, C. Klusmann, M. Mu¨ller, H. Kalt, *Optics Express* **2018**, *26* 3579.

[180] Y. Wei, X. Lin, C. Wei, W. Zhang, Y. Yan, Y. S. Zhao, *ACS Nano* **2017**, *11* 597.

[181] T. V. Nguyen, N. V. Pham, H. H. Mai, D. C. Duong, H. H. Le, R. Sapienza, V. duong Ta, *Soft Matter* **2019**, *15* 9721.

[182] Y. L. Sun, Z. S. Hou, S. M. Sun, B. Y. Zheng, J. F. Ku, W. F. Dong, Q. D. Chen, H. B. Sun, *Scientific Reports* **2015**, *5* 1.

[183] T. V. Nguyen, H. H. Mai, T. V. Nguyen, D. C. Duong, V. D. Ta, *Journal of Physics D: Applied Physics* **2020**, *53* 445104.

[184] H. H. Mai, T. T. Nguyen, K. M. Giang, X. T. Do, T. T. Nguyen, H. C. Hoang, V. D. Ta, *Soft Matter* **2020**, *16* 9069.

[185] M. Humar, S. H. Yun, *Nature Photonics* **2015**, *9* 572.

[186] M. Schubert, K. Volckaert, M. Karl, A. Morton, P. Liehm, G. B. Miles, S. J. Powis, M. C. Gather, *Scientific Reports* **2017**, *7* 1.

[187] M. Humar, A. Dobravec, X. Zhao, S. H. Yun, *Optica* **2017**, *4* 1080.

[188] M. Manzo, O. Cavazos, E. Ramirez-Cedillo, H. R. Siller, *Journal of Engineering and Science in Medical Diagnostics and Therapy* **2020**, *3* 1.

[189] V. D. Ta, T. V. Nguyen, Q. V. Pham, T. V. Nguyen, *Optics Communications* **2020**, *459* 124925.

[190] N. Toropov, F. Vollmer, *Light: Science & Applications* **2021**, *10* 6.

[191] S. J. Tang, P. H. Dannenberg, A. C. Liapis, N. Martino, Y. Zhuo, Y. F. Xiao, S. H. Yun, *Light: Science and Applications* **2021**, *10*.

[192] Z. Yuan, X. Tan, X. Gong, C. Gong, X. Cheng, S. Feng, X. Fan, Y. C. Chen, *Nanoscale* **2021**, *13* 1608.

[193] X. Li, Y. Qin, X. Tan, Y. cheng Chen, Q. Chen, W. hung Weng, *ACS Photonics* **2019**, *6* 531.

[194] Q. Chen, X. Zhang, Y. Sun, M. Ritt, S. Sivaramakrishnan, X. Fan, *Lab on a Chip* **2013**, 1–3.

[195] F. Azeem, L. S. Trainor, A. Gao, M. Isarov, D. V. Strekalov, H. G. L. Schwefel, *Advanced Optical Materials* **2022**, *10* 2102137.

[196] S. K. Vanga, A. A. Bettiol, *Nuclear Instruments and Methods in Physics Research, Section B: Beam Interactions with Materials and Atoms* **2015**, *348* 209.

[197] R. Duan, Z. Zhang, L. Xiao, X. Zhao, Y. T. Thung, L. Ding, Z. Liu, J. Yang, V. D. Ta, H. Sun, *Advanced Materials* **2022**, *34* 2108884.

[198] Y. Wang, H. Li, L. Zhao, Y. Liu, S. Liu, J. Yang, *Applied Physics Letters* **2016**, *109* 0.

[199] N. B. Tomazio, L. D. Boni, C. R. Mendonca, *Scientific Reports* **2017**, *7* 1.

[200] Y. Lv, Y. J. Li, J. Li, Y. Yan, J. Yao, Y. S. Zhao, *Journal of the American Chemical Society* **2017**, *139* 11329.

[201] Q. Liao, K. Hu, H. Zhang, X. Wang, J. Yao, H. Fu, *Advanced Materials* **2015**, *27* 3405.

[202] Y. Kawabe, C. Spiegelberg, A. Schu¨lzgen, M. F. Nabor, B. Kippelen, E. A. Mash, P. M. Allemand, M. Kuwata-Gonokami, K. Takeda, N. Peyghambarian, *Applied Physics Letters* **1998**, *72* 141.

[203] A. P. Tarasov, A. S. Lavrikov, L. A. Zadorozhnaya, V. M. Kanevsky, *JETP Letters* **2022**, *115* 502.

[204] H. M. Dong, Y. H. Yang, G. W. Yang, *Scientific Reports* **2015**, *5* 1.

[205] Y. Wang, F. Qin, J. Lu, J. Li, Z. Zhu, Q. Zhu, Y. Zhu, Z. Shi, C. Xu, *Nano Research* **2017**, *10* 3447.

[206] Z. Chen, B. Lai, J. Zhang, G. Wang, S. Chu, *Nanotechnology* **2014**, *25*.

[207] A. Kiraz, Q. Chen, X. Fan, *ACS Photonics* **2015**, *2* 707.

[208] Y. Wang, V. D. Ta, K. S. Leck, B. H. I. Tan, Z. Wang, T. He, C. D. Ohl, H. V. Demir, H. Sun, *Nano Letters* **2017**, *17* 2640.

[209] Q. Zhang, R. Su, X. Liu, J. Xing, T. C. Sum, Q. Xiong, *Advanced Functional Materials* **2016**, *26* 6238.

[210] K. Wang, S. Sun, C. Zhang, W. Sun, Z. Gu, S. Xiao, Q. Song, *Materials Chemistry Frontiers* **2017**, *1* 477.

[211] S. Chen, X.-D. Wang, M.-P. Zhuo, G.-Q. Wei, J.-J. Wu, L.-S. Liao, *Advanced Optical Materials* **2022**, *10* 2101931.

[212] T. Wang, C. K. Siu, H. Yu, Y. Wang, S. Li, W. Lu, J. Hao, H. Liu, J. H. Teng, D. Y. Lei, X. Xu, S. F. Yu, *Inorganic Chemistry* **2018**, *57* 8200.

[213] S. I. Shopova, G. Farca, A. T. Rosenberger, W. M. S. Wickramanayake, N. A. Kotov, *Applied Physics Letters* **2004**, *85* 6101.

[214] M. Maqbool, K. Main, M. Kordesch, *Optics letters* **2010**, *35* 3637.

[215] A. E. Zhukov, N. V. Kryzhanovskaya, E. I. Moiseev, M. V. Maximov, *Light: Science and Applications* **2021**, *10*.

[216] L. Yang, D. K. Armani, K. J. Vahala, *Applied Physics Letters* **2003**, *83* 825.

[217] L. Yang, K. J. Vahala, *Optics Letters* **2003**, *28* 592.

[218] A. Pal, S. Y. Chen, R. Sen, T. Sun, K. T. V. Grattan, *Laser Physics Letters* **2013**, *10*.

[219] X. Fu, X. Fu, Y. Chen, L. Qin, H. Peng, R. Shi, F. Li, Q. Zhou, Y. Wang, Y. Zhou, Y. Ning, *Journal of Physical Chemistry Letters* **2020**, *11* 541.

[220] J. Ruan, D. Guo, B. Niu, K. Ge, T. Zhai, *NPG Asia Materials* **2022**, *14* 62.

[221] C. Shi, S. Soltani, A. M. Armani, *Nano Letters* **2013**, *13* 5827.

[222] M. P. Nezhad, A. Simic, O. Bondarenko, B. Slutsky, A. Mizrahi, L. Feng, V. Lomakin, Y. Fainman, *Nature Photonics* **2010**, *4* 395.

[223] C. Xu, F. Qin, Q. Zhu, J. Lu, Y. Wang, J. Li, Y. Lin, Q. Cui, Z. Shi, A. G. Manohari, *Nano Research* **2018**, *11* 3050.

[224] Y. Wang, G. Zhu, J. Mei, C. Tian, H. Liu, F. Wang, D. Zhao, *AIP Advances* **2017**, *7*.

[225] E. I. Moiseev, N. Kryzhanovskaya, Y. S. Polubavkina, M. V. Maximov, M. M. Kulagina, Y. M. Zadiranov, A. A. Lipovskii, I. S. Mukhin, A. M. Mozharov, F. E. Komissarenko, Z. F. Sadrieva, A. E. Krasnok, A. A. Bogdanov, A. V. Lavrinenko, A. E. Zhukov, *ACS Photonics* **2017**, *4* 275.

[226] S. Arnold, M. Khoshsima, I. Teraoka, *Optics Letters* **2003**, *28* 272.

[227] F. Vollmer, S. Arnold, D. Keng, *Proceedings of the National Academy of Sciences of the United States of America* **2008**, *105* 20701.

[228] F. Vollmer, S. Roy, *Journal of the Indian Institute of Science* **2012**, *92* 233.

[229] J. Zhu, S. K. Ozdemir, Y. F. Xiao, L. Li, L. He, D. R. Chen, L. Yang, *Nature Photonics* **2010**, *4* 46.

[230] F. Vollmer, D. Braun, A. Libchaber, M. Khoshsima, I. Teraoka, S. Arnold, *Applied Physics Letters* **2002**, *80* 4057.

[231] M. Noto, M. Khoshsima, D. Keng, I. Teraoka, V. Kolchenko, S. Arnold, *Applied Physics Letters* **2005**, *87* 1.

[232] I. Teraoka, S. Arnold, *Journal of Applied Physics* **2007**, *102* 1.

[233] M. Noto, D. Keng, I. Teraoka, S. Arnold, *Biophysical Journal* **2007**, *92* 4466.

[234] H. Zhu, I. M. White, J. D. Suter, P. S. Dale, X. Fan, *Optics Express* **2007**, *15* 9139.

[235] C. E. Soteropulos, K. M. Zurick, M. T. Bernards, H. K. Hunt, *Langmuir* **2012**, *28* 15743.

[236] Y. Wang, H. Zhang, S. Duan, W. Lin, B. Liu, J. Wu, *IEEE Sensors Journal* **2021**, *21* 9148.

[237] H. A. Huckabay, S. M. Wildgen, R. C. Dunn, *Biosensors and Bioelectronics* **2013**, *45* 223.

[238] R. Duan, Y. Li, H. Li, J. Yang, *Optics Express* **2019**, *27* 35427.

[239] M. Schubert, L. Woolfson, I. R. M. Barnard, A. M. Dorward, B. Casement, A. Morton, G. B. Robertson, P. L. Appleton, G. B. Miles, C. S. Tucker, S. J. Pitt, M. C. Gather, *Nature Photonics* **2020**, *14* 452.

[240] Y. Wang, L. Zhao, A. Xu, L. Wang, L. Zhang, S. Liu, Y. Liu, H. Li, *Sensors and Actuators, B: Chemical* **2018**, *258* 1090.

[241] J. Zhu, S¸ahin Kaya Ozdemir, L. He, D.-R. Chen, L. Yang, *Optics Express* **2011**, *19* 16195.

[242] S. Frustaci, F. Vollmer, *Current Opinion in Chemical Biology* **2019**, *51* 66.

[243] A. Fran¸cois, T. Reynolds, T. M. Monro, *Sensors (Switzerland)* **2015**, *15* 1168.

[244] T. Beck, M. Mai, T. Grossmann, T. Wienhold, M. Hauser, T. Mappes, H. Kalt, *Applied Physics Letters* **2013**, *102*.

[245] E. Ozgur, P. Toren, O. Aktas, E. Huseyinoglu, M. Bayindir, *Scientific Reports* **2015**, *5* 1.

[246] F. Wang, M. Anderson, M. T. Bernards, H. K. Hunt, *Sensors (Switzerland)* **2015**, *15* 18040.

[247] E. A. Tcherniavskaia, V. A. Saetchnikov, *Journal of Applied Spectroscopy* **2010**, *77* 692.

[248] A. K. Ajad, M. J. Islam, M. R. Kaysir, J. Atai, *Optik* **2021**, *226*.

[249] W. Kim, S. K. Ozdemir, J. Zhu, F. Monifi, C. Coban, L. Yang, *Optics Express* **2012**, *20* 29426.

[250] R. Duan, Y. Li, B. Shi, H. Li, J. Yang, *Talanta* **2020**, *209* 1.

[251] P. Toren, E. Ozgur, M. Bayindir, *Analytical Chemistry* **2015**, *87* 10920.

[252] F. Vollmer, S. Arnold, D. Braun, I. Teraoka, A. Libchaber, *Biophysical Journal* **2003**, *85* 1974.

[253] E. Nuhiji, P. Mulvaney, *Small* **2007**, *3* 1408.

[254] H.-C. Ren, F. Vollmer, S. Arnold, A. Libchaber,*Optics Express* **2007**, *15* 17410.

[255] M. Charlebois, A. Paquet, L. S. Verret, K. Boissinot, M. Boissinot, M. G. Bergeron, C. N. Allen, *IEEE Sensors Journal* **2013**, *13* 229.

[256] H. Ghali, P. Bianucci, Y. A. Peter, *Sensing and Bio-Sensing Research* **2017**, *13* 9.

[257] M. E. Anderson, E. C. O'Brien, E. N. Grayek, J. K. Hermansen, H. K. Hunt, *Biosensors* **2015**, *5* 562.

[258] S. T. Hsieh, J. E. Cheeney, X. Ding, N. V. Myung, E. D. Haberer, *Sensors and Actuators B: Chemical* **2022**, *367* 132062.

[259] H.-J. Chen, *Journal of Experimental and Theoretical Physics* **2018**, *126* 712.

[260] J. Chan, T. Thiessen, S. Lane, P. West, K. Gardner, A. Fran¸cois, A. Meldrum, *IEEE Sensors Journal* **2015**, *15* 3467.

[261] L. He, S¸ahin Kaya Ozdemir, J. Zhu, W. Kim, L. Yang, *Nature Nanotechnology* **2011**, *6* 428.

[262] M. Himmelhaus, A. Francois, *Biosensors and Bioelectronics* **2009**, *25* 418.

[263] N. M. Hanumegowda, I. M. White, H. Oveys, X. Fan, *Sensor Letters* **2005**, *3* 315.

[264] P. Toren, E. Ozgur, M. Bayindir, *ACS Sensors* **2018**, *3* 352.

[265] K. A. Wilson, C. A. Finch, P. Anderson, F. Vollmer, J. J. Hickman, *Biomaterials* **2015**, *38* 86.

[266] M. Noto, F. Vollmer, D. Keng, I. Teraoka, S. Arnold, *Optics Letters* **2005**, *30* 510.

[267] R. Duan, X. Hao, Y. Li, H. Li, *Sensors and Actuators, B: Chemical* **2020**, *308* 127672.

[268] R. Duan, Y. Li, Y. He, Y. Yuan, H. Li, *Analyst* **2020**, *145* 7595.

[269] V. R. Dantham, S. Holler, C. Barbre, D. Keng, V. Kolchenko, S. Arnold, *Nano Letters* **2013**, *13* 3347.

[270] Z. Ma, M. Xu, S. Zhou, W. Shan, D. Zhou, Y. Yan, W. Sun, Y. Liu, *Optics Letters* **2022**, *47* 381.

[271] J. Lu, C. Xu, H. Nan, Q. Zhu, F. Qin, A. G. Manohari, M. Wei, Z. Zhu, Z. Shi, Z. Ni, *Applied Physics Letters* **2016**, *109*.

[272] D. Keng, S. R. McAnanama, I. Teraoka, S. Arnold, *Applied Physics Letters* **2007**, *91* 2007.

[273] J. Su, A. F. G. Goldberg, B. M. Stoltz, *Light: Science and Applications* **2016**, *5* 2.

[274] S. Arnold, R. Ramjit, D. Keng, V. Kolchenko, I. Teraoka, *Faraday Discussions* **2007**, *137* 65.

[275] M. A. Santiago-Cordoba, S. V. Boriskina, F. Vollmer, M. C. Demirel, *Applied Physics Letters* **2011**, *99* 1.

[276] S. Arnold, V. R. Dantham, C. Barbre, B. A. Garetz, X. Fan, *Optics Express* **2012**, *20* 26147.

[277] J. D. Swaim, J. Knittel, W. P. Bowen, *Applied Physics Letters* **2011**, *99* 3.

[278] S. I. Shopova, R. Rajmangal, S. Holler, S. Arnold, *Applied Physics Letters* **2011**, *98*.

[279] M. A. Santiago-Cordoba, M. Cetinkaya, S. V. Boriskina, F. Vollmer, M. C. Demirel, *Journal of Biophotonics* **2012**, *5* 629.

[280] E. Arbabi, S. M. Kamali, S. Arnold, L. L. Goddard, *Applied Physics Letters* **2014**, *105* 1.

[281] H. Nadgaran, M. A. Garaei, *Journal of Applied Physics* **2015**, *118* 1.

[282] V. R. Dantham, S. Holler, V. Kolchenko, Z. Wan, S. Arnold, *Applied Physics Letters* **2012**, *101*.

[283] I. Brice, K. Grundsteins, A. Atvars, J. Alnis, R. Viter, A. Ramanavicius, *Sensors and Actuators, B: Chemical* **2020**, *318* 128004.

[284] M. D. Baaske, M. R. Foreman, F. Vollmer, *Nature Nanotechnology* **2014**, *9* 933.

[285] E. Kim, M. D. Baaske, I. Schuldes, P. S. Wilsch, F. Vollmer, *Science Advances* **2017**, *3*.

[286] E. Kim, M. D. Baaske, F. Vollmer, *Advanced Materials* **2016**, *28* 9941.

[287] N. M. Hanumegowda, I. M. White, X. Fan, *Sensors and Actuators, B: Chemical* **2006**, *120* 207.

[288] F. Wu, Y. Wu, Z. Niu, F. Vollmer, *Sensors (Switzerland)* **2016**, *16*.

[289] H. Chen, J. E. Saunders, S. Borjian, X. Wu, C. M. Crudden, D.-X. Xu, H.-P. Loock, *Advanced Sustainable Systems* **2019**, *3* 1800084.

[290] L. Fu, Q. Lu, X. Liu, X. Chen, X. Wu, S. Xie, *Talanta* **2020**, *213* 120815.

[291] R. Duan, Y. Li, H. Li, J. Yang, *Biomedical Optics Express* **2019**, *10* 6073.

[292] C. C. Chiang, J. C. Chao, *Journal of Nanomaterials* **2013**, *2013*.

[293] A. Schweinsberg, S. Hocd´e, N. N. Lepeshkin, R. W. Boyd, C. Chase, J. E. Fajardo, *Sensors and Actuators, B: Chemical* **2007**, *123* 727.

[294] M. Agarwal, I. Teraoka, *Analytical Chemistry* **2015**, *87* 10600.

[295] P. Song, C. Chen, J. Qu, P. Ou, M. H. T. Dastjerdi, Z. Mi, J. Song, X. Liu, *Nanotechnology* **2018**, *29*.

[296] R. Duan, Y. Li, Y. Yuan, L. Liu, H. Li, *Liquid Crystals* **2020**, *47* 1708.

[297] A. K. Mallik, G. Farrell, D. Liu, V. Kavungal, Q. Wu, Y. Semenova, *Scientific Reports* **2018**, *8* 2.

[298] N. Lin, L. Jiang, S. Wang, L. Yuan, H. Xiao, Y. Lu, H. Tsai, *Applied Optics* **2010**, *49* 6463.

[299] Q. Lu, X. Chen, L. Fu, S. Xie, X. Wu, *Nanomaterials* **2019**, *9* 1.

[300] R. I. Stoian, B. K. Lavine, A. T. Rosenberger, *Talanta* **2019**, *194* 585.

[301] M. Humar, I. Muˇseviˇc, *Optics Express* **2011**, *19* 19836.

[302] X. Feng, G. Zhang, L. K. Chin, A. Q. Liu, B. Liedberg, *ACS Sensors* **2017**, *2* 955.

[303] G. Yang, I. M. White, X. Fan, *Sensors and Actuators, B: Chemical* **2008**, *133* 105.

[304] S. Lee, J. H. Moon, G. Kim, *Analytical Methods* **2012**, *4* 1041.

[305] Y. Sun, X. Fan, *Optics Express* **2008**, *16* 10254.

[306] V. M. N. Passaro, F. Dell'Olio, F. D. Leonardis, *Sensors* **2007**, *7* 2741.

[307] N. A. Yebo, S. P. Sree, E. Levrau, C. Detavernier, Z. Hens, J. A. Martens, R. Baets, *Optics Express* **2012**, *20* 11855.

[308] B. Yao, C. Yu, Y. Wu, S. W. Huang, H. Wu, Y. Gong, Y. Chen, Y. Li, C. W. Wong, X. Fan, Y. Rao, *Nano Letters* **2017**, *17* 4996.

[309] M. Eryu¨rek, Y. Karadag, N. Tas, N. Kılınc, *Sensors and Actuators, B* **2015**, *212* 78.

[310] R. I. Stoian, K. V. Bui, A. T. Rosenberger, *Journal of Optics (United Kingdom)* **2015**, *17* 125011.

[311] Y. Sun, S. I. Shopova, G. Frye-Mason, X. Fan, *Optics Letters* **2008**, *33* 788.

[312] N. A. Yebo, P. Lommens, Z. Hens, R. Baets, *Optics Express* **2010**, *18* 11859.

[313] V. D. Ta, R. Chen, D. M. Nguyen, H. D. Sun, *Applied Physics Letters* **2013**, *102* 4.

[314] M. Gao, C. Wei, X. Lin, Y. Liu, F. Hu, Y. S. Zhao, *Chemical Communications* **2017**, *53* 3102.

[315] Y. Sun, J. Liu, G. Frye-Mason, S. J. Ja, A. K. Thompson, X. Fan, *Analyst* **2009**, *134* 1386.

[316] K. Scholten, X. Fan, E. T. Zellers, *Lab on a Chip* **2014**, *14* 3873.

[317] S. Panich, K. A. Wilson, P. Nuttall, C. K. Wood, T. Albrecht, J. B. Edel, *Analytical Chemistry* **2014**, *86* 6299.

[318] F. Gu, L. Zhang, Y. Zhu, H. Zeng, *Laser and Photonics Reviews* **2015**, *9* 682.

[319] T. H. Stievater, M. W. Pruessner, D. Park, W. S. Rabinovich, R. A. McGill, D. A. Kozak, R. Furstenberg, S. A. Holmstrom, J. B. Khurgin, *Optics Letters* **2014**, *39* 969.

[320] H. Wang, L. Yuan, C. W. Kim, X. Lan, J. Huang, Y. Ma, H. Xiao, *Sensors and Actuators, B: Chemical* **2015**, *216* 332.

[321] K. Scholten, W. R. Collin, X. Fan, E. T. Zellers, *Nanoscale* **2015**, *7* 9282.

[322] R. Ahmed, A. A. Rifat, A. K. Yetisen, M. S. Salem, S. H. Yun, H. Butt, *RSC Advances* **2016**, *6* 56127.

[323] S. I. Shopova, I. M. White, Y. Sun, H. Zhu, X. Fan, G. Frye-mason, A. Thompson, S. jyh Ja, *Analytical Chemistry* **2008**, *80* 2232.

[324] M. Gregor, C. Pyrlik, R. Henze, A. Wicht, A. Peters, O. Benson, *Applied Physics Letters* **2010**, *96* 3.

[325] H. Wang, L. Yuan, C.-W. Kim, Q. Han, T. Wei, X. Lan, H. Xiao, *Optics Letters* **2012**, *37* 94.

[326] D. Zhivotkov, D. Risti´c, S. T. Thalakkal, V. Gaˇspari´c, E. Romanova, M. Ivanda, *Optical Materials* **2022**, *129* 112544.

[327] S. H. Huang, S. Sheth, E. Jain, X. Jiang, S. P. Zustiak, L. Yang, *Optics Express* **2018**, *26* 51.

[328] M. D. Baaske, F. Vollmer, *Nature Photonics* **2016**, *10* 733.

[329] S. Subramanian, H. B. L. Jones, S. Frustaci, S. Winter, M. W. V. D. Kamp, V. L. Arcus, C. R. Pudney, F. Vollmer, *ACS Applied Nano Materials* **2021**, *4*.

[330] S. Vincent, S. Subramanian, F. Vollmer, *Nature Communications* **2020**, *11*.

[331] O. Gaathon, J. Culic-Viskota, M. Mihnev, I. Teraoka, S. Arnold, *Applied Physics Letters* **2006**, *89* 3.

[332] A. Fran¸cois, N. Riesen, K. Gardner, T. M. Monro, A. Meldrum, *Optics Express* **2016**, *24* 12466.

[333] L. Wan, H. Chandrahalim, J. Zhou, Z. Li, C. Chen, S. Cho, H. Zhang, T. Mei, H. Tian, Y. Oki, N. Nishimura, X. Fan, L. J. Guo, *Optics Express* **2018**, *26* 5800.

[334] C. He, H. Sun, J. Mo, C. Yang, G. Feng, H. Zhou, S. Zhou, *Laser Physics* **2018**, *28* 76202.

[335] L. Xu, X. Jiang, G. Zhao, D. Ma, H. Tao, Z. Liu, F. G. Omenetto, L. Yang, *Optics Express* **2016**, *24* 20825.

[336] C. W. Wu, K. C. Liu, C. C. Chiang, *IEEE Sensors Journal* **2017**, *17* 5444.

[337] M. Manzo, T. Ioppolo, U. K. Ayaz, V. Lapenna, M. V. Otu¨gen, *Review of Scientific Instruments* **2012**, *83* 1.

[338] M. Manzo, T. Ioppolo, *Optics Letters* **2015**, *40* 2257.

[339] K. Soler-Carracedo, I. R. Martin, M. Runowski, L. L. Mart´ın, F. Lahoz, A. D. Lozano-Gorr´ın, F. Paz-Buclatin, *Advanced Optical Materials* **2020**, *8* 1.

[340] K. Soler-Carracedo, P. Est´evez-Alonso, I. R. Martin, S. Rios, *Journal of Luminescence* **2021**, *238*.

[341] M. Eryu¨rek, Z. Tasdemir, Y. Karadag, S. Anand, N. Kilinc, B. E. Alaca, A. Kiraz, *Sensors and Actuators B: Chemical* **2017**, *242* 1115.

[342] Z. Liu, W. Liu, C. Hu, Y. Zhang, X. Yang, J. Zhang, J. Yang, L. Yuan, *Optics Express* **2019**, *27* 21946.

[343] M. H. Jali, H. R. A. Rahim, M. A. M. Johari, A. Ahmad, H. H. M. Yusof, S. H. Johari, M. F. Baharom, S. Thokchom, K. Dimyati, S. W. Harun, *Optics & Laser Technology* **2021**, *143* 107356.

[344] H. A. Zain, M. Batumalay, M. A. M. Johari, K. Dimyati, S. W. Harun, *Optoelectronics Letters* **2021**, *17* 328.

[345] T. Ioppolo, U. Ayaz, M. V. Otu¨gen,¨ *Optics Express* **2009**, *17* 16465.

[346] K. H. Kim, W. Luo, C. Zhang, C. Tian, L. J. Guo, X. Wang, X. Fan, *Scientific Reports* **2017**, *7* 1.

[347] S. Zhu, L. Shi, N. Liu, X. Xu, X. Zhang, *Microfluidics and Nanofluidics* **2017**, *21* 1.

[348] V. D. Ta, R. Chen, L. Ma, Y. J. Ying, H. D. Sun, *Laser and Photonics Reviews* **2013**, *7* 133.

[349] T. Ma, J. Yuan, L. Sun, Z. Kang, B. Yan, X. Sang, K. Wang, Q. Wu, H. Liu, J. Gao, C. Yu, *IEEE Photonics Journal* **2017**, *9* 1.

[350] Z. Ding, P. Liu, J. Chen, D. Dai, Y. Shi, *Optics Express* **2019**, *27* 28649.

[351] Z. Xu, T. Zhai, X. Shi, J. Tong, X. Wang, J. Deng, *ACS Applied Materials and Interfaces* **2021**, *13* 19324.

[352] N. M. Hanumegowda, C. J. Stica, B. C. Patel, I. White, X. Fan, *Applied Physics Letters* **2005**, *87* 1.

[353] H. Zhu, I. M. White, J. D. Suter, M. Zourob, X. Fan, *Analytical Chemistry* **2007**, *79* 930.

[354] I. M. White, H. Oveys, X. Fan, *Optics Letters* **2006**, *31* 1319.

[355] I. M. White, X. Fan, *Optics Express* **2008**, *16* 1020.

[356] T. Tang, X. Wu, L. Liu, L. Xu, *Applied Optics* **2016**, *55* 395.

[357] Y. Q. Kang, A. Fran¸cois, N. Riesen, T. M. Monro, *Sensors (Switzerland)* **2018**, *18* 1.

[358] L. Ren, X. Wu, M. Li, X. Zhang, L. Liu, L. Xu, *Optics Letters* **2012**, *37* 3873.

[359] L. Ren, X. Zhang, X. Guo, H. Wang, X. Wu, *IEEE Photonics Technology Letters* **2017**, *29* 639.

[360] R. Chen, V. D. Ta, H. Sun, *ACS Photonics* **2014**, *1* 11.

[361] N. Zhou, P. Wang, Z. X. Shi, Y. X. Gao, Y. X. Yang, Y. P. Wang, Y. Xie, D. W. Cai, X. Guo, L. Zhang, J. R. Qiu, L. M. Tong, *Optics Express* **2019**, *27* 8180.

[362] T. Y. Kang, W. Lee, H. Ahn, D. M. Shin, C. S. Kim, J. W. Oh, D. Kim, K. Kim, *Scientific Reports* **2017**, *7* 1.

[363] G. Guan, S. Arnold, M. V. Otugen, *AIAA Journal* **2006**, *44* 2385.

[364] C. H. Dong, L. He, Y. F. Xiao, V. R. Gaddam, S. K. Ozdemir, Z. F. Han, G. C. Guo, L. Yang, *Applied Physics Letters* **2009**, *94* 3.

[365] B. B. Li, Q. Y. Wang, Y. F. Xiao, X. F. Jiang, Y. Li, L. Xiao, Q. Gong, *Applied Physics Letters* **2010**, *96* 2008.

[366] T. Ioppolo, M. Manzo, *Applied Optics* **2014**, *53* 5065.

[367] R. Fan, Z. Mu, J. Li, *Journal of Physics and Chemistry of Solids* **2019**, *129* 307.

[368] X. Xu, X. Jiang, G. Zhao, L. Yang, *Optics Express* **2016**, *24* 25905.

[369] Z. Wang, A. K. Mallik, F. Wei, Z. Wang, A. Rout, Q. Wu, Y. Semenova, *Optics Express* **2021**, *29* 23569.

[370] L. Zhao, Y. Wang, Y. Yuan, Y. Liu, S. Liu, W. Sun, J. Yang, H. Li, *Optics Communications* **2017**, *402* 181.

[371] R. Luo, H. Jiang, H. Liang, Q. Lin, *Optics Letters* **2017**, *42* 1281.

[372] S. Bhattacharya, A. V. Veluthandath, G. S. Murugan, P. B. Bisht, *Journal of Luminescence* **2020**, *221* 1.

[373] T. Ioppolo, M. Kozhevnikov, V. Stepaniuk, M. V. Otu¨gen, V. Sheverev,¨ *Applied Optics* **2008**, *47* 3009.

[374] A. H. Zamanian, T. Ioppolo, *Applied Optics* **2015**, *54* 7124.

[375] A. Bianchetti, A. Federico, S. Vincent, S. Subramanian, F. Vollmer, *Optics Communications* **2017**, *394* 152.

[376] B. Bhola, P. Nosovitskiy, H. Mahalingam, W. H. Steier, *IEEE Sensors Journal* **2009**, *9* 740.

[377] S. Mehrabani, P. Kwong, M. Gupta, A. M. Armani, *Applied Physics Letters* **2013**, *102* 241101.

[378] Q. lan Huang, H. liang Xu, M. tian Li, Z. shan Hou, C. Lv, X. peng Zhan, H. long Li, H. Xia, *Journal of Lightwave Technology* **2018**, *36* 819.

[379] L. Labrador-Paez, K. Soler-Carracedo, M. Hernandez-Rodriguez, I. R. Martin, T. Carmon, L. L. Martin, *Optics Express* **2017**, *25* 1165.

[380] D. Armani, B. Min, A. Martin, K. J. Vahala, *Applied Physics Letters* **2004**, *85* 5439.

[381] N. Riesen, Z. Q. Peterkovic, B. Guan, A. Franc¸ois, D. G. Lancaster, C. Priest, Caged-sphere optofluidic sensors: Whispering gallery resonators in wicking microfluidics, **2022**.

[382] Y. W. Hu, B. B. Li, Y. X. Liu, Y. F. Xiao, Q. Gong, *Optics Communications* **2013**, *291* 380.

[383] W. Zhang, O. J. F. Martin, *ACS Photonics* **2015**, *2* 144.

[384] M. Zhang, G. Wu, D. Chen, B. Liu, *Journal of Optics* **2019**, *48* 308.

[385] H. Fan, C. Xia, L. Fan, L. Wang, M. Shen, *Optics Communications* **2018**, *410* 668.

[386] M. Zhang, B. Liu, G. Wu, D. Chen, *Optics Communications* **2016**, *380* 6.

[387] D. Urbonas, A. Balˇcytis, M. Gabalis, K. Vaˇskeviˇcius, G. Naujokaite˙, S. Juodkazis, R. Petruˇskeviˇcius, *Optics Letters* **2015**, *40* 2977.

[388] X. Sun, D. Dai, L. Thyl´en, L. Wosinski, *Photonics* **2015**, *2* 1116.

[389] C. Wang, X. Zhang, J. Ma, K. Xie, J. Zhang, Z. Hu, *Photonic Sensors* **2022**, *12* 220307.

[390] S. H. Nam, S. Yin, *IEEE Photonics Technology Letters* **2005**, *17* 2391.

[391] Z. Liu, L. Liu, Z. Zhu, Y. Zhang, Y. Wei, X. Zhang, E. Zhao, Y. Zhang, J. Yang, L. Yuan, *Optics Letters* **2016**, *41* 4649.

[392] M. S. Nawrocka, T. Liu, X. Wang, R. R. Panepucci, *Applied Physics Letters* **2006**, *89*.

[393] R. Madugani, Y. Yang, V. H. Le, J. M. Ward, S. N. Chormaic, *IEEE Photonics Technology Letters* **2016**, *28* 1134.

[394] Y. Yang, S. Saurabh, J. M. Ward, S. N. Chormaic, *Optics Express* **2016**, *24* 294.

[395] J. Yan, D. N. Wang, Y. Ge, Y. Guo, B. Xu, *Journal of Lightwave Technology* **2022**, *40* 2651.

[396] E. Yacoby, Y. Meshorer, Y. London, *Optics & Laser Technology* **2022**, *151* 108019.

[397] C. Zhang, S. Pu, Z. Hao, B. Wang, M. Yuan, Y. Zhang, Magnetic field sensing based on whispering gallery mode with nanostructured magnetic fluid-infiltrated photonic crystal fiber, **2022**.

[398] B.-B. Li, G. Brawley, H. Greenall, S. Forstner, E. Sheridan, H. RubinszteinDunlop, W. P. Bowen, *Photonics Research* **2020**, *8* 1064.

[399] B. B. Li, D. Bulla, V. Prakash, S. Forstner, A. Dehghan-Manshadi, H. Rubinsztein-Dunlop, S. Foster, W. P. Bowen, *APL Photonics* **2018**, *3* 1.

[400] J. Zhu, G. Zhao, I. Savukov, L. Yang, *Scientific Reports* **2017**, 1–5.

[401] S. Forstner, E. Sheridan, J. Knittel, C. L. Humphreys, G. A. Brawley, H. Rubinsztein-Dunlop, W. P. Bowen, *Advanced Materials* **2014**, *26* 6348.

[402] M. Chamanzar, M. Soltani, B. Momeni, S. Yegnanarayanan, A. Adibi, *Applied Physics B: Lasers and Optics* **2010**, *101* 263.

[403] X. Ma, S. Fan, H. Wei, Z. Zuo, S. Krishnaswamy, J. Fang, *Optics Express* **2019**, *27* 33051.

[404] Y. Yin, T. Nie, M. Ding, *IEEE Sensors Journal* **2020**, *20* 9871.

[405] M. S. Kwon, B. Ku, Y. Kim, *Scientific Reports* **2016**, *6* 1.

[406] V. B. Braginsky, M. L. Gorodetsky, V. S. Ilchenko, *Physics Letters A* **1989**, *137* 393.

[407] H. Parsamyan, K. Sahakyan, K. Nerkararyan, *Journal of Physics D: Applied Physics* **2022**, *55* 165102.

[408] T. J. Kippenberg, S. M. Spillane, D. K. Armani, K. J. Vahala, *Optics Letters* **2004**, *29* 1224.

[409] P. Del'Haye, S. A. Diddams, S. B. Papp, *Applied Physics Letters* **2013**, *102*.

[410] S. M. Spillane, T. J. Kippenberg, K. J. Vahala, *Nature* **2002**, *415* 621.

[411] S. Kumar, S. K. Biswas, *Journal of the Optical Society of America B* **2016**, *33* 1677.

[412] S. K. Biswas, S. Kumar, *Optics Express* **2015**, *23* 26738.

[413] J. H. Chen, X. Shen, S. J. Tang, Q. T. Cao, Q. Gong, Y. F. Xiao, *Physical Review Letters* **2019**, *123* 173902.

[414] J. Moore, M. Tomes, T. Carmon, M. Jarrahi, *Optics Express* **2011**, *19* 24139.

[415] V. S. Ilchenko, A. B. Matsko, A. A. Savchenkov, L. Maleki, *Journal of the Optical Society of America B* **2003**, *20* 1304.

[416] D. Haertle, *Journal of Optics A: Pure and Applied Optics* **2010**, *12*.

[417] Y. Zhang, H. Zhou, S. W. Liu, Z. R. Tian, M. Xiao, *Nano Letters* **2009**, *9* 2109.

[418] J. U. Fu¨rst, D. V. Strekalov, D. Elser, M. Lassen, U. L. Andersen, C. Marquardt, G. Leuchs, *Physical Review Letters* **2010**, *104* 1.

[419] D. Farnesi, A. Barucci, G. C. Righini, S. Berneschi, S. Soria, G. N. Conti, *Physical Review Letters* **2014**, *112* 2.

[420] L. S. Trainor, F. Sedlmeir, C. Peuntinger, H. G. L. Schwefel, *Physical Review Applied* **2018**, *9* 24007.

[421] J. L. Dominguez-Juarez, G. Kozyreff, J. Martorell, *Nature Communications* **2011**, *2*.

[422] Y. Feng, Y. Zheng, F. Zhang, J. Yang, T. Qin, W. Wan, *Applied Physics Letters* **2019**, *114*.

[423] M. G. Suh, Q. F. Yang, K. Y. Yang, X. Yi, K. J. Vahala, *Science* **2016**, *354* 600.

[424] Q. Ai, L. Gui, D. Paone, B. Metzger, M. Mayer, K. Weber, A. Fery, H. Giessen, *Nano Letters* **2018**, *18* 5576.

[425] R. Castro-Beltr´an, V. M. Diep, S. Soltani, E. Gungor, A. M. Armani, *ACS Photonics* **2017**, *4* 2828.

[426] S. Soltani, V. M. Diep, R. Zeto, A. M. Armani, *ACS Photonics* **2018**, *5* 3550.

[427] W. Liang, A. A. Savchenkov, V. S. Ilchenko, D. Eliyahu, D. Seidel, A. B. Matsko, L. Maleki, *Optics Letters* **2014**, *39* 2920.

[428] V. Brasch, M. Geiselmann, T. Herr, G. Lihachev, M. H. P. Pfeiffer, M. L. Gorodetsky, T. J. Kippenberg, *Science* **2016**, *351* 357.

[429] P. Marin-Palomo, J. N. Kemal, M. Karpov, A. Kordts, J. Pfeifle, M. H. P. Pfeiffer, P. Trocha, S. Wolf, V. Brasch, M. H. Anderson, R. Rosenberger, K. Vijayan, W. Freude, T. J. Kippenberg, C. Koos, *Nature* **2017**, *546* 274.

[430] V. Brasch, E. Lucas, J. D. Jost, M. Geiselmann, T. J. Kippenberg, *Light: Science and Applications* **2017**, *6* 16202.

[431] J. R. Stone, T. C. Briles, T. E. Drake, D. T. Spencer, D. R. Carlson, S. A. Diddams, S. B. Papp, *Physical Review Letters* **2018**, *121* 63902.

[432] P. Trocha, M. Karpov, D. Ganin, M. H. P. Pfeiffer, A. Kordts, S. Wolf, J. Krockenberger, P. Marin-Palomo, C. Weimann, S. Randel, W. Freude, T. J. Kippenberg, C. Koos, *Science* **2018**, *359* 887.

[433] J. Liu, A. S. Raja, M. Karpov, B. Ghadiani, M. H. P. Pfeiffer, A. Lukashchuk, N. J. Engelsen, H. Guo, M. Zervas, T. J. Kippenberg, *Optica* **2018**, *5* 1347.

[434] L. Stern, J. R. Stone, S. Kang, D. C. Cole, M. G. Suh, C. Fredrick, Z. Newman, K. Vahala, J. Kitching, S. A. Diddams, S. B. Papp, *Science* **2020**, *6* 1.

[435] Q. F. Yang, B. Shen, H. Wang, M. Tran, Z. Zhang, K. Y. Yang, L. Wu, C. Bao, J. Bowers, A. Yariv, K. Vahala, *Science* **2019**, *363* 965.

[436] M. Yu, Y. Okawachi, A. G. Griffith, M. Lipson, A. L. Gaeta, *Optics Letters* **2017**, *42* 4442.

[437] Y. Pu, R. Grange, C. L. Hsieh, D. Psaltis, *Physical Review Letters* **2010**, *104* 1.

[438] M. L. Ren, W. Liu, C. O. Aspetti, L. Sun, R. Agarwal, *Nature Communications* **2014**, *5*.

[439] N. Weber, S. P. Hoffmann, M. Albert, T. Zentgraf, C. Meier, *Journal of Applied Physics* **2018**, *123*.

[440] F. Ren, H. Takashima, Y. Tanaka, H. Fujiwara, K. Sasaki, *Optics Express* **2015**, *23* 21730.

[441] P. L. Stiles, J. A. Dieringer, N. C. Shah, R. P. V. Duyne, *Annual Review of Analytical Chemistry* **2008**, *1* 601.

[442] R. M. Sto¨ckle, Y. D. Suh, V. Deckert, R. Zenobi, *Chemical Physics Letters* **2000**, *318* 131.

[443] C. Thomsen, S. Reich, *Physical Review Letters* **2000**, *85* 5214.

[444] V. R. Dantham, P. B. Bisht, C. K. R. Namboodiri, *Journal of Applied Physics* **2011**, *109* 30.

[445] R. J. Hopkins, R. Symes, R. M. Sayer, J. P. Reid, *Chemical Physics Letters* **2003**, *380* 665.

[446] R. Symes, R. J. J. Gilham, R. M. Sayer, J. P. Reid, *Physical Chemistry Chemical Physics* **2005**, *7* 1414.

[447] R. Symes, J. P. Reid, *Physical Chemistry Chemical Physics* **2006**, *8* 293.

[448] T. C. Preston, J. P. Reid, *Journal of the Optical Society of America B* **2013**, *30* 2113.

[449] A. Rafferty, T. C. Preston, *Physical Chemistry Chemical Physics* **2018**, *20* 17038.

[450] L. K. Ausman, G. C. Schatz, *Journal of Chemical Physics* **2008**, *129*.

[451] M. S. Anderson, *Applied Physics Letters* **2010**, *97* 8.

[452] S. H. Huang, X. Jiang, B. Peng, C. Janisch, A. Cocking, S¸ahin Kaya Ozdemir,¨ Z. Liu, L. Yang, *Photonics Research* **2018**, *6* 346.

[453] T. Ogura, *Scientific Reports* **2022**, *12* 5346.

[454] V. R. Dantham, P. B. Bisht, P. S. Dobal, *Journal of Raman Spectroscopy* **2011**, *42* 1373.

[455] P. A. Dmitriev, D. G. Baranov, V. A. Milichko, S. V. Makarov, I. S. Mukhin, A. K. Samusev, A. E. Krasnok, P. A. Belov, Y. S. Kivshar, *Nanoscale* **2016**, *8* 9721.

[456] A. V. Veluthandath, P. B. Bisht, *Journal of Luminescence* **2017**, *187* 255.

[457] I. Alessandri, *Journal of the American Chemical Society* **2013**, *135* 5541.

[458] N. Bontempi, L. Carletti, C. D. Angelis, I. Alessandri, *Nanoscale* **2016**, *8* 3226.

[459] B. Vennes, A. Rafferty, T. C. Preston, *Journal of the Optical Society of America B* **2021**, *38* 893.

[460] T. Inoue, J. Y. Kohno, *Chemistry Letters* **2021**, *50* 68.

[461] H. C. Lai, Y. J. Wang, C. A. Dai, C. H. Hsueh, S. J. Wang, J. H. Li, *IEEE Journal of Selected Topics in Quantum Electronics* **2021**, *27*.

[462] C. Xing, Y. Yan, C. Feng, J. Xu, P. Dong, W. Guan, Y. Zeng, Y. Zhao, Y. Jiang, *ACS Applied Materials and Interfaces* **2017**, *9* 32896.

[463] J. Zhang, J. Li, S. Tang, Y. Fang, J. Wang, G. Huang, R. Liu, L. Zheng, X. Cui, Y. Mei, *Scientific Reports* **2015**, *5* 1.

[464] J. F. Cardenas, *Journal of Raman Spectroscopy* **2013**, *44* 540.

[465] I. M. White, X. Fan, *Chemical and Biological Sensors for Industrial and Environmental Security* **2005**, *5994* 59940G.

[466] I. M. White, J. Gohring, X. Fan, *Optics Express* **2007**, *15* 17433.

[467] M. Wang, Y. Yan, Y. Mi, Y. Jiang, *Journal of Raman Spectroscopy* **2022**, *53* 1238.

[468] J. Qian, Z. Zhu, J. Yuan, Y. Liu, B. Liu, X. Zhao, L. Jiang, *Nanoscale Advances* **2020**, *2* 4682.

[469] C. R. Simovski, *Physical Review B - Condensed Matter and Materials Physics* **2009**, *79* 1.

[470] S. J. Oldenburg, S. L. Westcott, R. D. Averitt, N. J. Halas, *Journal of Chemical Physics* **1999**, *111* 4729.

[471] J. B. Jackson, S. L. Westcott, L. R. Hirsch, J. L. West, N. J. Halas, *Applied Physics Letters* **2003**, *82* 257.

[472] H. Yang, B. Q. Li, X. Jiang, J. Shao, *Nanoscale* **2019**, *11* 13484.

[473] M. Runowski, I. R. Mart´ın, V. N. Sigaev, V. I. Savinkov, G. Y. Shakhgildyan, S. Lis, *Journal of Rare Earths* **2019**, *37* 1152.

[474] Y. Mi, Y. Yan, M. Wang, L. Yang, J. He, Y. Jiang, *Nanophotonics* **2022**.

[475] S. Hou, J. Wang, C. Wang, Y. Yuan, X. Zhang, Y. Huang, S. Yan, *Spectrochimica Acta Part A: Molecular and Biomolecular Spectroscopy* **2022**, *264* 120252.